\documentclass[trackchanges,preprint2]{aastex7}
\usepackage{amsmath}
\usepackage{soul}
\usepackage{longtable} % Uzun tablolar için gerekli paket
\usepackage{caption}   % Daha gelişmiş başlık kontrolü
\usepackage{subcaption}
\usepackage{graphicx}
\usepackage{float}
\usepackage{booktabs} % daha şık çizgiler için (opsiyonel)
\usepackage{hyperref}
\usepackage{gensymb}
\usepackage{natbib}
\definecolor{red}{rgb}{1.0, 0.0, 0.0}
\shorttitle{TTVs in HAT-P-16, TOI-1516, TOI-2046}
\shortauthors{Sonbas et al.}
\begin{document}

\title{A Combined TESS and Ground-based Study of Transit Timing Variations in HAT-P-16, TOI-1516 and TOI-2046 Systems}

\author[orcid=0000-0002-6909-192X,sname='Sonbas']{E. Sonbas}
\affiliation{Department of Physics, Ad{\i}yaman University, Ad{\i}yaman 02040, T\"{u}rkiye}
\affiliation{Department of Physics, The George Washington University, Washington, DC 20052, USA}
\email[show]{edasonbas@gmail.com}  

\author[orcid=0000-0003-0631-1961,sname='Kaplan']{K. Kaplan} 
\affiliation{Istasyon Gaziantep, Gaziantep Metropolitan Municipality, TR-27560, T\"{u}rkiye}
\email{}

\author[orcid=0000-0002-3263-9680,sname='Tanriver']{M. Tanriver}
\affiliation{Department of Astronomy and Space Science, Faculty of Science, Erciyes University, Kayseri TR-38039, T\"{u}rkiye}
\affiliation{Erciyes University, Astronomy and Space Science Observatory Application and Research Center, TR-38039, Kayseri, T{\"u}rkiye}
\email{}

\author[orcid=0000-0002-9314-0648,sname='Keskin']{A. Keskin}
\affiliation{Department of Astronomy and Space Science, Faculty of Science, Erciyes University, Kayseri TR-38039, T\"{u}rkiye}
\email{}

\author[orcid=0000-0002-5422-4873,sname='Dhuga']{K. Dhuga}
\affiliation{Department of Physics, The George Washington University, Washington, DC 20052, USA}
\email{}

\author[orcid=0000-0002-7215-926X,sname='Bulut']{A. Bulut}
\affiliation{Department of Physics, Faculty of Sciences, Çanakkale Onsekiz Mart University, Terzioglu Campus, TR-17020, Çanakkale, T{\"u}rkiye}
\affiliation{Astrophysics Research Center and Observatory, Canakkale Onsekiz Mart University, Terzioglu Kampusu, TR-17020, Canakkale, T\"{u}rkiye}
\email{}

\author[orcid=0000-0002-5274-6790,sname='G\"o\u{g}\"u\c{s}']{E. G\"o\u{g}\"u\c{s}}
\affiliation{Sabanc\i~University, Orhanl\i~- Tuzla, Istanbul 34956, T\"{u}rkiye}
\email{}

\author[orcid=0000-0002-6293-9940,sname='Og{\l}oza']{W. Og{\l}oza}
\affiliation{Mt.Suhora Observatory, University of National Education Commission, Cracow Poland}
\email{}

%\collaboration{all}{The Terra Mater collaboration}

\begin{abstract}

We present new results for the hot Jupiters HAT-P-16b, TOI-1516b, and TOI-2046b, based on photometric observations collected using both space- and ground-based facilities. Ground-based data 
were collected in the 2020-2024 time span with the 0.6 m telescope (ADYU60) located at the Adiyaman University Application and Research Center (Adiyaman, T\"{u}rkiye) and the 1.0 m telescope at the T\"{u}rkiye National Observatory (TUG, T\"{u}rkiye). Through a combination of fits to our ground-based data, the mid-transit times data from TESS and additional data taken from the literature, we present an updated linear ephemeris for each system. Transit timing variations (TTVs) were analyzed using linear, orbital decay, and apsidal precession models. The resulting BIC($\Delta$BIC) values indicate that the orbital decay model is statistically favored for HAT-P-16b and TOI-1516b, while the constant period model is preferred for TOI-2046b. False alarm probabilities (FAPs) were computed to assess the significance of any periodic signals. TOI-1516b displays a strong TTV signal with a FAP (of 0.0001) well below the 0.01 threshold, suggesting a likely dynamical origin that warrants further investigation. The higher FAP value (0.0055) for HAT-P-16b suggests that the case of a possible presence of an additional body in the system is less convincing. In contrast, the much higher FAP value (0.0196) for TOI-2046b implies that there are no statistically significant TTVs.

\end{abstract}

\keywords{techniques: photometric – stars: individual}


\section{Introduction}

\noindent Hot Jupiters, known as giant planets with orbital periods shorter than 10 days, have captivated researchers since the discovery of 51 Pegasi b in 1995 \citep{mayor1995jupiter}. These celestial bodies, situated in close proximity to their parent stars, present significant challenges to conventional theories of planetary formation and evolution (\cite{pollack1996formation}). Various hypotheses have been proposed, including in situ formation (\cite{batygin2016situ}) and migration processes (\cite{lin1996orbital}, \cite{dawson2018origins}); however, the precise mechanisms governing their current orbital configurations remain unclear. Although their occurrence is relatively
low, estimated at less than 1\% among solar-type stars (\cite{wright2012frequency}), hot Jupiters are more readily detectable due to their substantially large sizes and short orbital periods, making them key subjects in exoplanet research (\cite{howard2012planet}). It is worth noting that the occurrence rate of hot Jupiters decreases toward low-mass stars; for example the occurrence rate around M stars is significantly lower than 1\%, at approximately 0.2\% \citep{2023MNRAS.521.3663B,2023AJ....166..165G}.\\
\\
Transit timing variations (TTVs) are deviations from the strictly periodic transit times of a planet caused by gravitational interactions with other bodies in the system. These variations manifest as shifts in the expected timing of a planet’s transit across its host star, providing a sensitive probe of additional planetary companions, including those that do not transit and are therefore undetectable by direct photometric observations. TTV analysis enables the measurement of masses and orbital parameters of non-transiting planets, offering critical insights into the dynamical architecture and evolution of planetary systems.\\ 
\\
TTVs in hot Jupiters have been the focus of several recent studies: 
\citet{Ivshina2022}, \citet{Alvarado2024}, and \citet{Yalcinkaya2024} have shown that while TTVs are less common in hot Jupiters compared to multi-planet systems with compact configurations, the detection of even subtle variations can signal the presence of non-transiting companions or tidal evolution. Notably, systems such as WASP-12b and TOI-2109b exhibit compelling evidence for orbital decay, likely driven by tidal interactions with their host stars \citep{Yee2020, Wong2022, 2021AJ....162..256W}, while apsidal precession has been detected or proposed in systems with mild orbital eccentricity \citep{Patra2020}. In some cases, TTVs have led to the discovery of additional planets, as seen in TOI-1130 \citep{Huang2020}, TOI-1408 \citep{2024ApJ...971L..28K}, and TOI-2818 \citep{2025ApJ...981..106M}. These findings collectively underscore the utility of TTV analysis not only in identifying hidden planetary companions but also in constraining orbital evolution mechanisms such as tidal dissipation and dynamical interactions. Our study builds upon this framework by providing new TTV measurements for TOI-1516b and TOI-2046b---systems previously unexplored in this context---and by re-evaluating the orbital behavior of HAT-P-16b in light of recent work suggesting potential orbital decay \citep{Sun2023}.\\ 
\\
HAT-P-16b is a transiting hot Jupiter discovered in 2010 by the HATNet survey \citep{buchhave2010hat}. It orbits an F8V-type star with a visual magnitude of V = 10.8, a stellar mass of 1.22 M$_{\odot}$, and a radius of 1.24 R$_{\odot}$. The planet itself has a mass of 4.19 $\pm$ 0.09 M$_{J}$ and a radius of 1.36 $\pm$ 0.03 R$_{J}$. The planet has a period of 2.77 days. Since its discovery, HAT-P-16b has been the subject of multiple follow-up studies using spectroscopic and photometric methods to refine the system parameters and investigate potential TTVs. These studies have included ground- and space-based observations aimed at detecting dynamical perturbations that might indicate additional planetary companions or other orbital effects (e.g., \cite{sada2012extrasolar}; \cite{ciceri2013simultaneous}; \cite{pearson2014photometric}; \cite{davoudi2020light}; \cite{aladaug2021analysis}). In particular, see \cite{Sun2023}, who note the observation of TTVs, and fit with the orbital decay and apsidal precession models.\\
\\
Identified by the TESS space mission in 2022 \citep{kabath2022toi}, TOI-1516b and TOI-2046b are classified as inflated hot Jupiters. TOI-1516b has a mass of 3.16 ($\pm$ 0.12) M$_{\mathrm{J}}$ and a radius of 1.36 ($\pm$ 0.03) R$_{\mathrm{J}}$, and orbits an F8V-type star with V = 11.55 $\pm$ 0.02 mag, a mass of 1.09 M$_{\odot}$, and a radius of 1.25 R$_{\odot}$. Its orbital period is 2.05 days \citep{kabath2022toi, akhand2023ground}. TOI-2046b, also reported by \citet{kabath2022toi}, was characterized using combined TESS and ground-based photometric and spectroscopic observations. It has a mass of 2.3 ($\pm$ 0.28) M$_{\mathrm{J}}$, a radius of 1.44 ($\pm$ 0.11) R$_{\mathrm{J}}$, and orbits an F8V-type star with V = 10.91 mag, a mass of 1.15 M$_{\odot}$, and a radius of 1.24 R${\odot}$, with a period of 2.79 days. Although both systems were included in the discovery paper by \citet{kabath2022toi}, no detailed analysis of TTV was reported in that paper, highlighting the need for further investigation of potential dynamical interactions.\\
\\
\noindent In this article, we present the findings from recent transit observations of three hot Jupiter-type exoplanets: TOI-1516b, TOI-2046b and HAT-P-16b derived from data collected between the 2020-2024 observation period.\\
\section{Observation and Data Reduction}
\subsection{Photometric Observation}
\noindent HAT-P-16b, TOI-1516b and TOI-2046b were recently observed  using ADYU60 (0.6-meter) and T100 (1.0 meter) telescopes. ADYU60 located at the Adiyaman University Application and Research Center in Adiyaman, Turkey. ADYU60 is operated remotely from the Adiyaman University Observatory. The ADYU60 is equipped with a 1 x 1k$^2$ Andor iKon-M934 CCD with a pixel size of 13 × 13 $\mu$m$^2$ and an image scale of 0.67$^{''}$/pixel; ADYU60 has been effectively used in previous exoplanet studies \citep{sonbas2022probing}.The 1.0 m Ritchey–Chr\'etien (RC) telescope (T100) is located at the Bakirlitepe Mountain and is currently operated remotely from T\"{u}rkiye National Observatories located in Antalya, T\"{u}rkiye. The T100 telescope houses a 4k $\times$ 4k SI 1100 CCD camera operating at -90$^{\circ}$C. The pixel size, overall field of view and image scale of the CCD camera are 15$\mu$m $\times$ 15$\mu$m, $21.5^{'}\times21.5^{'}$ and 0.31$^{''}$/pixel respectively. Photometric observations for ADYU60 were made with the $Johnson$ R filter and for T100 we used the $Bessel$ R and V filters. \\
\\
We observed a total of 11 transits for HAT-P-16b (between November 2021 - December 2024), 16 transits for TOI-1516b (between September 2020 - December 2024) and 13 transits for TOI-2046b (between October 2020 - December 2024) with ADYU60  and three transits for TOI-2046b (between December 2021 - August 2023) with T100. Except in a few cases the exposure was kept fixed while acquiring data with consistent time intervals during all phases of the transits. The summaries of our observations are given in Table \ref{tab:log_of_observations}.\\
\\
%%%%%%%%%%%%
\startlongtable
\begin{deluxetable*}{lccccccccc}
\tabletypesize{\scriptsize}
\tablewidth{\textwidth}
\tablecaption{Log of observations. \label{tab:log_of_observations}}
\tablehead{
\colhead{Date of} & \colhead{Object}  & \colhead{Telescope} & \colhead{Filter} & \colhead{Exposure  (s)} & \colhead{Number of} & \colhead{Airmass} & \colhead{RMS} & \colhead{Transit} & \colhead{Global Fit}\\
 \colhead{observations} &\colhead{} & \colhead{} &\colhead{} &\colhead{} & \colhead{data points}&\colhead{} & \colhead{mmag} & \colhead{Coverage} & \colhead{}
}
\startdata
08.11.2021 & HAT-P-16b & ADYU60  & R & 120 & 161 & 1.01 - 2.36 & 1.16 & Full transit & Yes\\
11.11.2021 & HAT-P-16b & ADYU60  & R & 120 & 131 & 1.35 - 1.03 & 1.56 & Full transit & Yes \\
22.11.2021 & HAT-P-16b & ADYU60  & R & 120 & 150 & 1.00 - 1.38 & 1.34 & Full transit & Yes\\
06.12.2021 & HAT-P-16b & ADYU60  & R & 90-60 & 207 & 1.00 - 1.21 & 1.28 & Full transit & Yes\\
24.10.2022 & HAT-P-16b & ADYU60  & R & 90 & 142 & 1.56 - 1.01 & 1.68 & Egress & Yes\\
04.11.2022 & HAT-P-16b & ADYU60  & R & 90 & 178 & 1.20 - 1.39 & 1.63 & Full transit & Yes\\
24.12.2022 & HATP-16b & ADYU60  & R & 60 & 290 & 1.00 - 1.81 & 1.79 & Full transit & Yes\\
11.11.2023 & HAT-P-16b & ADYU60  & R & 90-60 & 221 & 1.15 - 1.00 & 1.94 & Full transit & Yes\\
31.08.2024 & HAT-P-16b & ADYU60  & R & 90 & 193 & 1.01 - 1.21 & 2.33 & Ingress & No\\
14.09.2024 & HAT-P-16b & ADYU60  & R & 90 & 244 & 1.42 - 1.00 & 1.89 & Full transit & Yes\\
25.09.2024 & HAT-P-16b & ADYU60  & R & 90 & 211 & 1.00 - 1.74 & 1.83 & Full transit & Yes\\
\hline
03.09.2020 & TOI-1516b & ADYU60  & R & 120 & 144 & 1.36 - 1.18 & 2.55 & Full transit & Yes\\
09.09.2020 & TOI-1516b & ADYU60  & R & 120 & 138 & 1.19 - 1.62 & 1.63 & Full transit & Yes\\
06.10.2020 & TOI-1516b & ADYU60  & R & 120 & 129 & 1.31 - 1.18 & 2.21 & Full transit & Yes\\
08.10.2020 & TOI-1516b & ADYU60  & R & 120 & 162 & 1.30 - 1.18 & 1.55 & Full transit & Yes\\
12.10.2020 & TOI-1516b & ADYU60  & R & 30 & 552 & 1.18 - 1.65 & 2.86 & Full transit & No\\
14.10.2020 & TOI-1516b & ADYU60  & R & 60 & 286 & 1.23 - 1.95 & 2.59 & Full transit & Yes\\
12.11.2020 & TOI-1516b & ADYU60  & R & 60 & 274 & 1.18 - 1.39 & 1.73 & Full transit & Yes\\
16.11.2020 & TOI-1516b & ADYU60  & R & 60 & 229 & 1.30 - 1.95 & 2.27 &Egress & Yes\\
25.06.2022 & TOI-1516b & ADYU60  & R & 120 & 123 & 2.13 - 1.24 & 2.97 & Full transit & No\\
14.10.2022 & TOI-1516b & ADYU60  & R & 90 & 142 & 1.17 - 1.44 & 4.03 & Full transit & No\\
16.10.2022 & TOI-1516b & ADYU60  & R & 60 -90 & 183 & 1.18 - 1.29 & 2.22 & Ingress & Yes\\
14.11.2022 & TOI-1516b & ADYU60  & R & 60 & 210 & 1.18 - 1.28 & 2.71 & Egress & No\\
02.07.2024 & TOI-1516b & ADYU60  & R & 90& 259 & 2.37 - 1.33 & 3.06 & Full transit & No\\
08.08.2024 & TOI-1516b & ADYU60  & R & 90 & 272 & 1.92 - 1.25 & 2.38 & Full transit & Yes\\
10.08.2024 & TOI-1516b & ADYU60  & R & 120 & 127 & 1.43 - 1.25 & 3.19 & Full transit & No\\
16.09.2024 & TOI-1516b & ADYU60  & R & 120 & 144 & 1.18 - 1.72 & 1.65 & Full transit & Yes\\
\hline
18.10.2020 & TOI-2046b & ADYU60  & R & 120 & 137 & 1.56 - 1.25 &  2.64 & Full transit & Yes\\
19.12.2022 & TOI-2046b & ADYU60  & R & 60 & 217 & 1.20 - 1.68 & 3.22 & Full transit & No\\
22.12.2022 & TOI-2046b & ADYU60  & R & 60 & 253 & 1.25 - 1.40 & 4.22 & Full transit & No\\
28.12.2022 & TOI-2046b & ADYU60  & R & 60 & 212 & 1.25 - 1.38 & 4.14 & Full transit & No\\
03.01.2023 & TOI-2046b & ADYU60  & R & 60 & 257 & 1.25 - 1.42 & 3.09 & Full transit & Yes\\
09.01.2023 & TOI-2046b & ADYU60  & R & 60 & 211 & 1.00 - 1.52 & 2.73 & Full transit & Yes\\
21.01.2023 & TOI-2046b & ADYU60  & R & 60 & 199 & 1.32 - 1.96 & 3.69 & Egress & No\\
24.01.2023 & TOI-2046b & ADYU60  & R & 60 & 196 & 1.26 - 1.51 & 2.99 & Egress & Yes\\
15.10.2024 & TOI-2046b & ADYU60  & R & 90 & 198 & 1.01 - 2.18 & 2.33 & Full transit & Yes\\
27.10.2024 & TOI-2046b & ADYU60  & R & 90 & 170 & 1.29 - 1.73 & 1.92 & Full transit & Yes\\
30.10.2024 & TOI-2046b & ADYU60  & R & 90 & 108 & 1.31 - 1.56 & 2.03 & Egress & Yes\\
11.11.2024 & TOI-2046b & ADYU60  & R & 120 & 130 & 1.28 - 1.70 & 2.52 & Full transit & Yes\\
14.11.2024 & TOI-2046b & ADYU60  & R & 120 & 127 & 1.27 - 1.72 & 1.91 & Full transit & Yes\\
17.12.2021 & TOI-2046b & TUG-T100  & R & 30 & 253 & 1.26 - 2.16 & 8.90 & Full transit & No\\
17.12.2021 & TOI-2046b & TUG-T100  & V & 45 & 253 & 1.26 - 2.16 & 9.77 & Full transit & No\\
03.07.2023 & TOI-2046b & TUG-T100  & R & 7 & 496 & 2.33 - 1.34 & 7.24 & Full transit & No\\
03.07.2023 & TOI-2046b & TUG-T100  & V & 10 & 498 & 2.33 - 1.34 & 6.46 & Full transit & No\\
05.08.2023 & TOI-2046b & TUG-T100  & R & 7 & 595 & 2.14 - 1.26 & 4.77 & Full transit & No\\
05.08.2023 & TOI-2046b & TUG-T100  & V & 10 & 606 & 2.14 - 1.26 & 4.52 & Full transit & No\\
\enddata
\end{deluxetable*}
%%%%%%%%%%%%%%%%
\begin{deluxetable*}{lllll}
\tablewidth{\textwidth}
\tablecaption{The information on comparison stars. The magnitudes are taken from the SIMBAD catalog$^{1}$. \label{tab:comparisons}}
\tablehead{
\colhead{Object} & \colhead{RA}  & \colhead{DEC} & \colhead{Kmag} &  
}
\startdata
 HAT-P-16 & 00 38 17.56	& +42 27 47.22 & 9.55 & \\
 Gaia DR3 381592519806816896 & 00 38 16.08 &	+42 28 03.86 & - & \\
 2MASS J00382247+4228526 & 00 38 22.47 &	+42 28 52.72 & 12.54 & \\
 TYC 2792-1778-1 &	00 38 18.38	& +42 25 17.18 & 9.24 & \\
 TYC 2792-1737-1 & 00 38 13.06	& +42 30 25.93 & 7.93 & \\
 \\
 TOI-1516 & 22 40 20.26 & +69 30 13.45 & 9.67 & \\
 TYC 4480-149-1 & 22 39 56.17 & +69 31 08.30 & 10.69 & \\
 TYC 4480-7-1 & 22 39 30.17 &	+69 28 53.73 & 10.22 & \\
 TYC 4480-268-1 & 22 39 26.22 &	+69 29 30.64 & 8.24 & \\
 TYC 4480-295-1 & 22 41 30.02 &	+69 29 37.15 & 10.04 & \\
 \\
 TOI-2046 & 01 04 44.36 & +74 19 52.86 & 10.09 & \\
 TYC 4308-891-1 & 01 04 49.92 &	+74 22 19.67 & 6.55 & \\
 TYC 4308-1009-1 & 01 04 22.92 & +74 17 08.65 & 8.49 & \\
 TYC 4308-726-1 & 01 03 50.38 &	+74 23 08.76 & - & \\
\enddata
\tablecomments{ $^1\href{http://simbad.u-strasbg.fr/simbad/} {http://simbad.u-strasbg.fr/simbad/}$}
\end{deluxetable*}

%%%%%%%

\subsection{Data Reduction}\label{sec:data_red}  
\noindent The software package AstroImageJ (AIJ) \citep{collins2017astroimagej} was deployed to reduce the data, perform calibration and the extraction of differential aperture photometry, as well as extract detrend parameters. AIJ deploys the Data Processor module to correct the raw CCD images, create Median-combined bias, and flat frames. Because of negligible dark counts in the frames, dark correction was not applied. The Multi-Aperture (MA) module was used to perform differential photometry. To extract differential magnitudes, we considered a number of comparison stars in the CCD frames and selected two or three with minimally variable light curves in the final analysis. For all observations of a given object, the same standard stars were used. The MA module was used to estimate the uncertainties in flux and CCD read-out noise. The details of the selected comparisons are given in Table \ref{tab:comparisons}. We estimated the aperture sizes of the target and comparison stars by using the "radial profile" feature of the AIJ tool to minimize the transit-modeling residuals. Depending on the quality of the data the aperture sizes were allowed to vary by 1.2 times the FWHM value in each image. The conversion from Julian Date (JD) to Barycentric Julian Date (BJD) was done both within $AIJ$ and by using online applets such as; $https://astroutils.astronomy.osu.edu/time/$ \citep{2010PASP..122..935E}. To correct systematic effects in the light curves and improve the overall fit quality, the $Multi-plot$ module in $AIJ$ was used to extract detrending parameters, including airmass, time, sky background, the FWHM of the average PSF, total comparison star counts, and the x and y centroid positions of the target on the detector. Relative flux, corresponding flux errors and detrend parameters were then used to create simultaneous detrended light curves in EXOFASTv2 \citep{eastman2013exofast, 2017ascl.soft10003E} as part of the global and individual fits for consistency. For each source, the same number of detrend parameters was taken into account. We used the additive detrending scheme of EXOFASTv2 in our analysis. The final light curves were obtained from the differential magnitudes and, for each light curve, the RMS was calculated to measure the quality of the data. The RMS was found to vary in the range $\sim (1.2 - 4.2)$ \textit{mmag}. We assume this reflects the count rates and the different conditions under which the light curves were measured.\\
\\
Observations from the TÜBİTAK National Observatory (TUG) T100 telescope were analyzed using the EXOTIC (EXOplanet Transit Interpretation Code) software \citep{2023ascl.soft02009Z, 2020PASP..132e4401Z}, optimized for the V and R filters. EXOTIC is a Python-3 based exoplanet transit modeling program capable of processing FITS files or prereduced light curves. Standard image calibrations—including bias, dark, and flat-field corrections—were applied to the FITS data to generate light curves using differential photometry. The selection of comparison stars was performed using the least squares minimization algorithm from the scipy.optimize.least\textunderscore squares package fitting the \cite{mandel2002analytic} transit model to the data \citep{virtanen2020scipy}. Following the selection of optimal comparison stars, airmass-corrected light curves were modeled using the Markov Chain Monte Carlo (MCMC) method \citep{Kass0105...1998, Gilks.0109.1998} for parameter estimation. The software retrieves system parameters automatically from the NASA Exoplanet Archive, while also allowing for user-defined input. \\
\subsection{TESS light curves}
\begin{figure*}[hbt!]
\centering
% First image (TOI-1516)
\begin{subfigure}[b]{0.31\textwidth}
%    \centering
%\hspace{-0.8cm}
    \includegraphics[width=\linewidth]{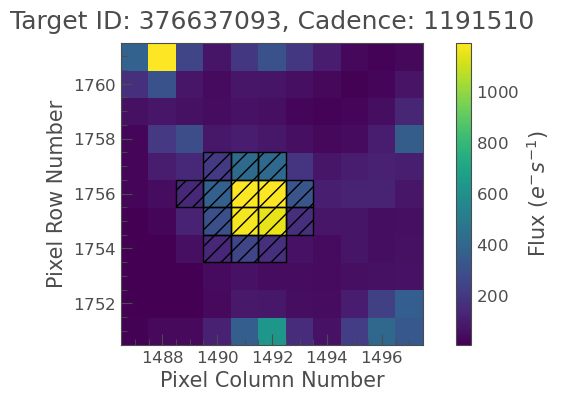}
    \caption{TOI-1516 in sector 58.}
    \label{fig:toi1516}
\end{subfigure}
%\hspace{-0.8cm} % Add space between figures
% Second image (TOI-2046)
\begin{subfigure}[b]{0.31\textwidth}
%    \centering
%\hspace{-0.8cm}
    \includegraphics[width=\linewidth]{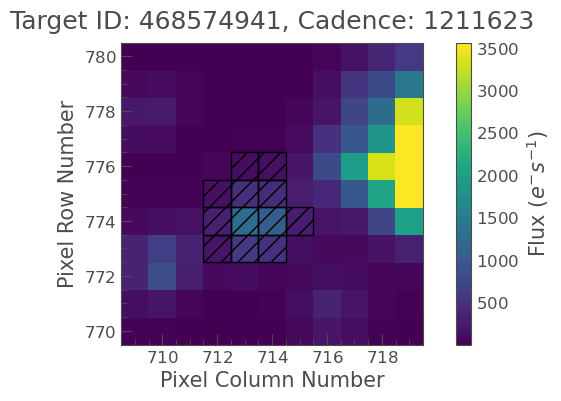}
    \caption{TOI-2046 in sector 58.}
    \label{fig:toi2046}
\end{subfigure}
%\hspace{-1em} % Add space between figures
% Third image (Hat-p-16)
\begin{subfigure}[b]{0.31\textwidth}
%    \centering
%\hspace{-0.8cm}
    \includegraphics[width=\linewidth]{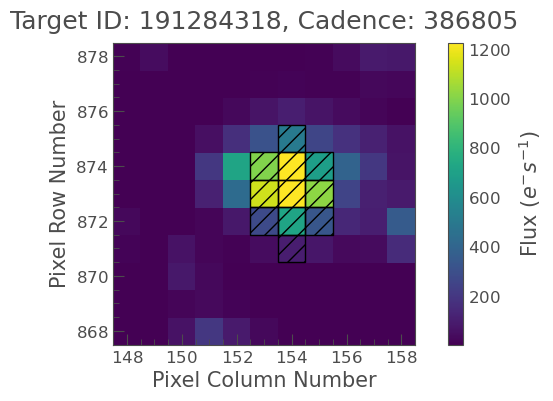}
    \caption{HAT-P-16 in sector 17.}
    \label{fig:hatp16}
\end{subfigure}
\caption{The plot of the target pixel files for TOI-1516 in sector 58 (top), TOI-2046 in sector 58 (middle), and HAT-P-16 in sector 17 (bottom). The black squares represent the chosen optimal photometric aperture utilized for the extraction of the SAP flux. It is important to note that the TESS pixel scale measures 21 arcseconds.}
\label{fig:targetpixel_combined}
\end{figure*}

\noindent TESS is a space mission that uses transit detection technique to identify exoplanets \citep{ricker2015transiting}. The spacecraft is equipped with four identical cameras, each containing 2000 $\times$ 2000 CCDs, which monitor a field of view of 24$^{\degree}$ $\times$ 24$^{\degree}$. TESS is designed to observe approximately 85 percent of the celestial sphere. By July 2020, the mission had successfully recorded the flux of more than 200,000 main-sequence stars at two-minute intervals. Furthermore, TESS acquired full-frame images (FFIs) of the entire field of view from its four cameras every 30 minutes. The hot Jupiters discussed in this article were seen as part of the Northern Hemisphere observation campaign conducted in 2019 and 2020. The identification of all three systems was facilitated by the TESS Quick Look Pipeline (QLP), as noted by \cite{huang2020photometry}.\\
\\
\noindent Observations of TOI-1516 and TOI-2046 were conducted by TESS, over eight sectors, with each sector featuring a 2-minute short cadence integration. In particular, the observations occurred in sectors 58, 59, 78, and 79 during the second extended mission of TESS. The light curves along with the target pixel files (TPFs) were sourced from the Mikulski Archive for Space Telescopes\footnote{\url{https://archive.stsci.edu/}} (MAST). We extracted the light curves using aperture masks derived from the pipeline-defined apertures, estimating and subtracting the background from dedicated background pixels. We then applied quality flag filtering and iterative sigma-clipping to remove outliers. In order to model the flux contamination from nearby stars that affects the target star’s light, we separately defined target pixels and background pixels when extracting the light curve from the TPFs. This separation allows for the correction of background flux contribution, ensuring that the light curves provided to EXOFAST are free from contamination and accurately represent the intrinsic variability of the target star. To evaluate the quality of the light curves, we measured the root-mean-square (RMS) scatter in the out-of-transit regions and compared it to the corresponding PDCSAP\_FLUX products from the TESS pipeline. We found that our curves have RMS values comparable to  the PDCSAP\_FLUX curves. This demonstrates that our extraction procedure yields light curves equivalent to those produced by 
the standard TESS pipeline products. 
\\
\\
\noindent The considerable dimensions and close distance of HAT-P-16b to its parent star make it an outstanding subject for a comprehensive study with the sensitive equipment employed by the TESS mission. It was observed during the second and fifth extended phases of TESS, specifically in sectors 17 and 57. We plotted the objects TOI-1516b and TOI-2046b from sector 58, and HAT-P-16b from sector 17 as TPFs in Figure \ref{fig:targetpixel_combined}, respectively.\\

%%%%%%%%%%%%%%%%
%%%%%%%%%%%%%%%%

\subsection{Global Fits}\label{sec:exofast} % for System Parameters
\noindent We used the transit modeling code, EXOFASTv2, to derive the stellar and planetary system parameters as well as the transit and limb darkening parameters, along with their uncertainties \citep{eastman2013exofast, 2017ascl.soft10003E, eastman2019exofastv2}. To constrain the stellar parameters robustly and recover the behavior of the original EXOFAST, both NOMIST and TORRES keywords were used. EXOFASTv2 facilitates the fitting of multiple sets of astrophysical data for a given number of planets with RV data; it does this by scaling the RV and light curve uncertainties.  In this work, the transit light curves are fitted along with published RV data. The HAT-P-16b RV data were taken from \citet{buchhave2010hat};  the TOI-1516b and the TOI-2046b RV data were taken from \citet{kabath2022toi}.\\ 
\\
For each source, a set of priors (both stellar and planetary) were taken from the literature (\cite{buchhave2010hat, bonomo2017gaps, Sun2023, kabath2022toi, fox2022neossat}). These prior parameters include; $T_{\rm eff}$, $[{\rm Fe/H}]$, $R_*$, $M_*$, linear and quadratic limb darkening coefficients, $R_P$, $M_P$, eccentricity, period, inclination, and transit impact parameter if available in the literature. Their uncertainties were included in our analysis (thus making the priors Gaussian: see \cite{eastman2019exofastv2}). The limb-darkening parameters for each system were either adopted from previous publications \cite{kabath2022toi}, including values available via the NASA Exoplanet Archive\footnote{\url{https://exoplanetarchive.ipac.caltech.edu}}, or through interpolating the coefficient tables of \cite{claret2004,claret2011} based on stellar atmospheric parameters. If~limb darkening values from Claret are questionable in the regime of the source of interest, we impose a generous uniform prior, as suggested in EXOFASTv2 see Table \ref{tab:priors}.\\
\\
In order to perform a global fit, the transit and RV data were independently fitted with EXOFAST and the errors were scaled to find the maximum likelihood with the lowest $\chi^2$ for every best-fit model. We used default EXOFASTv2 statistics as Tz $>$ 1000 and GR (Gelman–Rubin statistic (\citet{gelman1992inference}) $<$ 1.01 for each parameter to derive the stellar parameters from global fits. 
The median values of the posterior distributions for the system parameters are listed in Tables \ref{tab:syspar_hatp16}, \ref{tab:syspar_toi1516}, and \ref{tab:syspar_toi2046} . The $1\sigma$-level uncertainties are also given.  

%%%%%%%%%%%%%%%%%%%%%%%%%%%%%%%
\begin{deluxetable}{lllllllll}
\tablewidth{\textwidth}
\tablecaption{Prior parameters of the exoplanetary systems HAT-P-16b, TOI-1516b, and TOI-2046b. \label{tab:priors}}
\tablehead{
\colhead{Parameters} & \colhead{Units}  & \colhead{HAT-P-16} & \colhead{Ref.} &\colhead{TOI-1516} & \colhead{Ref.}& \colhead{TOI-2046} & \colhead{Ref.} &  
}
\startdata
%\multicolumn{8}{l}{\textbf{Stellar Parameters}} \\
$T_{eff}$ & Effective Temperature (K) & 6140$\pm$72& 2& 6520$\pm$90& 4 & 6160$\pm$100& 4\\
\ [Fe/H] & Metallicity & 0.17$\pm$0.08& 1& -0.14$\pm$0.12& 4& 0.04$\pm$0.12& 4\\
$R_*$ & Radius (R$_\odot$) &1.237$\pm$0.054 &1 & 1.245$\pm$0.031& 4 & 1.21$\pm$0.07& 4\\
$M_*$ & Mass (M$_\odot$) & 1.218$\pm$0.039& 1& 1.085$\pm$0.066& 4& 1.13$\pm$0.19& 4\\
$\log g_*$ & Surface gravity (cgs) &4.34$\pm$0.03 & 1& 4.25$\pm$0.18& 4& 4.38$\pm$0.18& 4\\
$\rho_*$ & Density (g/cm$^3$) & 0.784$\pm$0.040& 2& 1.090$\pm$31& 4& 890$\pm$98 & 4\\
%\multicolumn{8}{l}{\textbf{Planetary Parameters}} \\
$R_p$ & Radius (R$_J$) & 1.324$\pm$0.037& 2& -& -& -& -\\
%\multicolumn{8}{l}{\textbf{Primary Transit Parameters}} \\
$i$ & Inclination ($^\circ$) &86.6$\pm$0.7 & 1& 90.0$\pm$0.4 & 4& 83.6$\pm$0.9& 4\\
$e$ & Eccentricity & 0.036$\pm$0.004& 2& -& -& -& -\\
$P$ & Period (d) & 2.775960$\pm$0.000003& 1& 2.056014$\pm$0.000002& 4& 1.497184$\pm$0.000003& 4\\
%\multicolumn{8}{l}{\textbf{RV Parameters}} \\
K & $m~s^{-1}$ & 531.1$\pm$11& 1& 460.7$\pm$9.0&4 & 374.7$\pm$8.0& 4\\
$\omega$ & ($^\circ$)& 214 $\pm$8& 1& -& - & -& -\\
$b$ & Impact parameter &0.43$\pm$0.06 & 1& -& -& -& -\\
$u_1$ & Limb-darkening (linear)& ${^*}$& 3& 0.10$\pm$0.08& 4& 0.39$\pm$0.24& 4\\
$u_2$ & Limb-darkening (quad.)& ${^*}$& 3& 0.81$\pm$0.19& 4&  0.30$\pm$0.74& 4\\
$T_0$ & BJD & 2455027.592930$\pm$0.00031& 1& 2458765.3250$\pm$0.0001& 4& 2457792.2767$\pm$0.0024& 4\\
\enddata
\tablecomments{$^{1}$\cite{buchhave2010hat}, $^{2}$\cite{ciceri2013simultaneous}, ${^*}$ Imposed a generous uniform prior based on Claret tables ($^{3}$\cite{claret2004,claret2011}), $^{4}$\cite{kabath2022toi}}
\end{deluxetable}
%%%%%%%%%%%%%%%%
%%%%%
\subsection{TTV Modeling}
\label{TTV Modeling}
\noindent For each individual transit light curve, observed mid-transit times and their associated uncertainties were determined by using EXOFASTv2 and EXOTIC packages. Along with our own data, we also retrieved the published mid-times from the literature. For each system, the observed mid-transit times, their uncertainties, and the mid-times from the literature are provided in the (online) Appendix; see Tables \ref{tab:oc_hatp16b}, \ref{tab:oc_toi1516b}, and \ref{tab:oc_toi2046b}, respectively.\\ 
\\
To investigate the existence of TTVs, we analyzed the differences between the observed (O) and calculated (C) transit times using ExoPdot software \citep{Hagey..2022AJ....164..220H}. ExoPdot is a Python-based package specifically designed to analyze exoplanet TTV data and determine the orbital parameters of exoplanetary systems.
It incorporates a variety of models and fitting functions that facilitate precise adjustments to the observed data. The program utilizes an MCMC approach to navigate the parameter space and identify the most suitable model. A custom Metropolis-Hastings MCMC, with a Gibbs sampler \citep{Ford..2006ApJ...642..505F}, has been implemented to allow greater control over the algorithm. The latter approach is used for all the results presented here.\\
\\
We considered the following cases in which the planet is a) orbiting with a constant period, b) in a circular but decaying orbit, and finally c) in an orbit that is precessing. We set the initial parameters for all models in our analytical process. The ExoPdot program then refined these parameters to minimize the differences between the observed and calculated transit times. 
The fitting procedure involved the execution of multiple MCMC chains to ensure the convergence and reliability of the results. The posterior probability distributions of the parameters were examined to evaluate the quality of the fit.\\
\\
In the first case (i.e. the orbit with a constant period), the mid-transit times are set to increase linearly with the transit epoch E i.e.
\begin{align}
t_{tra} &= t_0 + PE    \label{eq:first_1} \\
t_{occ} &= t_0 + \frac{P}{2} + PE  \label{eq:first_2}
\end{align}
where $t_0$ is the reference epoch, $P$ is the period, and $t_{tra}$ and $t_{occ}$ are the expected transit and occultation times, respectively. \\
\\
In the second case, a steady change in the orbital period is modeled as a quadratic term in the expected mid-transit times. A negative sign for the period derivative, $\frac{dP}{dE}$, implies a decaying orbit. The model is given by
\begin{align}
t_{tra} &= t_0 + PE + \frac{1}{2}\frac{dP}{dE}E^2 \label{eq:second_1} \\
t_{occ} &= t_0 + \frac{P}{2} + PE + \frac{1}{2}\frac{dP}{dE}E^2 \label{eq:second_2}
\end{align}
\\
In the third scenario, apsidal precession is assumed to be the underlying cause of the timing variations; this requires the eccentricity of the orbit to be non-zero. Measurable effects are expected for even small eccentricities. It is assumed that the pericenter argument $\omega$ varies at a constant rate to capture the system's behavior in this model, resulting in sinusoidal trends in the timing data. Following \citet{Patra..2017AJ....154....4P}, the transit and occultation times can be expressed as:
\begin{align}
t_{tra} &= t_0 + P_s E - \frac{eP_a}{\pi}  \cos\omega \label{eq:third_1} \\
t_{occ} &= t_0 + \frac{P_s}{2} + P_sE \frac{eP_a}{\pi}  \cos\omega \label{eq:third_2} \\
\omega(E) &= \omega_0 + \frac{d\omega}{dE}E \label{eq:third_3} \\
P_s &= P_a\left(1 - \frac{d\omega/dE}{2\pi} \right)  \label{eq:third_4}
\end{align}
for argument of pericenter $\omega$, phase $\omega_0$, and precession rate $\frac{d\omega}{dE}$. $P_s$ represents the planet's sidereal period; %(a fixed parameter);
an additional term, $P_a$, referred to as the 'anomalistic' period, represents the signal due to precession.\\
\\
The posterior distributions for the model parameters are obtained with MCMC. In the first case, there are two free parameters: the reference epoch and the period. The second case has three parameters, adding the $\frac{dP}{dE}$ term. Lastly, the precession model has five parameters: the reference epoch, the sidereal period, eccentricity, phase constant, and the precession rate. \\
\\

To allow for a statistical comparison of models, we deploy the Bayesian Information Criterion (BIC). This metric balances the trade-off between the goodness of fit and model complexity by incorporating both the precision of the fit and a penalty for the number of parameters. BIC is defined as:
\begin{equation}
\mathrm{BIC} = \chi^2 + k \ln(n) \label{eq:9}
\end{equation}
where $n$ is the number of data points and $k$ is the number of free parameters in the model. The first term, $\chi^2$, quantifies how well the model fits the data, while the second term, $k \ln(n)$, penalizes additional complexity. This complexity penalty is essential for mitigating the risk of overfitting.\\
\\
A model with a lower BIC value is generally preferred, as it indicates a better trade-off between goodness of fit and simplicity. Not surprisingly, models with more parameters often fit the data better, but this does not necessarily imply they are more accurate. Furthermore, over-parameterized models may fit the training data well but can perform poorly on unseen data, hence the justification for a controlling mechanism such as the complexity penalty. BIC is clearly not an absolute measure of model correctness, but it is effective in comparing the relative performance of different models applied to the same dataset.\\
\\
Interpreting BIC values in isolation can sometimes be misleading. Therefore, to enhance the robustness of our model selection, we computed the difference in BIC values between models ($\Delta$BIC) to evaluate their relative statistical support. This is defined as:

\begin{equation}
\Delta\mathrm{BIC} = \mathrm{BIC}_{i} - \mathrm{BIC}_{min} 
\end{equation}

\noindent According to the interpretation guidelines by 
\citet{kass1995bayes}:

\begin{itemize}
\item  $\Delta$BIC $<$ 2: very weak evidence against the higher-BIC model,
\item 2 $\leq$ $\Delta$BIC $<$ 6: positive evidence against the higher-BIC model,
\item 6 $\leq$ $\Delta$BIC $<$ 10: strong evidence against the higher-BIC model,
\item $\Delta$BIC $\geq$ 10: very strong evidence against the higher-BIC model
\end{itemize}

\noindent Thus, BIC not only serves as a measure of model adequacy, but when used in conjunction with $\Delta$BIC, it provides a more nuanced and statistically rigorous framework for model comparison, particularly in the context of transit timing variation analyses.\\

\noindent The residuals of the TTV (O-C) data were frequency-analyzed using the Generalized Lomb-Scargle (GLS) periodogram to search for the presence of additional bodies connected to the planetary system. The GLS periodogram gives an analytic solution for the generalization to a full sine wave fit that allows for the inclusion of offsets and weights ($\chi^{2}$ fitting). It also implements this generalization for the evaluation of the Keplerian periodogram that searches for the period of the best-fitting Keplerian orbit to radial velocity data. The main algorithm is capable of detecting eccentric orbits, which makes it an important tool in the search for the orbital periods of exoplanets \citep{Zechmeister--2009A&A...496..577Z}. One of the metrics used in the algorithm is the False-Alarm Probability (FAP). As noted in the literature, FAP thresholds such as FAP $<$ 0.01 and FAP $<$ 0.001 are commonly adopted to define high and very high levels of statistical confidence, respectively \citep{Baluev--2008MNRAS.385.1279B, Zechmeister--2009A&A...496..577Z, VanderPlas--2018ApJS..236...16V}. If the FAP value is below 0.001, the periodicity and the detection of an additional body in the system are considered highly reliable. However, if the FAP value is between 0.001 and 0.01, the results are considered less robust. FAP values in the range of 0.01 - 0.1 imply that the case for periodicity and detection is very weak. An FAP value greater than 0.1 is considered to be background noise.\\
\\
While our main focus is the extraction and interpretation of TTVs, we mention here the tidal quality factor $Q$, a parameter that is potentially useful in probing the evolution of orbital dynamics of stellar systems. We calculate this factor for each host star using the following expression \citep{GoldreichSoter1966, Ogilvie2014}:
\begin{equation}
Q = -\frac{9}{2} \cdot k_2 \cdot \left( \frac{M_p}{M_*} \right) \cdot \left( \frac{R_*}{a} \right)^5 \cdot \frac{n}{\frac{1}{a} \cdot \frac{da}{dt}},
\end{equation}
where $k_2$ is the stellar Love number (taken as 0.014 in this study), $M_p$ and $M_*$ are the planetary and stellar masses, $R_*$ is the stellar radius, $a$ is the semimajor axis, $n = 2\pi/P$ is the orbital mean motion, and $\frac{da}{dt}$ is the rate of change in orbital separation derived from $\frac{dP}{dt}$ in the decay model. The parameter $k_2$, known as the stellar Love number, quantifies the quadrupolar tidal deformability of a star in response to external gravitational fields, a concept introduced by the British mathematician Augustus Edward H. Love \citep{love1911}\footnote{Augustus Edward H. Love introduced the idea of tidal deformability coefficients, now known as Love numbers, in his work on the elasticity of spheres and the Earth's response to tidal forces in the early 20\textsuperscript{th} century}.\\

%%%%%%%%%%%%%%%%%%%%%%%%%%%%%%%
\begin{longrotatetable}
\begin{deluxetable*}{ccccccc}
\tabletypesize{\scriptsize}
\tablewidth{\textwidth}
\tablecaption{System Parameters for HAT-P-16 \label{tab:syspar_hatp16}}
\tablehead{
\colhead{Parameters} & \colhead{Units} & \colhead{\cite{buchhave2010hat}} & \colhead{\cite{bonomo2017gaps}} & \colhead{\cite{Sun2023}} & \colhead{This study} &\colhead{This study} \\
\colhead{} & \colhead{} & \colhead{} & \colhead{} & \colhead{} & \colhead{(TESS)} & \colhead{(ADYU60)} \\
}
\startdata
\multicolumn{7}{l}{\textbf{Stellar Parameters}} \\
$M_*$ & Mass (M$_\odot$) & 1.218 $\pm$ 0.039 & 1.218 $\pm$ 0.039 & -- & 1.22 $\pm$ 0.03 & 1.23 $\pm$ 0.03 \\
$R_*$ & Radius (R$_\odot$) & 1.237 $\pm$ 0.054 & 1.237 $\pm$ 0.054 & 1.157 $\pm$ 0.030 & 1.25 $\pm$ 0.01 & 1.25 $\pm$ 0.01 \\
$L_*$ & Luminosity (L$_\odot$) & 1.97 $\pm$ 0.22 & -- & -- & 1.90 $\pm$ 0.10 & 2.00 $\pm$ 0.10 \\
$\rho_*$ & Density (g/cm$^3$) & -- & -- & -- & 0.89 $\pm$ 0.01 & 0.88 $\pm$ 0.02 \\
$\log g_*$ & Surface gravity (cgs) & 4.34 $\pm$ 0.03 & -- & -- & 4.33 $\pm$ 0.01 & 4.334 $\pm$ 0.01 \\
$T_{eff}$ & Effective Temperature (K) & 6158 $\pm$ 80 & 6158 $\pm$ 80 & 6158 $\pm$ 79 & 6110 $\pm$ 67 & 6131 $\pm$ 68 \\
\ [Fe/H] & Metallicity & 0.17 $\pm$ 0.08 & 0.170 $\pm$ 0.080 & -- & 0.18 $\pm$ 0.07 & 0.18 $\pm$ 0.08 \\%[0.5em]
\hline
\multicolumn{7}{l}{\textbf{Planetary Parameters}} \\
$e$ & Eccentricity & 0.036 $\pm$ 0.004 & 0.0462$_{-0.0024}^{+0.0027}$ & 0 (fixed) & 0.035 $\pm$ 0.003 & 0.03 $\pm$ 0.03 \\
$P$ & Period (d) & 2.775960$_{-0.000003}^{+0.000003}$ & 2.7759712$_{-0.000001}^{+0.000001}$ & 2.7759682$_{-0.0000002}^{+0.0000002}$ & 2.7759676$_{-0.0000002}^{+0.0000002}$ & 2.7759674$_{-0.0000002}^{+0.0000002}$ \\
$a$ & Semi-major axis (au) & 0.0413 $\pm$ 0.0004 & 0.04134$_{-0.00044}^{+0.00045}$ & 0.0405 $\pm$ 0.0002 & 0.0413 $\pm$ 0.0003 & 0.0415 $\pm$ 0.0003 \\
$M_p$ & Mass (M$_J$) & 4.193 $\pm$ 0.094 & 4.221 $\pm$ 0.092 & -- & 4.18 $\pm$ 0.07 & 4.21 $\pm$ 0.07 \\
$R_p$ & Radius (R$_J$) & 1.289 $\pm$ 0.066 & 1.289 $\pm$ 0.066 & 1.194 $\pm$ 0.037 & 1.30 $\pm$ 0.01 & 1.25 $\pm$ 0.01 \\
$\rho_p$ & Density (g/cm$^3$) & 2.42 $\pm$ 0.35 & 2.45$_{-0.34}^{+0.42}$ & -- & 2.35 $\pm$ 0.04 & 2.64 $\pm$ 0.06 \\
$\log g_p$ & Surface gravity (cgs) & 3.80 $\pm$ 0.04 & -- & -- & 3.79 $\pm$ 0.01 & 3.82 $\pm$ 0.01 \\
$T_{eq}$ & Equilibrium Temperature (K) & 1626 $\pm$ 40 & -- & 1587 $\pm$ 26 & 1617 $\pm$ 18 & 1624 $\pm$ 19 \\[0.5em]
\hline
\multicolumn{7}{l}{\textbf{Primary Transit Parameters}} \\
$R_p$/$R_*$ & Radius of planet in stellar radii & 0.1071 $\pm$ 0.0014 & -- & -- & 0.1073 $\pm$ 0.0002 & 0.1029 $\pm$ 0.0006 \\
$a$/$R_*$ & Semi-major axis in stellar radii & 7.17 $\pm$ 0.28 & -- & -- & 7.14$_{-0.04}^{+0.03}$ & 7.12 $\pm$ 0.04 \\
$u_1$ ($R$ band) & Limb-darkening (linear) & -- & -- & -- & 0.344$_{-0.003}^{+0.006}$ & 0.36$_{-0.01}^{+0.02}$ \\
$u_2$ ($R$ band) & Limb-darkening (quad.) & -- & -- & -- & 0.17$_{-0.01}^{+0.02}$ & 0.28 $\pm$ 0.04 \\
$i$ & Inclination ($^\circ$) & 86 $\pm$ 0.7 & 86.6 $\pm$ 0.7 & 87.93 $\pm$ 0.62 & 86.60 $\pm$ 0.01 & 86.60 $\pm$ 0.01 \\
$b$ & Impact parameter & 0.439$_{-0.098}^{+0.065}$ & -- & 0.272 $\pm$ 0.008 & 0.43$_{-0.06}^{+0.05}$ & 0.43$_{-0.01}^{+0.01}$ \\
$T_0$ & BJD & 2455027.593 $\pm$ 0.0003 & 2455027.5928 $\pm$ 0.0004 & 2455027.593 $\pm$ 0.0002 & 2458905.6198 $\pm$ 0.0001 & 2458172.7641 $\pm$ 0.0002\\ 
\enddata
\end{deluxetable*}
\end{longrotatetable}

%%%%%%%%%%%%%%%%
%%%%%%%%%%%%%%%%%%%%%%%%%%%%%%%
\begin{rotatetable*}
\begin{deluxetable*}{ccccccc}
\tabletypesize{\scriptsize}
\tablewidth{\textwidth}
\tablecaption{System Parameters for TOI-1516 \label{tab:syspar_toi1516}}
\tablehead{
\colhead{Parameters} & \colhead{Units} & \colhead{\cite{kabath2022toi}} & \colhead{\cite{fox2022neossat}} & \colhead{This study} & \colhead{This study} \\
\colhead{} & \colhead{} & \colhead{} & \colhead{} & \colhead{(TESS)} & \colhead{(ADYU60)} \\
}
\startdata
\multicolumn{6}{l}{\textbf{Stellar Parameters}} \\
$M_*$ & Mass (M$_\odot$) & 1.085$_{-0.066}^{+0.061}$ & -- & 1.10 $\pm$ 0.04 & 1.16$_{-0.04}^{+0.04}$ \\
$R_*$ & Radius (R$_\odot$) & 1.245$_{-0.032}^{+0.031}$ & -- & 1.35 $\pm$ 0.02 & 1.26 $\pm$ 0.02 \\
$L_*$ & Luminosity (L$_\odot$) & -- & -- & 2.79 $\pm$ 0.18 & 2.54 $\pm$ 0.17 \\
$\rho_*$ & Density (g/cm$^3$) & 1.090 $\pm$ 0.031 & 0.901 $\pm$ 0.150 & 0.64 $\pm$ 0.03 & 0.81$_{-0.03}^{+0.05}$ \\
$\log g_*$ & Surface gravity (cgs) & 4.25 $\pm$ 0.15 & -- & 4.22 $\pm$ 0.02 & 4.29$_{-0.02}^{+0.02}$ \\
$T_{eff}$ & Effective Temperature (K) & 6520 $\pm$ 90 & -- & 6426$_{-89}^{+86}$ & 6487 $\pm$ 85 \\
\ [Fe/H] & Metallicity & -0.05 $\pm$ 0.1 & -- & -0.30 $\pm$ 0.11 & -0.23 $\pm$ 0.11 \\
\hline
\multicolumn{6}{l}{\textbf{Planetary Parameters}} \\
$e$ & Eccentricity & 0 & -- & 0.12$_{-0.03}^{+0.04}$ & 0.04$_{-0.03}^{+0.05}$ \\
$P$ & Period (d) & 2.056014$_{-0.000002}^{+0.000002}$ & 2.056034 $\pm$ 0.000012 & 2.05601399$_{-0.0000002}^{+0.0000002}$ & 2.0560118 $\pm$ 0.0000005 \\
$a$ & Semi-major axis (au) & -- & -- & 0.0322 $\pm$ 0.0004 & 0.0333 $\pm$ 0.0004 \\
$M_p$ & Mass (M$_J$) & 3.16 $\pm$ 0.12 & -- & 3.05 $\pm$ 0.09 & 3.17$_{-0.09}^{+0.09}$ \\
$R_p$ & Radius (R$_J$) & 1.36 $\pm$ 0.03 & -- & 1.57 $\pm$ 0.03 & 1.51 $\pm$ 0.03 \\
$\rho_p$ & Density ($\rho_J$) & -- & -- & 0.97 $\pm$ 0.05 & 1.14$_{-0.05}^{+0.06}$ \\
$\log g_p$ & Surface gravity (cgs) & -- & -- & 3.48 $\pm$ 0.02 & 3.54 $\pm$ 0.02 \\
$T_{eq}$ & Equilibrium Temperature (K) & -- & -- & 1988 $\pm$ 30 & 1927 $\pm$ 30 \\
\hline
\multicolumn{6}{l}{\textbf{Primary Transit Parameters}} \\
\textsl{}$R_p$/$R_*$ & Radius of planet in stellar radii & 0.1224$_{-0.0005}^{+0.0005}$ & 0.1177 $\pm$ 0.0019 & 0.1200 $\pm$ 0.0002 & 0.1228 $\pm$ 0.0007 \\
$a$/$R_*$ & Semi-major axis in stellar radii & 6.22$_{-0.077}^{+0.041}$ & 5.834 $\pm$ 0.331 & 5.22 $\pm$ 0.09 & 5.65$_{-0.08}^{+0.12}$ \\
$u_1$ ($R$ band) & Limb-darkening (linear) & -- & -- & 0.20 $\pm$ 0.02 & 0.21 $\pm$ 0.03 \\
$u_2$ ($R$ band) & Limb-darkening (quad.) & -- & -- & 0.22 $\pm$ 0.04 & 0.34 $\pm$ 0.04 \\
$i$ & Inclination ($^\circ$) & 90 $\pm$ 0.4 & 87.02 $\pm$ 1.85 & 86.0 $\pm$ 0.1 & 86.0 $\pm$ 0.1 \\
$b$ & Impact parameter & 0.09$_{-0.07}^{+0.10}$ & 0.302 $\pm$ 0.187 & 0.32 $\pm$ 0.01 & 0.40$_{-0.01}^{+0.02}$ \\
$T_0$ & BJD & 2458765.325 $\pm$ 0.1 & 2459392.410 $\pm$ 0.0004 & 2459310.1687$_{-0.0001}^{+0.0008}$ & 2459092.2310 $\pm$ 0.0001 \\
\enddata
\end{deluxetable*}
\end{rotatetable*}

%%%%%%%%%%%%%%%%
\begin{rotatetable*}
\begin{deluxetable*}{ccccccc}
\tabletypesize{\scriptsize}
\tablewidth{\textwidth}
\tablecaption{System Parameters for TOI-2046 \label{tab:syspar_toi2046}}
\tablehead{
\colhead{Parameters} & \colhead{Units} & \colhead{\cite{kabath2022toi}} & \colhead{\cite{fox2022neossat}} & \colhead{This study} & \colhead{This study} \\
\colhead{} & \colhead{} & \colhead{} & \colhead{} & \colhead{TESS} & \colhead{ADYU60} \\
}
\startdata
\multicolumn{6}{l}{\textbf{Stellar Parameters}} \\
$M_*$ & Mass (M$_\odot$) & 1.153$_{-0.093}^{+0.10}$ & -- & $1.18^{+0.06}_{-0.06}$ & $1.18^{+0.06}_{-0.06}$ \\
$R_*$ & Radius (R$_\odot$) & 1.237$_{-0.032}^{+0.036}$ & -- & $1.28^{+0.03}_{-0.03}$ & $1.24\pm0.03$ \\
$L_*$ & Luminosity (L$_\odot$) & -- & -- & $2.11^{+0.19}_{-0.17}$ & $1.98^{+0.18}_{-0.16}$ \\
$\rho_*$ & Density (g/cm$^3$) & 0.89 $\pm$ 0.098 & 0.808 $\pm$ 0.136 & $0.80^{+0.03}_{-0.03}$ & $0.86^{+0.04}_{-0.03}$ \\
$\log g_*$ & Surface gravity (cgs) & 4.30 $\pm$ 0.15 & -- & $4.30^{+0.01}_{-0.01}$ & $4.32^{+0.01}_{-0.01}$ \\
$T_{\mathrm{eff}}$ & Effective Temperature (K) & 6250 $\pm$ 140 & -- & $6152\pm97$ & $6146^{+98}_{-99}$ \\ 
\ [Fe/H] & Metallicity & -0.060 $\pm$ 0.150 & -- & $0.01\pm0.12$ & $0.03\pm0.12$ \\
\hline
\multicolumn{6}{l}{\textbf{Planetary Parameters}} \\
$e$ & Eccentricity & 0 & -- & $0.060^{+0.026}_{-0.024}$ & $0.02^{+0.03}_{-0.02}$ \\
$P$ & Period (d) & 1.4971842$_{-0.0000033}^{+0.0000031}$ & 1.49712564 $\pm$ 0.0000082 & $1.4971871^{+0.0000003}_{-0.0000003}$ & $1.4971884^{+0.0000006}_{-0.0000006}$ \\
$a$ & Semi-major axis (au) & -- & -- & $0.0270\pm0.0004$ & $0.0271^{+0.0004}_{-0.0004}$ \\
$M_p$ & Mass (M$_J$) & 2.30 $\pm$ 0.28 & -- & $2.35^{+0.09}_{-0.09}$ & $2.35^{+0.09}_{-0.09}$ \\
$R_p$ & Radius (R$_J$) & 1.44 $\pm$ 0.11 & -- & $1.51^{+0.03}_{-0.03}$ & $1.52\pm0.03$ \\
$\rho_p$ & Density ($\rho_J$) & -- & -- & $0.86^{+0.03}_{-0.04}$ & $0.84^{+0.03}_{-0.03}$ \\
$\log g_p$ & Surface gravity (cgs) & -- & -- & $3.41^{+0.01}_{-0.01}$ & $3.41\pm0.01$ \\
$T_\mathrm{eq}$ & Equilibrium Temperature (K) & -- & -- & $2038\pm35$ & $2007\pm34$ \\
\hline
\multicolumn{6}{l}{\textbf{Primary Transit Parameters}} \\
$R_p$/\textsl{R}$_*$ & Radius ratio & 0.1213$_{-0.0021}^{+0.0017}$ & 0.101 $\pm$ 0.002 & $0.1213\pm0.0004$ & $0.1255^{+0.0009}_{-0.0009}$ \\
$a$/$R_*$ & Scaled semi-major axis & 4.75$_{-0.17}^{+0.18}$ & 4.561 $\pm$ 0.252 & $4.56^{+0.06}_{-0.07}$ & $4.68^{+0.07}_{-0.05}$ \\
$u_1$ ($R$ band) & Limb-darkening (linear) & -- & -- & $0.21^{+0.03}_{-0.03}$ & $0.32\pm0.04$ \\
$u_2$ ($R$ band) & Limb-darkening (quad.) & -- & -- & $0.24^{+0.04}_{-0.04}$ & $0.31^{+0.05}_{-0.05}$ \\
$i$ & Inclination ($^\circ$) & 83.6 $\pm$ 0.9 & 86.62 $\pm$ 2.35 & $83.6\pm0.1$ & $83.6\pm0.1$ \\
$b$ & Impact parameter & 0.51$_{-0.07}^{+0.06}$ & 0.272 $\pm$ 0.187 & $0.48^{+0.02}_{-0.02}$ & $0.51^{+0.01}_{-0.01}$ \\
$T_0$ &  BJD & 2457792.2767 $\pm$ 0.0024 & 2459424.214 $\pm$ 0.00085 & $2460121.89760^{+0.00008}_{-0.00008}$ & $2460247.6612\pm0.0002$ \\
\enddata
\end{deluxetable*}
\end{rotatetable*}

%%%%%%%%%

%%%%%%%%%%%%%%%%
\section{Results and Discussion}
\subsection{System Parameters}
\begin{figure}[ht!]
%\centering
\hspace{-0.5cm}
\includegraphics[scale=0.6]{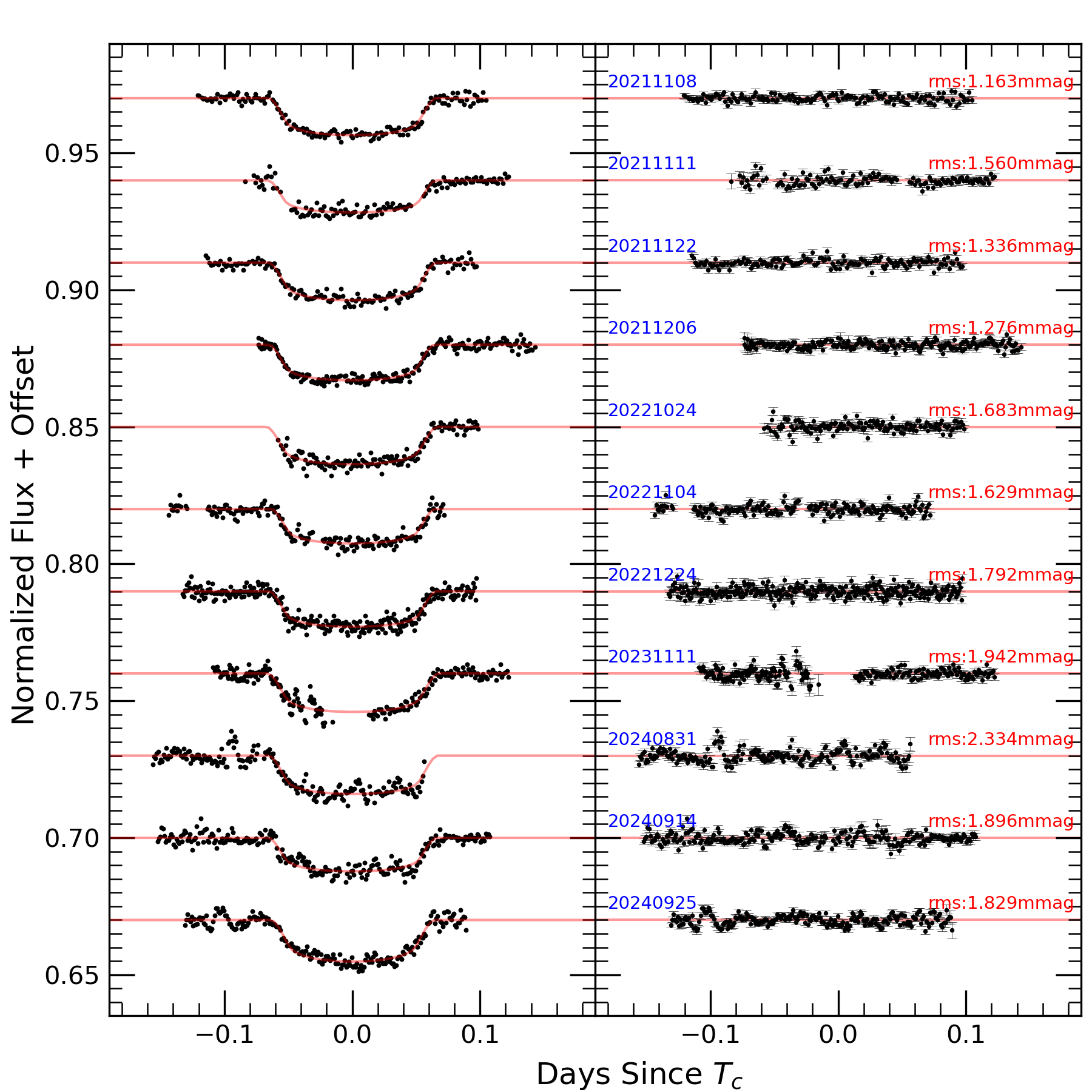}
\caption{The left panel shows eleven new transit light curves of HAT-P-16b observed with the ADYU60 telescope along with the transit model fits indicated as red lines. The y-scale is arbitrarily adjusted to provide visual clarity. The right panel displays the residuals along with the dates (in the format: yyyymmdd) and the RMS calculated from residuals (in mmag) for each light curve.}
\label{fig:hatp16adyu}
\end{figure}

\begin{figure}
\centering
\includegraphics[scale=0.9]{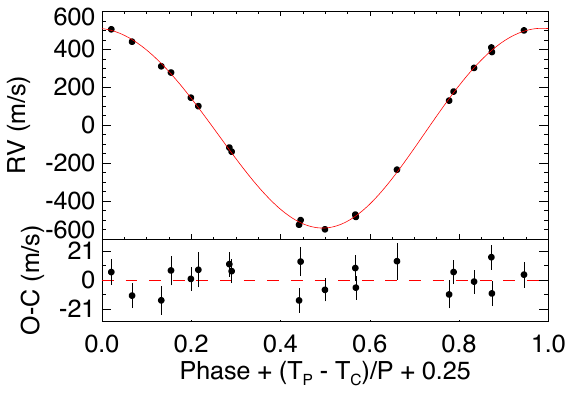}
\caption{Distribution of the radial velocities (RVs) for HAT-P-16b in the top panel and the residuals in the bottom panel (\cite{buchhave2010hat}).}
\label{fig:hatp16tess}
\end{figure}

\begin{figure*}[ht!]
\centering
\includegraphics[scale=0.5]{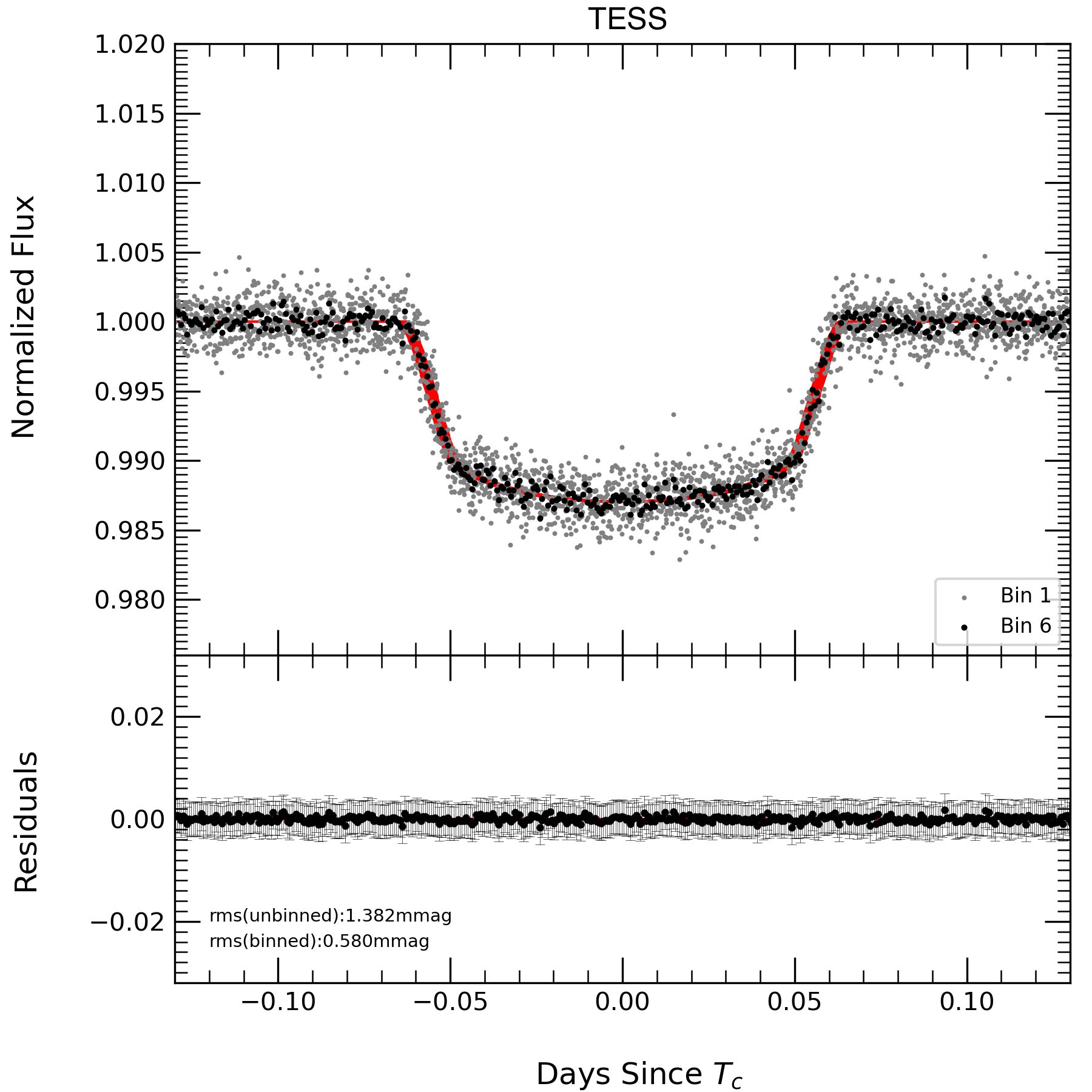}
\includegraphics[scale=0.5]{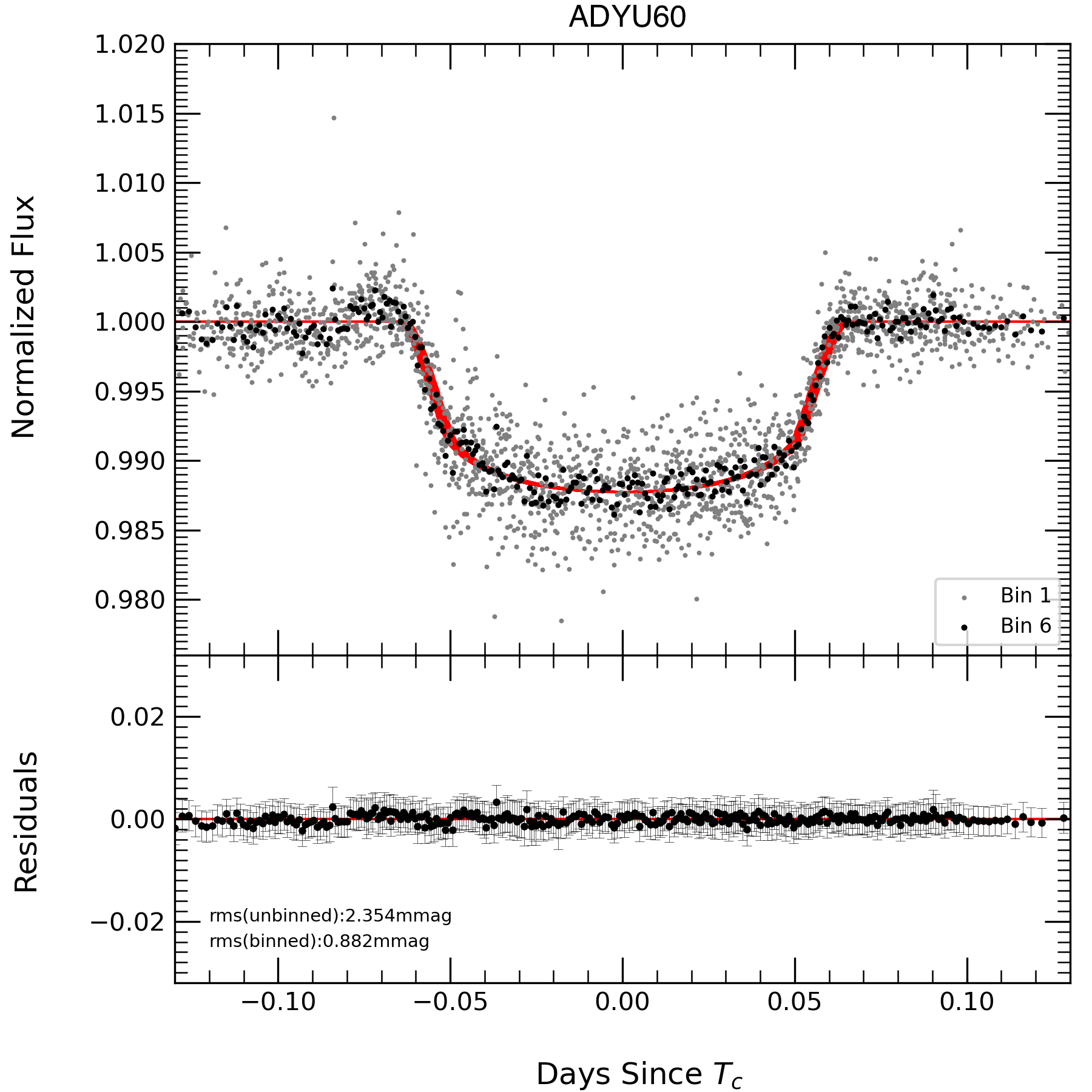}
\caption{Transit light curves of the exoplanet HAT-P-16b from two observational datasets. The left panel presents data from the TESS, and the right panel shows data from ADYU60. Each panel displays the normalized flux as a function of time, with unbinned data points shown as small black dots and binned data represented by a black line. The best-fit transit model is overlaid in red, highlighting the transit event. Below each flux plot, the residuals for the binned data are displayed with root mean square (RMS) values of 0.58 mmag for TESS and 0.88 mmag for ADYU60.}
\label{fig:hatp16globals}
\end{figure*}

\begin{figure}[ht!]
%\centering
\hspace{-0.5cm}
\includegraphics[scale=0.6]{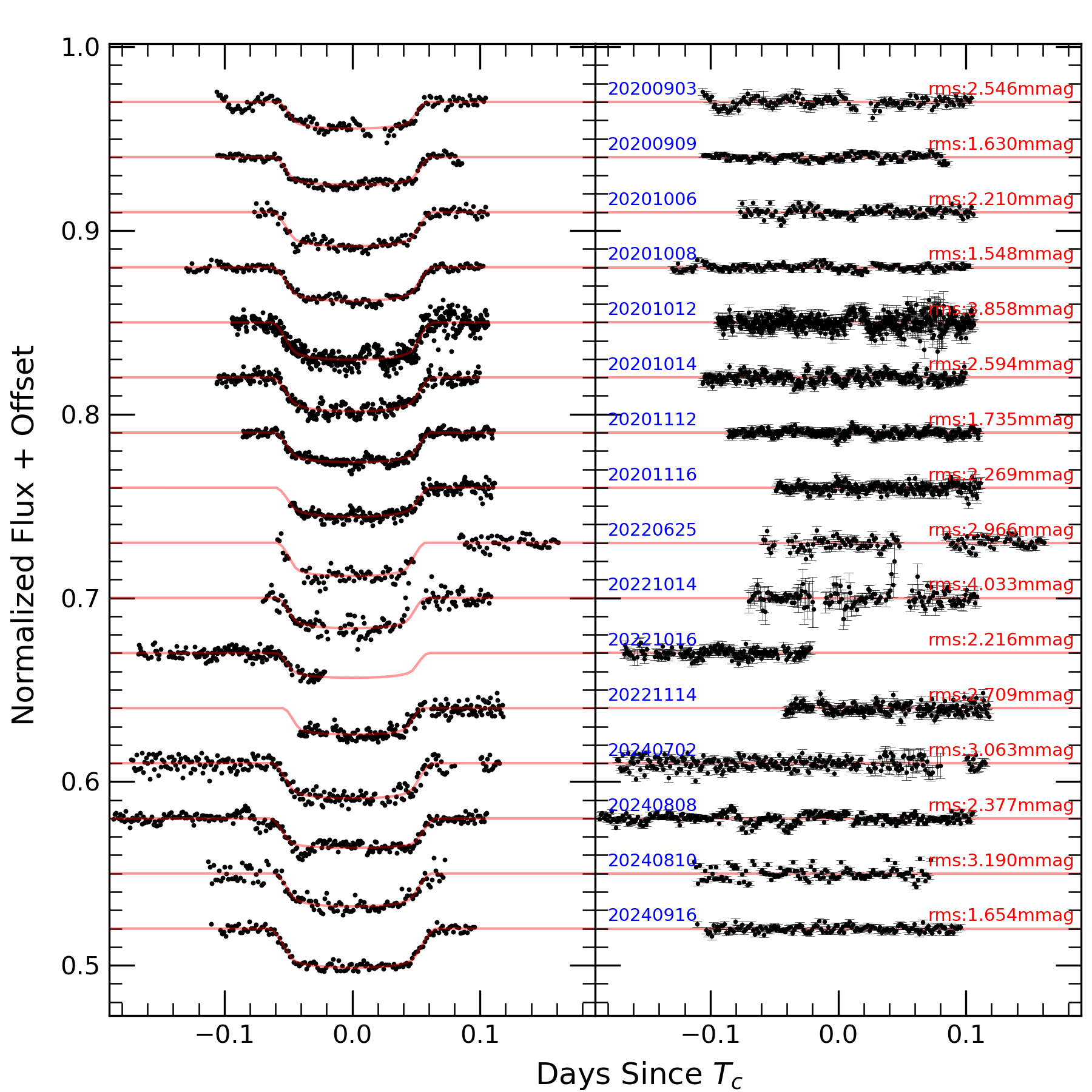}
\caption{Sixteen new transit light curves of TOI-1516b observed with the ADYU60 telescope.}
\label{fig:toi1516adyu}
\end{figure}

\begin{figure}[ht!]
\centering
\includegraphics[scale=0.9]{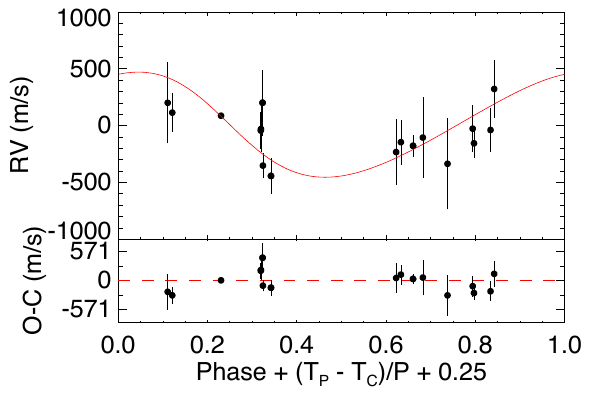}
\caption{Distribution of the radial velocities (RVs) for TOI-1516b in the top panel and the residuals in the bottom panel (\cite{kabath2022toi}).}
\label{fig:toi1516tess}
\end{figure}

\begin{figure*}[ht!]
\centering
\includegraphics[scale=0.5]{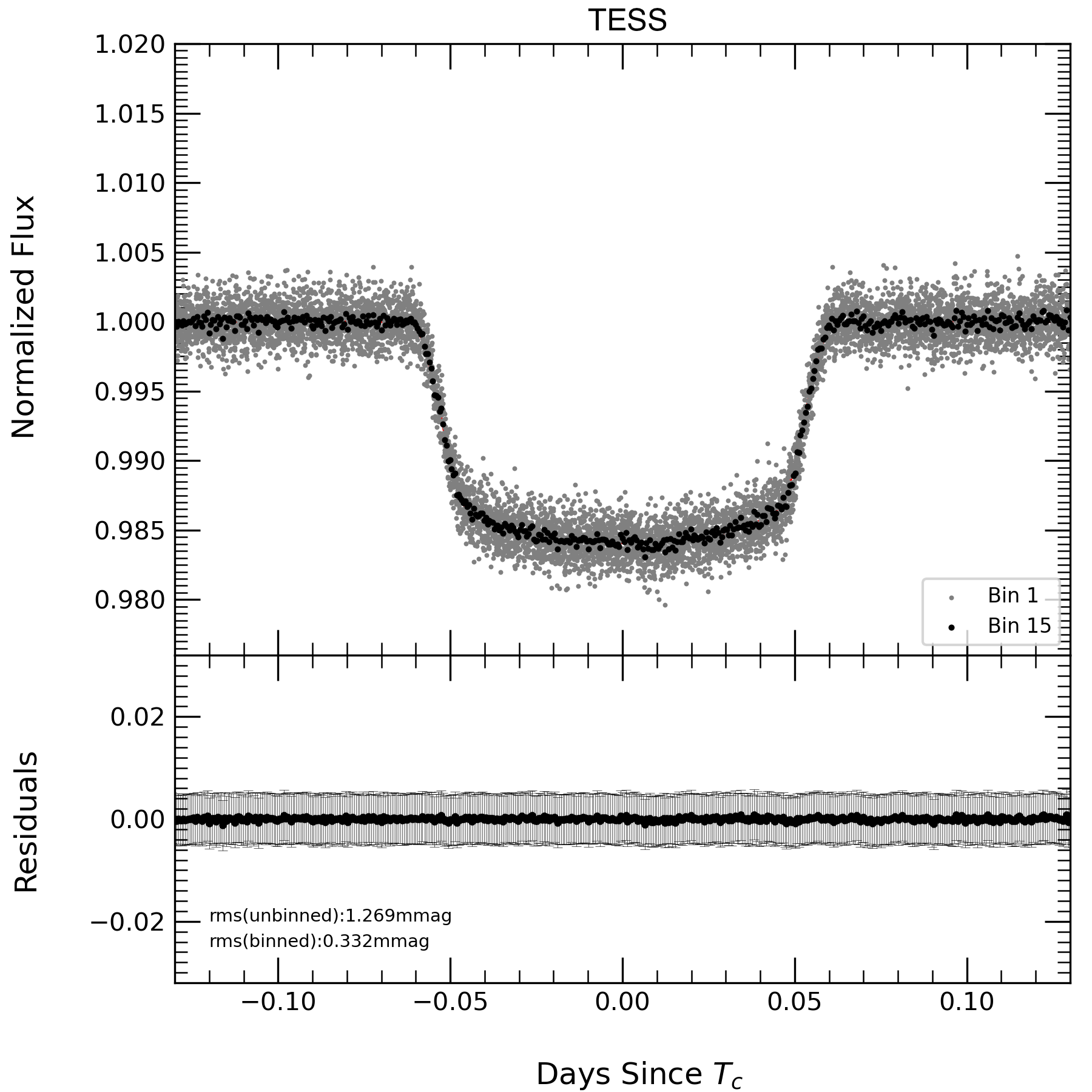}
\includegraphics[scale=0.5]{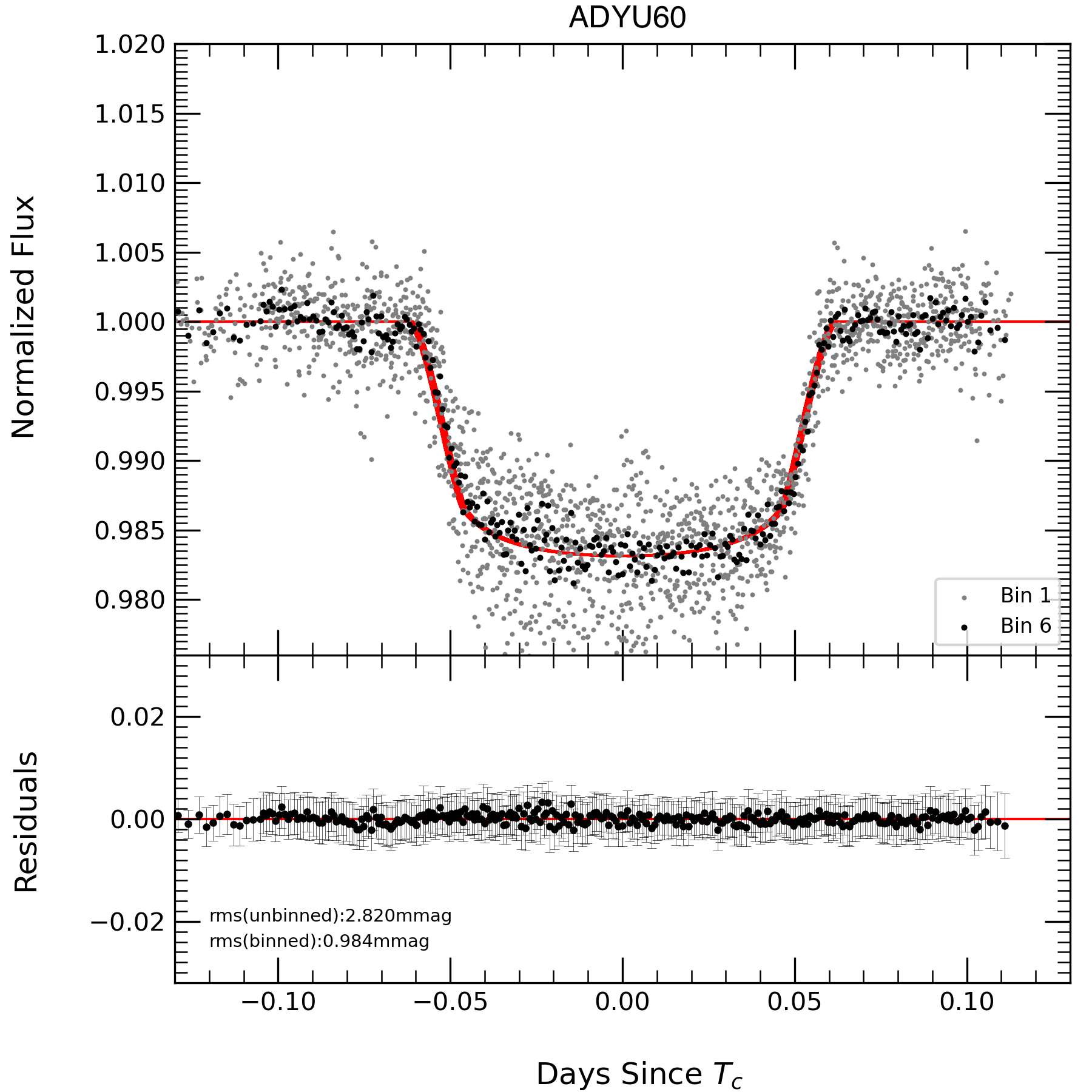}
\caption{Same as in Figure 4 but for TOI-1516b. Below each flux plot, the residuals for the binned data are displayed with root mean square (RMS) values of 0.33 mmag for TESS and 0.98 mmag for ADYU60}.
\label{fig:toi1516globals}
\end{figure*}

\begin{figure}[ht!]
%\centering
\hspace{-0.5cm}
\includegraphics[scale=0.6]{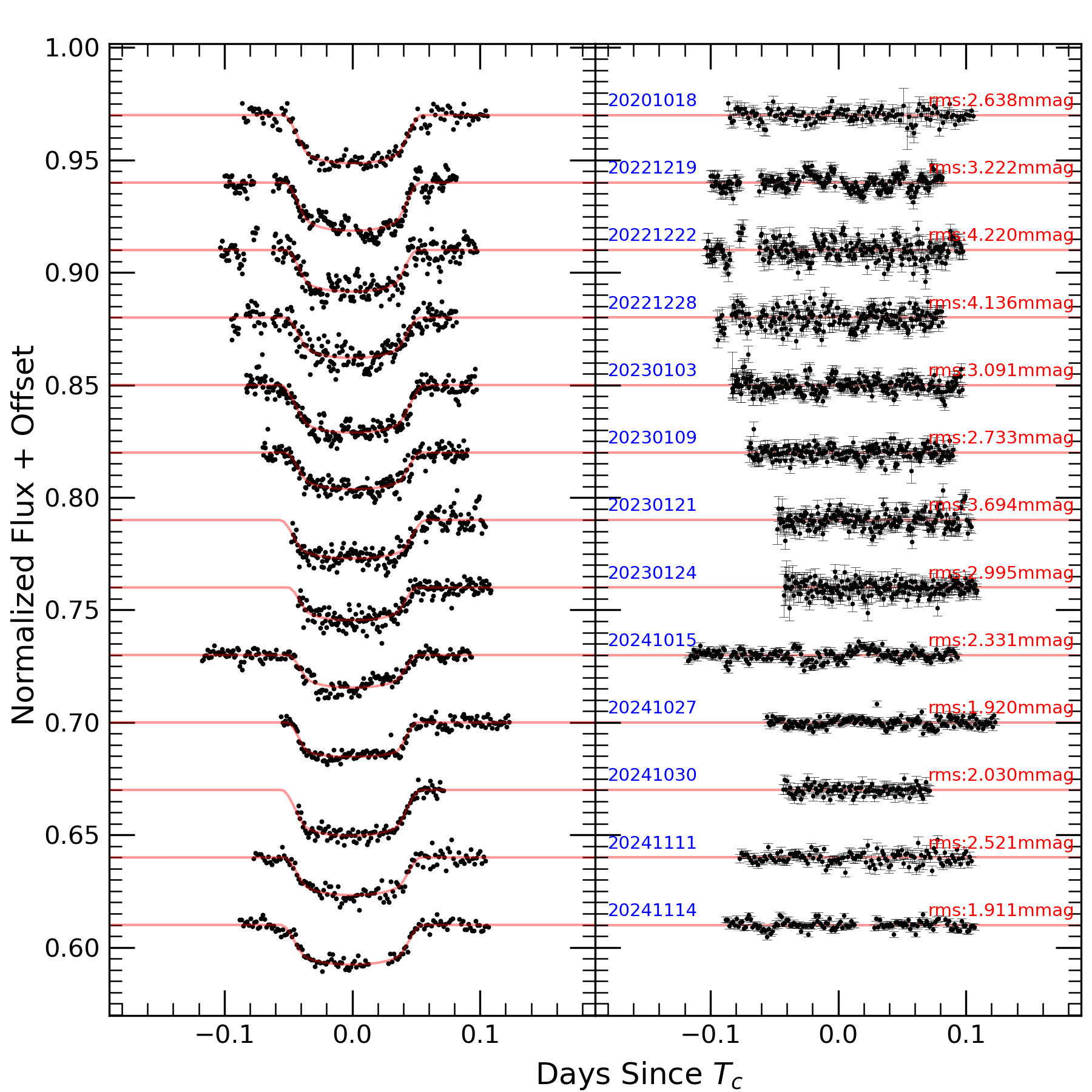}
\caption{Thirteen new transit light curves of TOI-2046b observed with the ADYU60 telescope.}
\label{fig:toi2046adyu}
\end{figure}

\begin{figure}[ht!]
\centering
\includegraphics[scale=0.9]{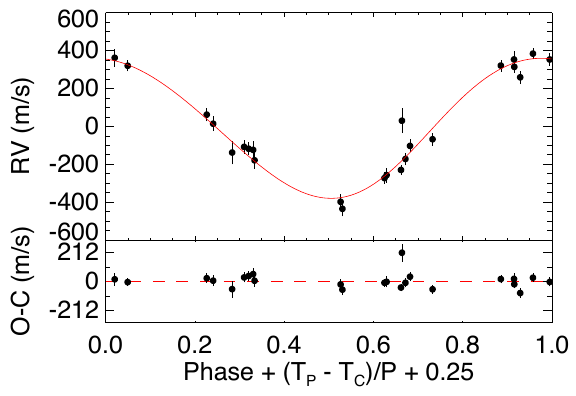}
\caption{Distribution of the radial velocities (RVs) for TOI-2046b in the top panel and the residuals in the bottom panel (\cite{kabath2022toi}).}
\label{fig:toi2046RV}
\end{figure}

\begin{figure*}[ht!]
\centering
%\hspace{-0.5cm}
\includegraphics[scale=0.25]{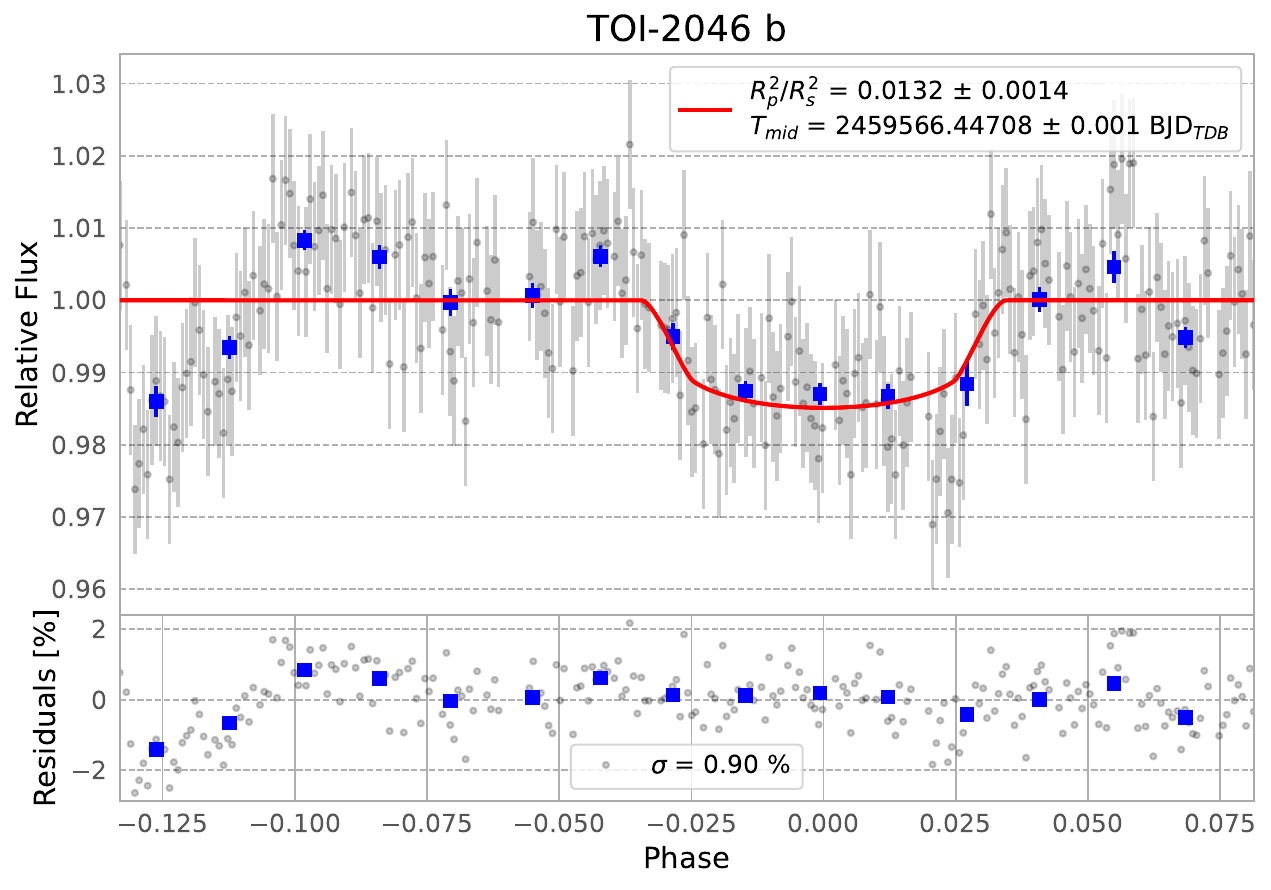}
\includegraphics[scale=0.25]{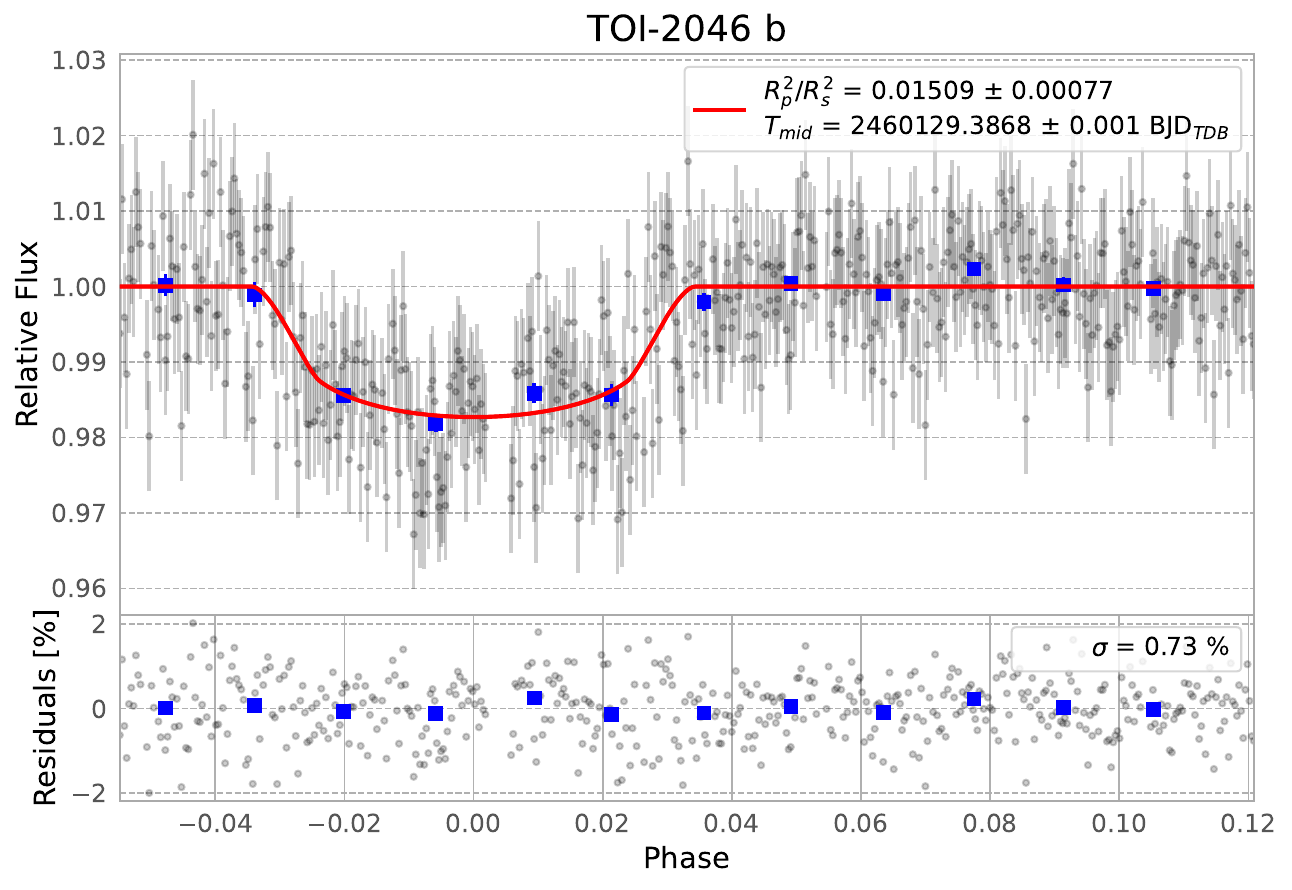}
\includegraphics[scale=0.25]{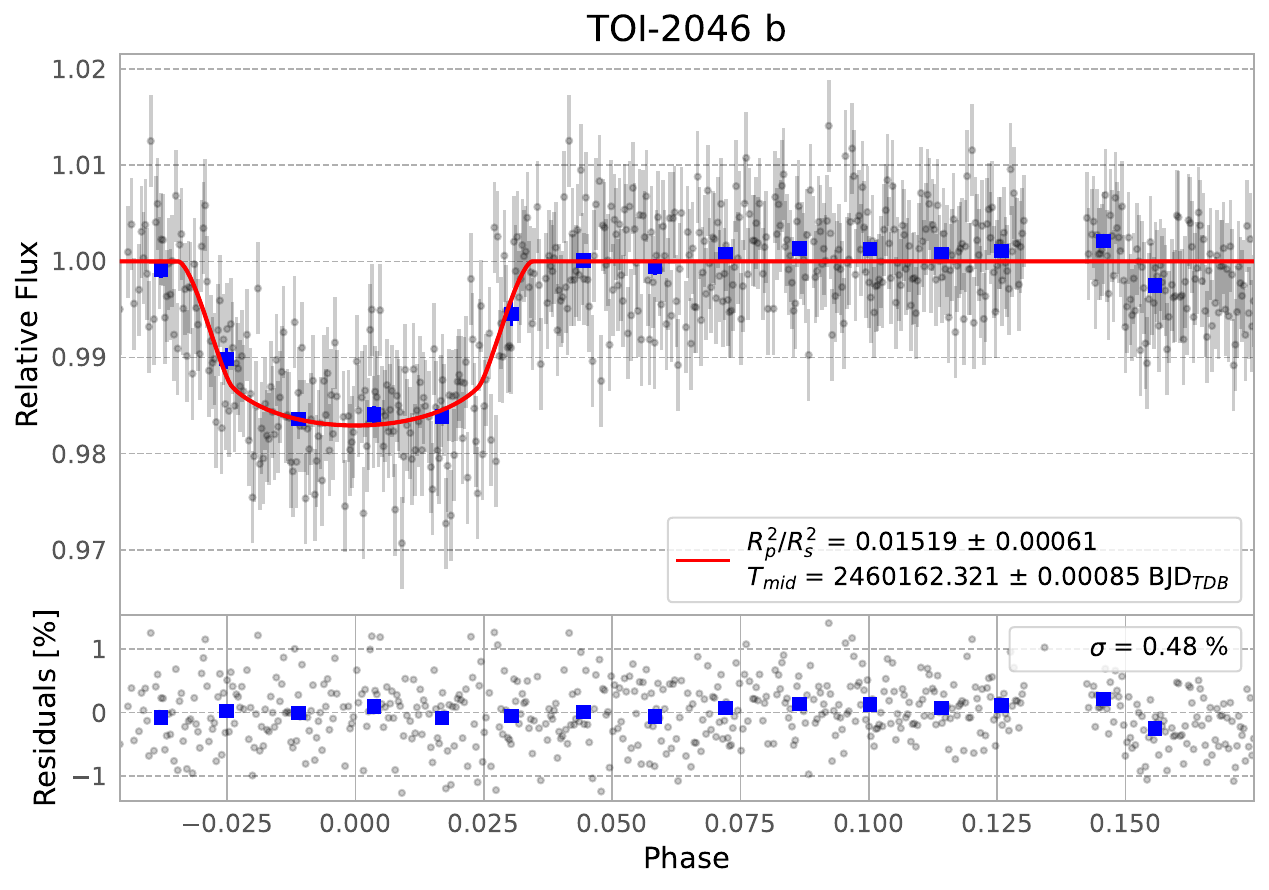}
\includegraphics[scale=0.25]{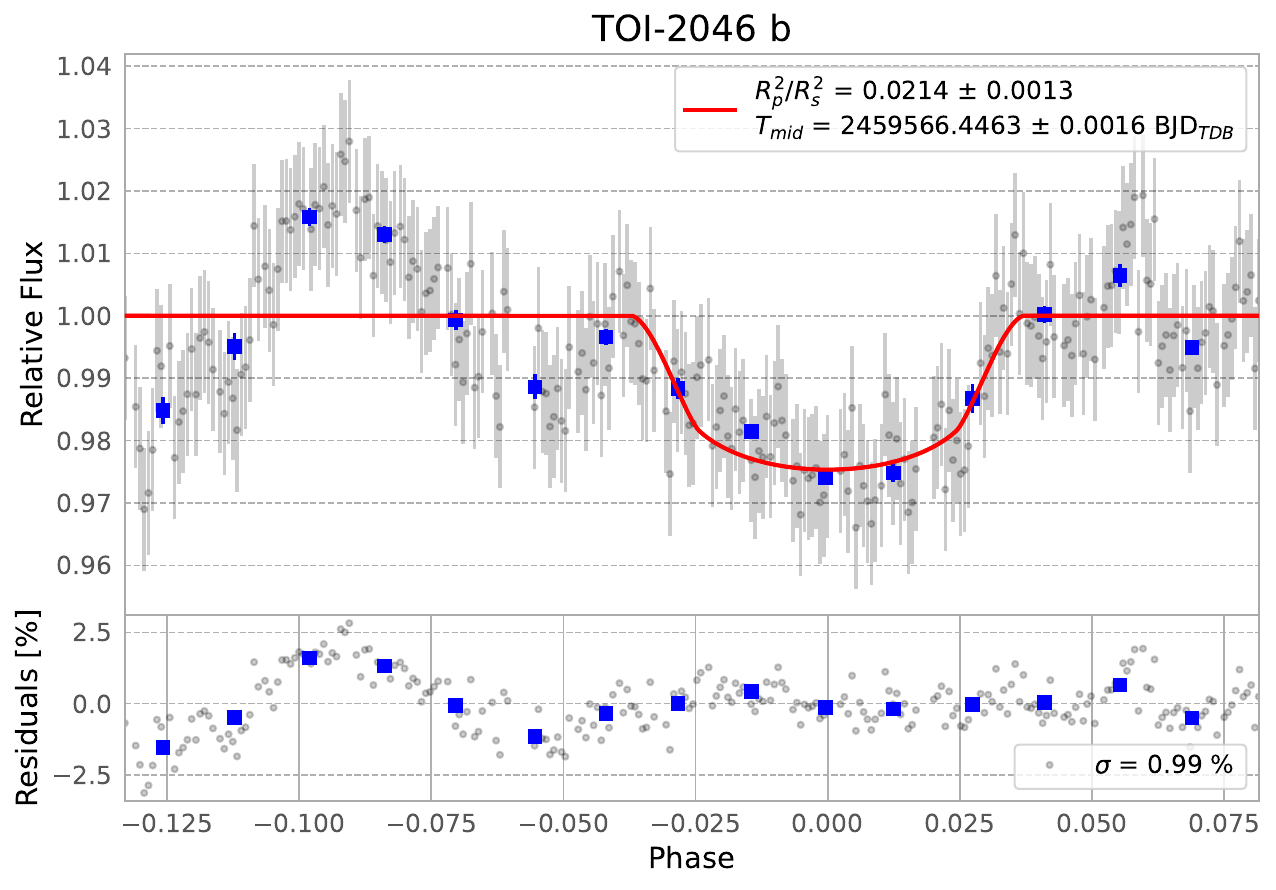}
\includegraphics[scale=0.25]{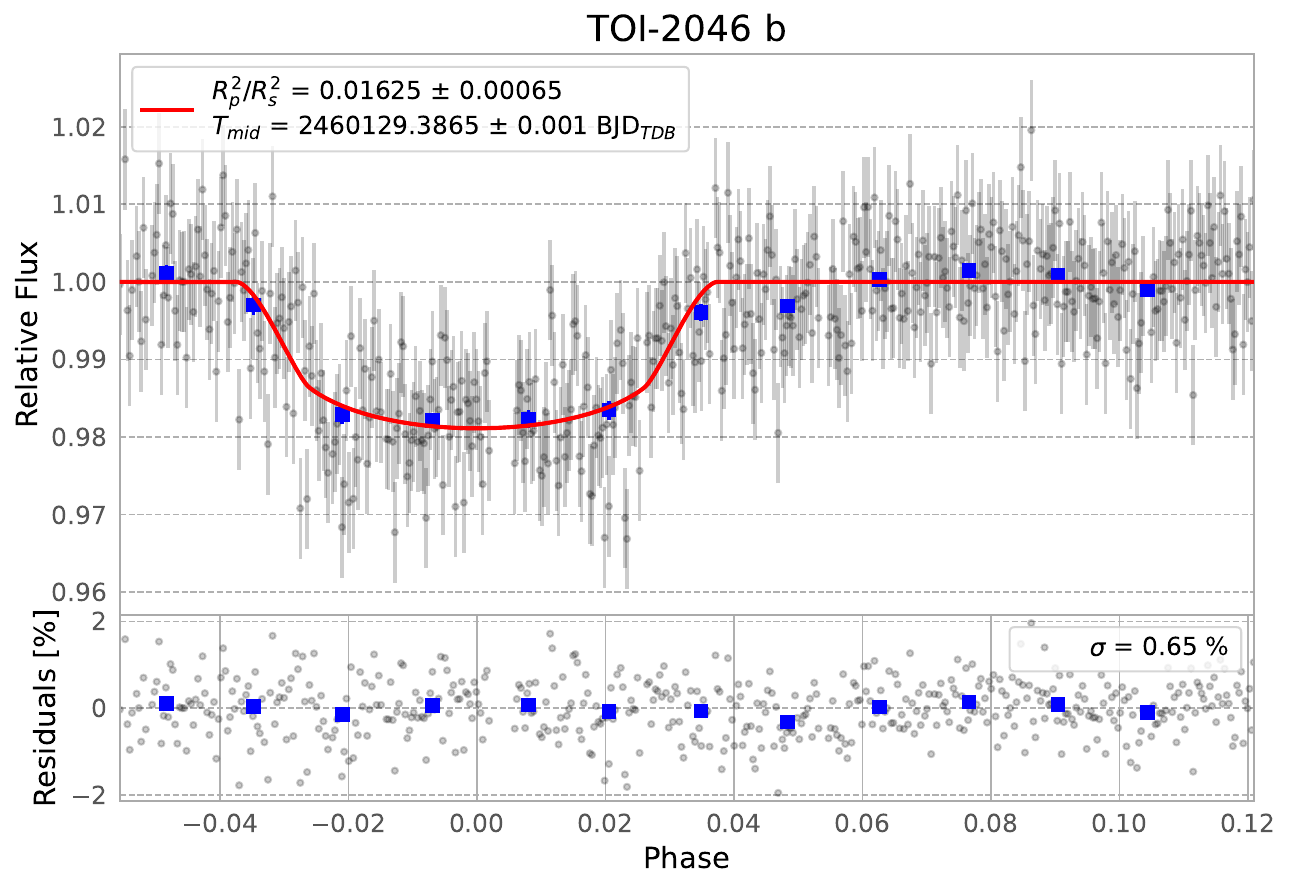}
\includegraphics[scale=0.25]{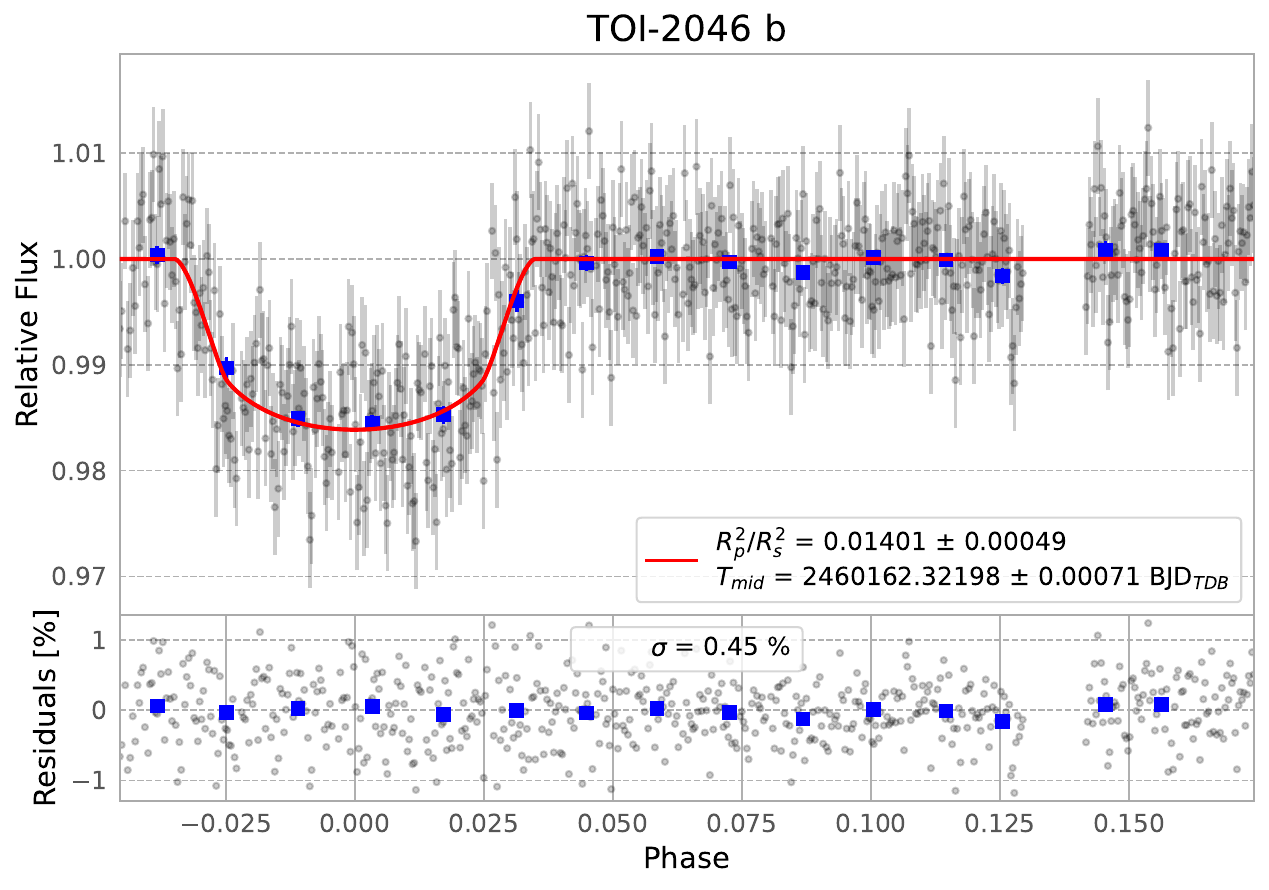}
\caption{Three new transit light curves of TOI-2046b observed with R (upper row) and V (lower row) filters on {\bf Left:} December 17, 2021; {\bf Middle:} July 3, 2023 and {\bf Right:} August 5, 2023 with the T100 telescope.}
\label{fig:toi2046fromT100}
\end{figure*}

\begin{figure*}[ht!]
\centering
\includegraphics[scale=0.5]{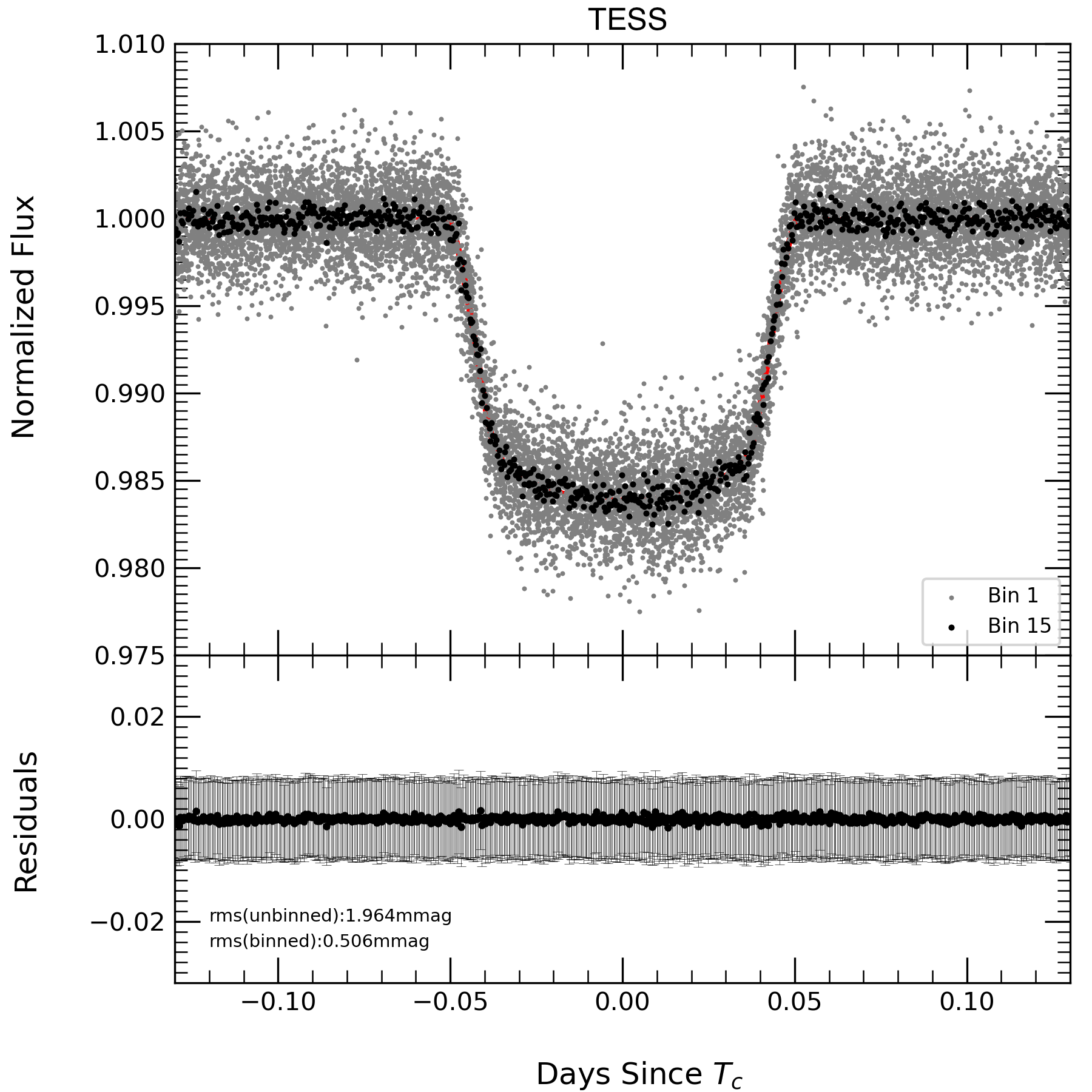}
\includegraphics[scale=0.5]{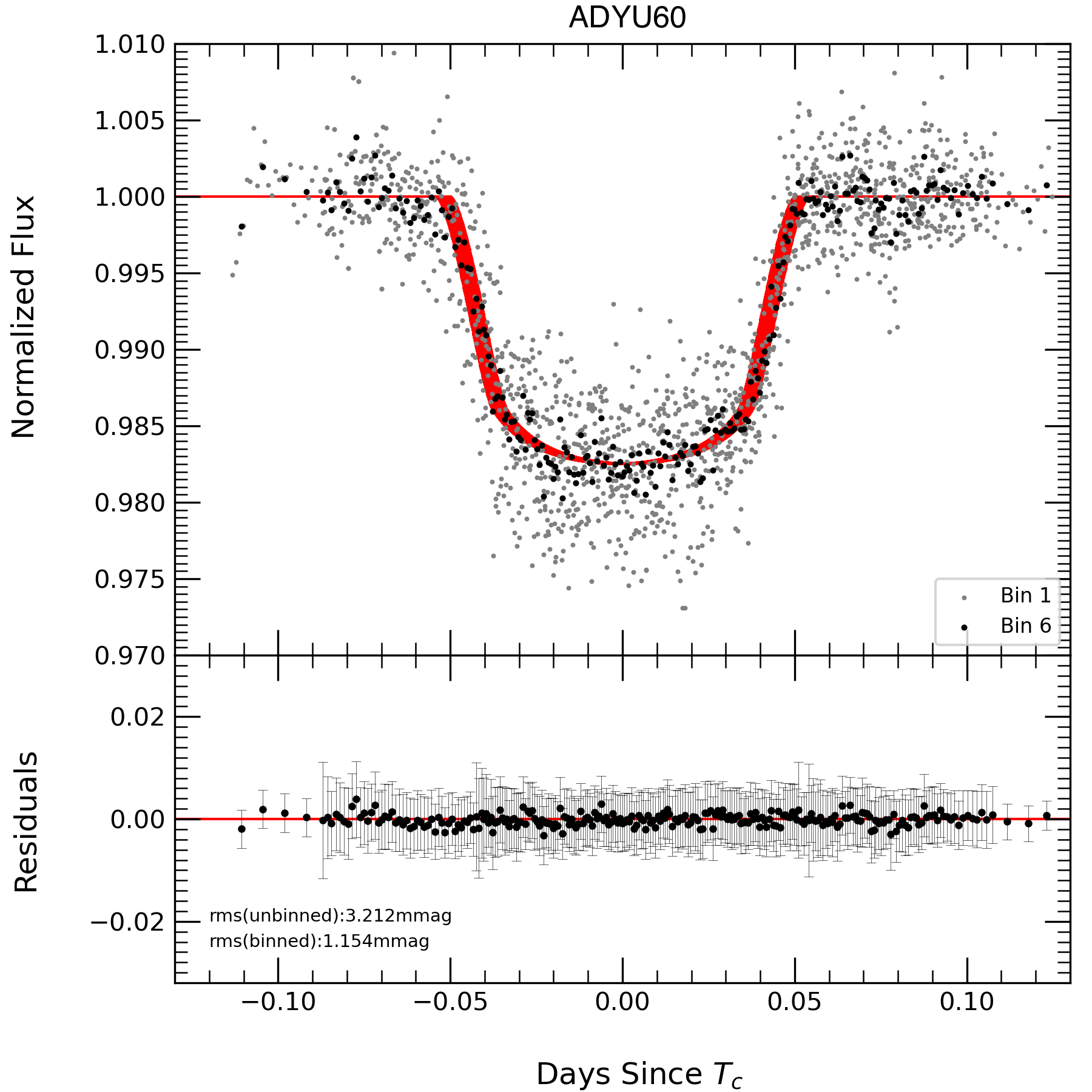}
\caption {Same as in Figure 4 but for TOI-2046b. Below each flux plot, the residuals for the binned data are displayed with root mean square (RMS) values of 0.51 mmag for TESS and 1.15 mmag for ADYU60}.

\label{fig:toi2046globals}
\end{figure*}

\noindent We begin by briefly outlining the totality of the analysis results before presenting the details for each individual system studied. \\
\\
The system parameters extracted from the global fits using EXOFASTv2 for each system are presented in Tables \ref{tab:syspar_hatp16}, 
\ref{tab:syspar_toi1516} and \ref{tab:syspar_toi2046} along with the parameters reported from previous studies. The final ADYU60  
light curves for HAT-P-16b, TOI-1516b, and TOI-2046b, along with the fits of the transit model, are shown in Figures \ref{fig:hatp16adyu},
\ref{fig:toi1516adyu}, and \ref{fig:toi2046adyu} respectively. Three transit curves extracted from the observations taken by TUG-T100 for TOI-2046b are given in Figure \ref{fig:toi2046fromT100}. The combined ADYU60 and TESS global transit fits for each system are shown in Figures \ref{fig:hatp16globals}, \ref{fig:toi1516globals}, and \ref{fig:toi2046globals}, respectively. The TESS and ADYU60 data are presented as separate figures to highlight differences in filter bandpasses (e.g., TESS vs. $R$), which can potentially affect the transit model fits in EXOFAST. The appropriate fits to the RV data are also shown in Figures \ref{fig:hatp16tess}, \ref{fig:toi1516tess} and \ref{fig:toi2046RV}. The final TESS light curves for HAT-P-16b, TOI-1516b, and TOI-2046b, along with the fits of the transit model are shown in the (online)
Appendix; see Figures \ref{fig:hatp16_tess}, \ref{fig:toi1516_tess},  \ref{fig:toi2046_tess}. \\
\\
For all the systems studied, we adopted a uniform approach in the final selection and analysis of the data. For instance, to ensure robustness and reproducibility of the extracted parameters, not all of the transit observations were utilized in the final modeling process, i.e., only those light curves that met specific quality criteria were considered for the fitting. These criteria included a threshold root mean square (rms) of scatter, the presence of both ingress and egress phases, and a well-defined baseline. Although incorporating as much data as possible is always preferable, imposing such a threshold helps mitigate the influence of highly scattered or incomplete observations. The chosen rms cutoff provides a balance between maximizing the available data set and maintaining a robust model fit.

%%%%%

%%%%%

%%%%

%%%%%

%%%%%

%%%%%%

%%%%%%%%
%%%%%
%%%%%

%%%%%

%%%%%%%%
%%%%%

\subsubsection{HAT-P-16b}

\begin{figure*}
\centering
\includegraphics[scale=0.5]{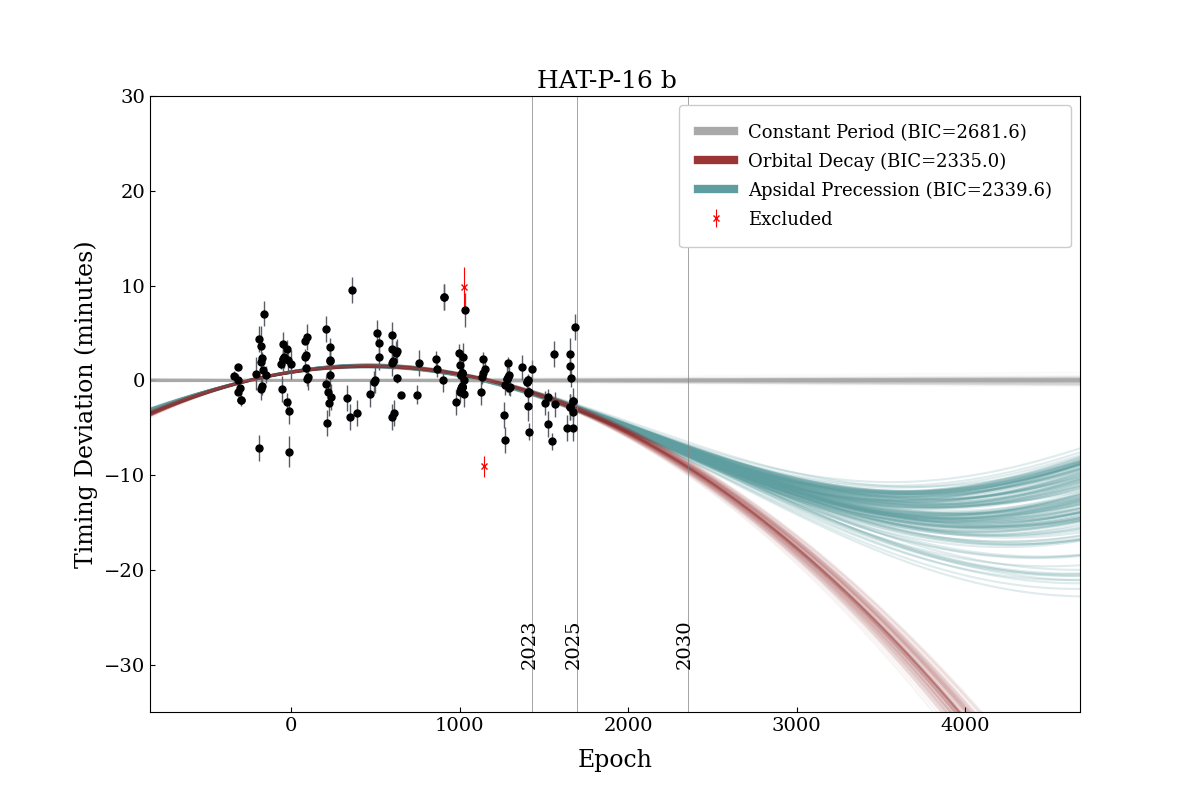}
\caption{Timing residuals of HAT-P-16 b with future projections of the three transit-timing models shown with 150 random draws from the MCMC posterior chains. Each data point is the difference between an observed time and the time predicted by the best-fit constant-period model. The size of the data points corresponds to the data quality index from 1 to 3, with the higher-quality transits being larger. The red crosses represent observations removed during the sigma-clipping process.}
\label{fig:hatp16b-expdot-oc}
\end{figure*}

\begin{figure*}
\vspace{-1.5cm}
\centering
\includegraphics[scale=1.0]{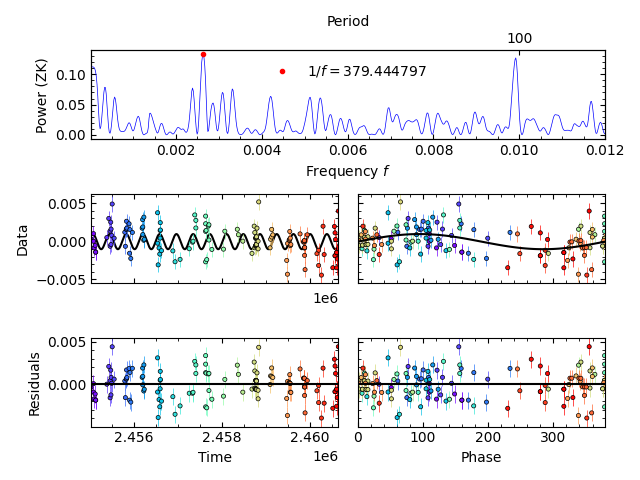}
\caption{The generalised Lomb-Scargle periodogram of HAT-P-16b. The top panel displays the power spectrum, the middle panel (left) shows the TTV distribution of the system, the bottom panel (left) presents the residuals, the middle panel (right) illustrates the variation of periodicity with phase, and the bottom panel (right) shows the residuals from the fit.}
\label{fig:hatp16b-GLS}
\end{figure*}

\noindent Figure \ref{fig:hatp16adyu} shows 11 normalized light curves with offsets, observed by ADYU60 over three years for HAT-P-16b. The left panel displays all transit observations, each tagged with specific observation dates, and the right panel presents the residuals of these curves best fitting model. As noted above, not all were utilized in the final modeling process; a threshold root mean square (rms) scatter of $ \leq 2.2$ mmag was applied in the selection. \\
\\
Figure \ref{fig:hatp16_tess} presents 14 transit light curves of HAT-P-16b extracted from TESS observations, along with the corresponding transit model fits (shown as red lines) and the residuals. Each data set is normalized with offsets to facilitate clear visualization. The space-based nature of TESS minimizes atmospheric interference, leading to improved photometric precision and lower rms values (e.g., rms: 1.42 mmag, rms: 1.38 mmag). This reduction in systematic noise improves the accuracy of transit parameter estimation, allowing for a more detailed investigation of the characteristics of the planetary system. Figure \ref{fig:hatp16tess} displays the radial velocity (RV) measurements of HAT-P-16b as a function of its orbital phase, highlighting the periodic motion induced by the orbit of the planet. The top panel shows the RV data points (black dots) with the best-fit model (red curve), indicating periodic motion. The bottom panel depicts the residuals (O-C) centered around zero, suggesting a well-fitting model that is crucial for determining the orbital parameters and confirming the presence of the orbiting planet.\\
\\
The global transit light curves of HAT-P-16b were obtained from two different telescopes: the space-based TESS and the ground-based ADYU60 telescope (left and right panels of Fig. \ref{fig:hatp16globals}, respectively). The best-fit transit models are represented with the red line, while the residuals are shown in the lower panels for each dataset.\\
\\
Despite the high scatter in the ADYU60 light curve, the transit event is clearly detected, with ingress and egress phases distinctly resolved. The elevated noise is attributed to atmospheric variability, fluctuating sky conditions, and instrumental limitations common to ground-based facilities. However, the close agreement between the ADYU60 data and the transit model supports the reliability of the observations to constrain the parameters of the system.\\ 
\\
\noindent For HAT-P-16b, the mid-transit times obtained from the literature, TESS observations, and ADYU60 observations were combined to construct the O–C (observed - calculated) curves, also known as transit timing variation (TTV) curves. Three model approximations were applied to the O–C variation curves using the ExoPdot software in combination with MCMC optimization (for details see Section \ref{TTV Modeling}).  The model fits obtained as a result of these approximations are given in Figure \ref{fig:hatp16b-expdot-oc} and Table \ref{tab:HAT-P-16b_exopdot_result}. The corner plots for each of the three models are shown in the (online)
Appendix; see Figures \ref{fig:hatp16b-linear-corner}-\ref{fig:hatp16b-decay-corner}-\ref{fig:hatp16b-precession_corner}. When the BIC and $\Delta$BIC values obtained from the analysis of the O-C diagram for this system are compared for the different models, it is evident that the orbital decay model (BIC = 2335.0; see Table \ref{tab:HAT-P-16b_exopdot_result}) provides the best fit. Our findings for HAT-P-16b can be placed in direct context with the recent analysis by \citet{Sun2023}, who, based on an analysis of BIC ($\Delta$BIC) values and Mengo maps, reported significant TTVs that the authors state are best explained with the apsidal precession model. We note that their BIC ($\Delta$BIC) values for the orbital decay model are not too dissimilar from those for the apsidal precession model. We too find BIC ($\Delta$BIC) values that are very close for the two models ($\Delta$BIC = 4.6) but in contrast to the findings of \citet{Sun2023}, we find slightly lower BIC ($\Delta$BIC) values for the orbital decay model. Interestingly they report very large $\Delta$BIC values (317.7 and 152.5 respectively), something that is clearly reflected in the fits they present in their Figure 8 --- the two models deviate from each other quite significantly. No such deviation is noticeable in our fits (see our Figure \ref{fig:hatp16b-expdot-oc}) which incorporate a significantly larger number of ground-based transits, including several from the past two years. The difference between our results and those of \citet{Sun2023} may arise from the expanded baseline in our dataset and improved treatment of partial transits. Continued monitoring of the system with high-precision photometry will be essential to definitively establish the cause of the observed TTVs.\\
\\
The residuals of the TTV (O-C) data were analyzed using the GLS periodogram package to extract the dominant frequencies and their amplitudes. The results are displayed in Figure \ref{fig:hatp16b-GLS}. The maximum frequency obtained from the power analysis performed in GLS is 0.002635 ($\pm$0.000032) Hz, with a power value of 0.132869. The best sine period obtained from this frequency is 379.45 ($\pm$4.66) days, and the amplitude is 0.00099 ($\pm$0.00020). The RMS error of the residuals is 0.001665. The FAP value obtained from the GLS analysis is 0.005524, a value that suggests that the extracted periodicity and the detection are not robust. Based on the period change rate of the decay model $(\mathrm{dP/dE} = -4.12 \times 10^{-9}days/epoch)$ and the astrophysical parameters of the system, the tidal dissipation factor $Q$ of approximately 47.1 was calculated. This suggests a moderate level of tidal dissipation in the host star and implies a substantial role for star-planet tidal interactions in the dynamical evolution of the system.

%%%%%%
\begin{deluxetable*}{lccccc}
\tablecaption{HAT-P-16b Model Results\label{tab:HAT-P-16b_exopdot_result}}
\tablehead{
\colhead{Parameter} & \colhead{Symbol} & \colhead{Unit} & \colhead{Value} & \colhead{Lower Error} & \colhead{Upper Error}
}
\startdata
\multicolumn{6}{l}{\textbf{Constant period}} \\
Transit center & $t_{0}$ & $\mathrm{BJD}_{\mathrm{TDB}}$ & 2455968.645838 & $1.93 \times 10^{-5}$ & $1.93 \times 10^{-5}$ \\
Period & $P_{0}$ & days & 2.775968069 & $3.30 \times 10^{-8}$ & $3.33 \times 10^{-8}$ \\
Chi-square statistic & $\chi^{2}$ & & 2671.7 & & \\
Bayesian information criterion & BIC ($\Delta$BIC)  & & 2681.6 (346.6) & \multicolumn{2}{l}{Strongly Rejected} \\
\\
\multicolumn{6}{l}{\textbf{Orbital decay}} \\
Transit center & $t_{0}$ & $\mathrm{BJD}_{\mathrm{TDB}}$ & 2455968.646434 & $2.95 \times 10^{-5}$ & $2.96 \times 10^{-5}$ \\
Period & $P_{0}$ & days & 2.775969971 & $7.88 \times 10^{-8}$ & $7.88 \times 10^{-8}$ \\
Decay rate & $dP/dE$ & days epoch$^{-1}$ & $-4.12 \times 10^{-9}$ & $1.55 \times 10^{-10}$ & $1.55 \times 10^{-10}$ \\
Transit timing shift & PdT & sec & $-46.84$ & $1.76$ & $1.76$ \\
Chi-square statistic & $\chi^{2}$ & & 2320.2 & & \\
Bayesian information criterion & BIC($\Delta$BIC) & & 2335.0 (0.0)& \multicolumn{2}{l}{\checkmark\ Best Fit} \\
\\
\multicolumn{6}{l}{\textbf{Apsidal precession}} \\
Transit center & $t_{0}$ & $\mathrm{BJD}_{\mathrm{TDB}}$ & 2455968.6414 & 0.0007 & 0.0011 \\
Sidereal period & $P_{s}$ & days & 2.775968089 & $3.31 \times 10^{-8}$ & $3.30 \times 10^{-8}$ \\
Eccentricity & $e$ & & 0.0062 & 0.0013 & 0.0007 \\
Argument of periastron & $\omega_{0}$ & rad & 2.74 & 0.04 & 0.03 \\
Precession rate & $\omega dE$ & rad epoch$^{-1}$ & 0.000906 & $6.50 \times 10^{-5}$ & $8.88 \times 10^{-5}$ \\
Chi-square statistic & $\chi^{2}$ & & 2314.9 & & \\
Bayesian information criterion & BIC($\Delta$BIC) & & 2339.6 (4.6)& \multicolumn{2}{l}{Positive Evidence} \\
\enddata
\tablecomments{$\Delta$BIC values are interpreted as evidence against each model relative to the model with the minimum BIC ($\Delta$BIC = 0). According to the BIC metric the Orbital Decay model is preferred but the Apsidal Precession model cannot be ruled out.} 
\end{deluxetable*}

%%%%%%
%%%

\subsubsection{TOI-1516b}
\begin{figure*}
\centering
\includegraphics[scale=0.5]{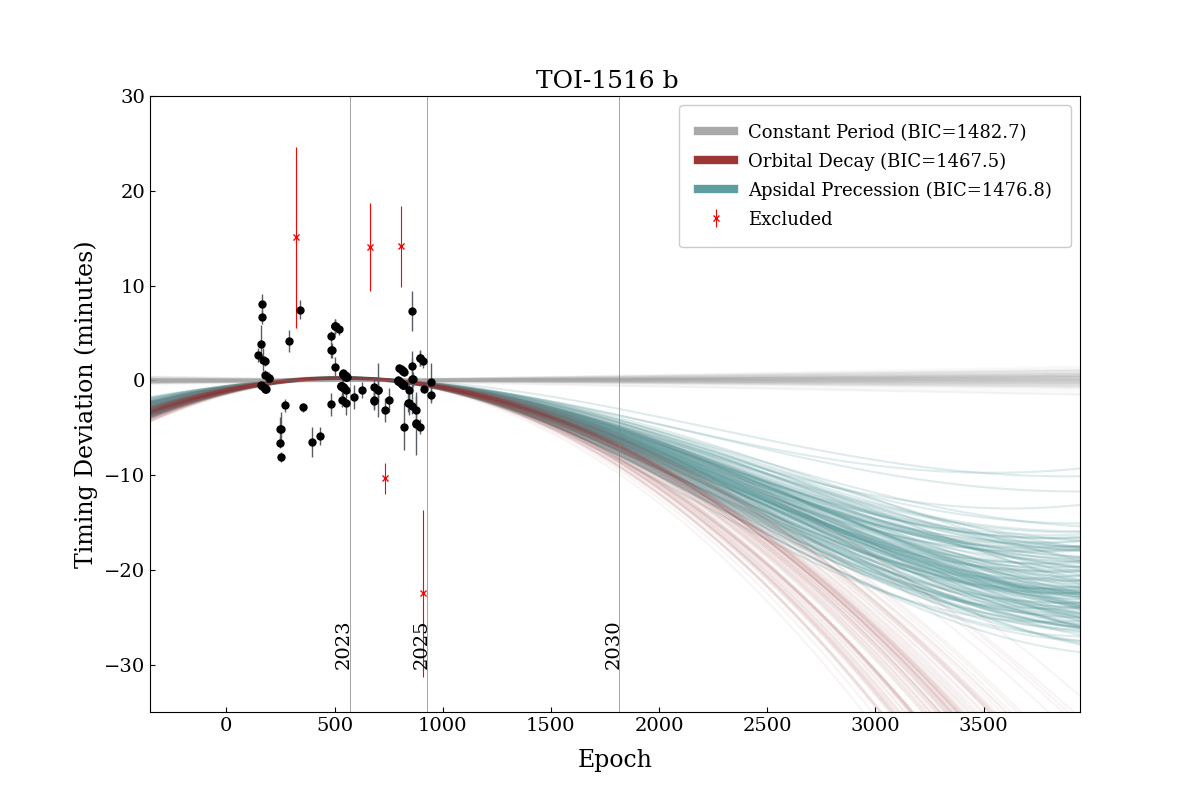}
\caption{Timing residuals of TOI-1516b with future projections of the three transit-timing models shown with 150 random draws from the MCMC posterior chains. Each data point is the difference between an observed time and the time predicted by the best-fit constant-period model. The size of the data points corresponds to the data quality index from 1 to 3, with the higher-quality transits being larger. The red crosses represent observations removed during the sigma-clipping process.}
\label{fig:toi1516b-expdot-oc}
\end{figure*}

\begin{figure*}
\vspace{-1.2cm}
\centering
\includegraphics[scale=1.0]{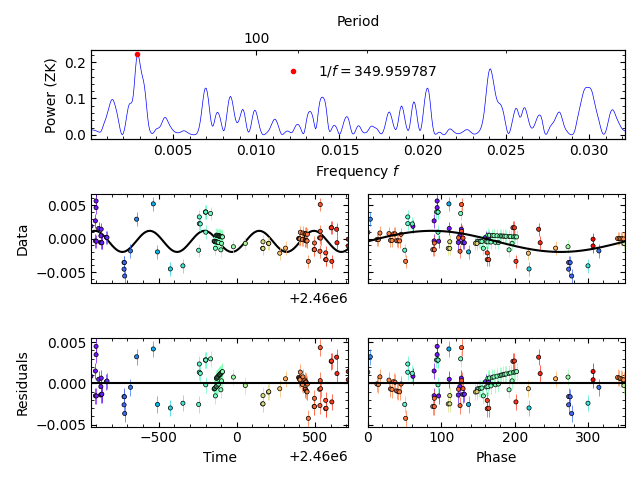}
\caption{The generalised Lomb-Scargle periodogram of TOI-1516b. The top panel displays the power spectrum, the middle panel (left) shows the TTV distribution of the system, the bottom panel (left) presents the residuals, the middle panel (right) illustrates the variation of periodicity with phase, and the bottom panel (right) shows the residuals from the fit.}
\label{fig:toi1516b-GLS}
\end{figure*}

\noindent Figure \ref{fig:toi1516adyu} presents a series of normalized light curves, offset for clarity, acquired during multiple transit events of TOI-1516b using the ADYU60 telescope. The left panel displays 16 transit observations, each annotated with its respective date, while the right panel illustrates the corresponding residuals following the best-fit transit model. The photometric quality of these observations varies between epochs due to factors noted above. An rms threshold of 2.6 mmag was adopted for the global fit of TOI-1516b, optimizing the balance between data inclusion and model reliability. \\
\\ 
Figure \ref{fig:toi1516_tess} shows the transit light curves of TOI-1516b, extracted from multiple TESS sectors. A comparative analysis between the four datasets reveals distinct variations in photometric precision across the different observation epochs. For each sector, the rms varies, and this variation suggests potential differences in instrumental systematics, observation conditions, or intrinsic stellar activity between the sectors. The transit depth and shape remain consistent across both datasets, indicating a stable planetary signal. However, individual transit events display minor deviations from the best-fit model, which could be attributed to uncorrected systematics or time-correlated noise. The data in Fig. \ref{fig:toi1516tess} show the RV distribution of TOI-1516b and the residuals of the model.  The light curves for TOI-1516b in Fig. \ref{fig:toi1516globals}, obtained from TESS and the ADYU60 telescope, reveal important information despite the varying precision levels. TESS data, characterized by high-precision photometry, exhibit smooth transit profiles and minimal residuals, enabling detailed parameter extraction. The ADYU60 telescope data provide complementary temporal coverage and independent validation, albeit with greater scatter and noise.\\
\\ 
The O-C results for TOI-1516b are presented in Figure \ref{fig:toi1516b-expdot-oc} and Table \ref{tab:TOI-1516-b_exopdot_result}. The corner plots for each model are shown in the (online)
Appendix; see Figures \ref{fig:toi1516b-linear-corner}, \ref{fig:toi1516b-decay-corner}, \ref{fig:toi1516b-precession_corner}. The BIC and $\Delta$BIC values were then computed for the different models and compared. The best-fit model was found to be orbital decay (BIC = 1467.5; see Table \ref{tab:TOI-1516-b_exopdot_result}). \\
\\
The GLS power analysis (see Figure \ref{fig:toi1516b-GLS}) yielded maximum power and frequency values of 0.220967 and 0.00286 ($\pm$0.00008) Hz, respectively. The corresponding period and amplitude of the best-fitting sine wave are 349.96 ($\pm$10.17) days and 0.00158 ($\pm$0.00024) respectively. The FAP value associated with the frequency solution is 0.00014. The FAP is significantly smaller than 0.001 and therefore strongly suggests the presence of a detection.\\

Using updated stellar and planetary parameters, the tidal dissipation factor $Q$ was estimated to be approximately $93.0$. Compared to HAT-P-16b, this suggests weaker tidal dissipation, although the presence of orbital decay remains supported by observational data.

%%%%%
\begin{deluxetable*}{lccccc}
\tablecaption{TOI-1516b Model Results\label{tab:TOI-1516-b_exopdot_result}}
\tablehead{
\colhead{Parameter} & \colhead{Symbol} & \colhead{Unit} & \colhead{Value} & \colhead{Lower Error} & \colhead{Upper Error}
}
\startdata
\multicolumn{6}{l}{\textbf{Constant period}} \\
Transit center & $t_{0}$ & $\mathrm{BJD}_{\mathrm{TDB}}$ & 2458765.325144 & $7.32 \times 10^{-5}$ & $7.30 \times 10^{-5}$ \\
Period & $P_{0}$ & days & 2.056013695 & $1.17 \times 10^{-7}$ & $1.17 \times 10^{-7}$ \\
Chi-square statistic & $\chi^{2}$ & & 1473.4 & & \\
Bayesian information criterion & BIC($\Delta$BIC) & & 1482.7 (15.2)& \multicolumn{2}{l}{Very Strongly Rejected} \\
\\
\multicolumn{6}{l}{\textbf{Orbital decay}} \\
Transit center & $t_{0}$ & $\mathrm{BJD}_{\mathrm{TDB}}$ & 2458765.3244 & 0.0001 & 0.0001 \\
Period & $P_{0}$ & days & 2.056017150 & $5.49 \times 10^{-7}$ & $5.47 \times 10^{-7}$ \\
Decay rate & $dP/dE$ & days epoch$^{-1}$ & $-6.47 \times 10^{-9}$ & $9.99 \times 10^{-10}$ & $9.99 \times 10^{-10}$ \\
Transit timing shift & PdT & sec & $-99.33$ & 15.34 & 15.35 \\
Chi-square statistic & $\chi^{2}$ & & 1453.5 & & \\
Bayesian information criterion & BIC($\Delta$BIC) & & 1467.5 (0.0)& \multicolumn{2}{l}{\checkmark\ Best Fit} \\
\\
\multicolumn{6}{l}{\textbf{Apsidal precession}} \\
Transit center & $t_{0}$ & $\mathrm{BJD}_{\mathrm{TDB}}$ & 2458765.3172 & 0.0018 & 0.0012 \\
Sidereal period & $P_{s}$ & days & 2.056013691 & $1.17 \times 10^{-7}$ & $1.18 \times 10^{-7}$ \\
Eccentricity & $e$ & & 0.0123 & 0.0018 & 0.0027 \\
Argument of periastron & $\omega_{0}$ & rad & 2.68 & 0.05 & 0.05 \\
Precession rate & $\omega dE$ & rad epoch$^{-1}$ & 0.000869 & $7.93 \times 10^{-5}$ & $8.85 \times 10^{-5}$ \\
Chi-square statistic & $\chi^{2}$ & & 1453.5 & & \\
Bayesian information criterion & BIC($\Delta$BIC) & & 1476.8 (9.3)& \multicolumn{2}{l}{Strongly Rejected} \\
\enddata
\tablecomments{According to the BIC metric, the Orbital Decay model is preferred and the others are rejected.}
\end{deluxetable*}

%%%%%%
%%%

\subsubsection{TOI-2046b}
\begin{figure*}
\centering
\includegraphics[scale=0.5]{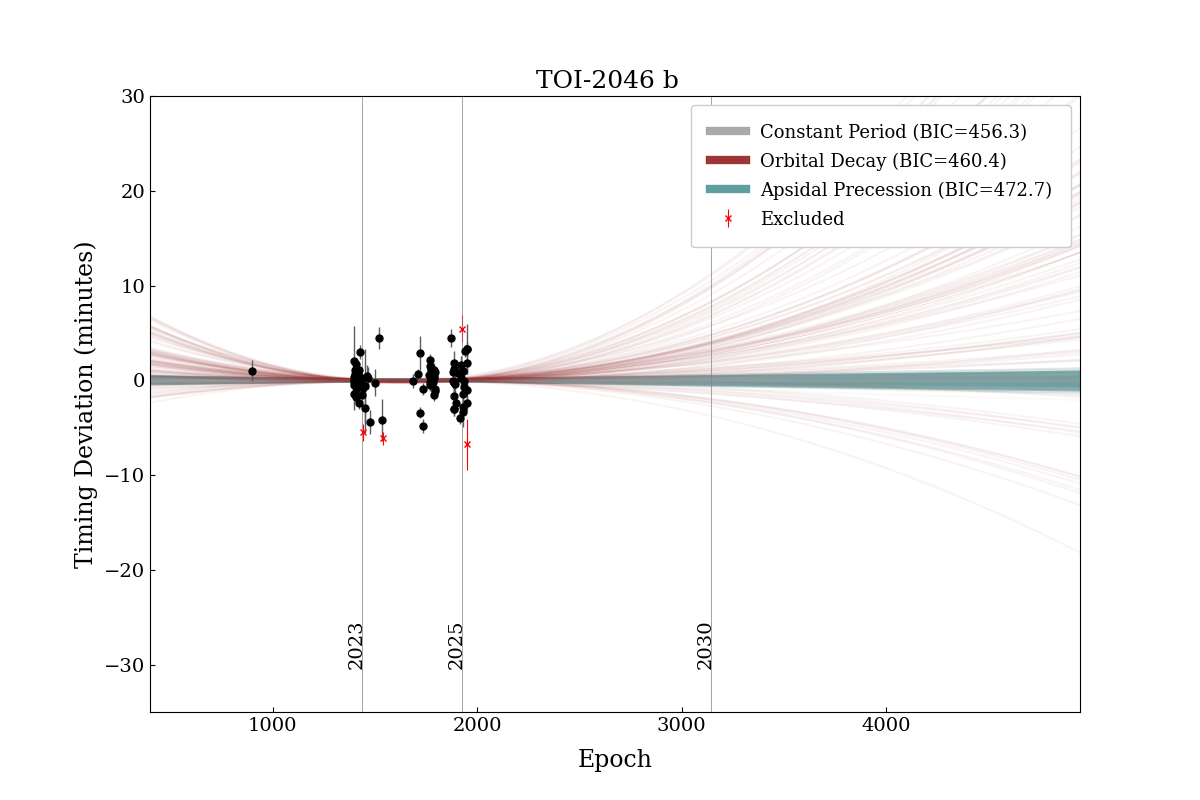}
\caption{Timing residuals of TOI-2046b with future projections of the three transit-timing models shown with 150 random draws from the MCMC posterior chains. Each data point is the difference between an observed time and the time predicted by the best-fit constant-period model. The size of the data points corresponds to the data quality index from 1 to 3, with the higher-quality transits being larger. The red crosses represent observations removed during the sigma-clipping process.}
\label{fig:toi2046b-expdot-oc}
\end{figure*}

\begin{figure*}
\vspace{-1.1cm}
\centering
\includegraphics[scale=1.0]{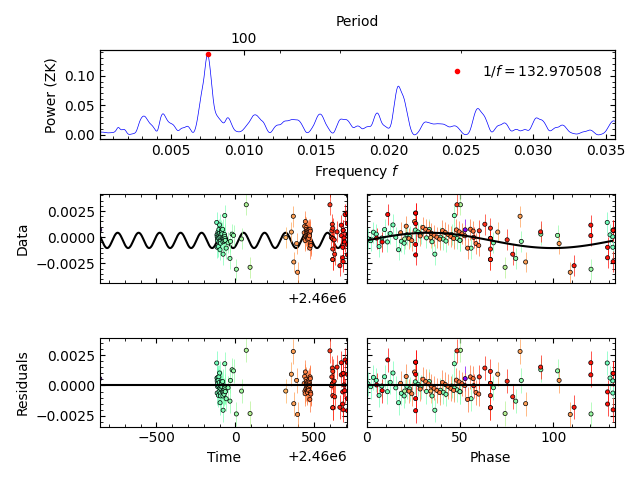}
\caption{The generalised Lomb-Scargle periodogram of TOI-2046b. The top panel displays the power spectrum, the middle panel (left) shows the TTV distribution of the system, the bottom panel (left) presents the residuals, the middle panel (right) illustrates the variation of periodicity with phase, and the bottom panel (right) shows the residuals from the fit.}
\label{fig:toi2046b-GLS}
\end{figure*}

\begin{figure*}
\centering
\includegraphics[scale=0.7]{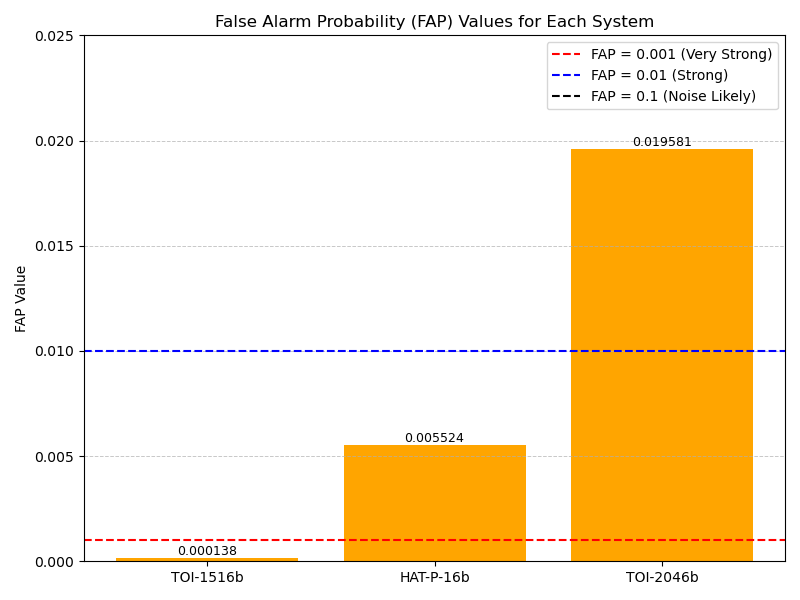}
\caption{False alarm probability (FAP) values for each system}
\label{fig:FAP_all}
\end{figure*}

\noindent Figure \ref{fig:toi2046adyu} presents thirteen newly obtained transit light curves of TOI-2046b, observed with the ADYU60 telescope. The scatter in the residuals varies between 1.9 and 4.2 mmag. The global fit is performed assuming an rms of 3.1 mmag, providing a representative uncertainty for parameter estimation. The consistency between observed transit profiles and the model fit suggests a well-constrained planetary transit signature, supporting the robustness of the derived system parameters. The transit light curves of TOI-2046b in Figs. \ref{fig:toi2046_tess}, obtained from sectors 58, 59, 78 and 79 of TESS, exhibit high precision photometry with consistently low rms values in all datasets. The well-defined transit profiles, observed in multiple sectors, indicate a stable detection of the planetary signal. The T100 telescope data from various observing times in Figs. \ref{fig:toi2046fromT100}, offer valuable complementary information with ADYU60 and TESS data. Although these data exhibit increased scatter and noise, they do however serve an important function of confirming the findings and extending the temporal coverage. The planetary radius (R$_p$) measurements obtained from both TESS and T100 show a close agreement, with TESS reporting a slightly smaller radius (1.35R$_p$) compared to T100's 1.40 R$_p$. The uncertainties in both measurements are comparable, suggesting that the differences are within acceptable limits of observational precision. Similarly, the stellar radius (R$_*$) values from TESS and T100 are in good concordance, with TESS reporting 1.10 R$_*$ and T100 slightly larger at 1.12 R$_*$.
 Fig. \ref{fig:toi2046RV} presents a comprehensive analysis of the radial velocities (RVs) for TOI-2046b, plotted vesus phase. In Fig. \ref{fig:toi1516tess} for TOI-1516 O-C reaching up to 570 m/s, this value for TOI-2046b peaks around 210 m/s, reflecting a tighter fit to the model and, consequently, higher precision in the RV measurements. The RV data points for TOI-1516b, exhibit considerable scatter and larger uncertainties. In contrast, the data points for TOI-2046b show less scatter, with more tightly clustered points and smaller error bars, indicating higher observational accuracy and less noise in the measurements. In Fig. \ref{fig:toi2046globals}, the light curve data for the star TOI-2046, as observed by the TESS telescope, are presented along with data obtained from ADYU60. The TESS data show a more pronounced dip in the normalized flux.  The unbinned residuals for TESS are 1.96 mmag, while the binned residuals are 0.51 mmag, indicating a relatively good fit to the model. For the ADYU60 data, the unbinned residuals are 2.21 mmag, and the binned residuals are 1.15 mmag, suggesting a slightly higher level of noise or less precise fitting compared to the TESS data.\\
\\
The O–C fit results for TOI-2046b are presented in Figure \ref{fig:toi2046b-expdot-oc} and Table \ref{tab:TOI-2046-b_exopdot_result}. The BIC and $\Delta$BIC values indicate that the constant period model (BIC=456.3; see Table \ref{tab:TOI-2046-b_exopdot_result}) is the preferred model for TOI-2046b. The O–C diagram showed no systematic deviations, and the decay model resulted in a positive value $\mathrm{PdE}$ and an unphysical negative value, the tidal dissipation factor $Q$ (the tidal dissipation factor) of approximately $-336.4$. These findings confirm that the system remains dynamically stable, with no significant observable tidal decay currently detected.
 The corner plots for each model are shown in the (online)
Appendix; see Figures \ref{fig:toi2046b-linear-corner}, \ref{fig:toi2046b-decay-corner}, and \ref{fig:toi2046b-precession_corner}. \\
\\
The results of the GLS analysis are presented in Figure \ref{fig:toi2046b-GLS}. The GLS power analysis yielded maximum power and frequency values of 0.1357 and 0.0075 ($\pm$0.0001) Hz, respectively. The period and amplitude of the best-fitting sine wave corresponding to this frequency are 132.97 ($\pm$2.28) days and 0.0007 ($\pm$0.0001), respectively. The FAP value associated with the frequency solution is 0.0196, a value that is consistent with background noise. The calculated FAP values for each planetary system are shown as a bar plot in Figure \ref{fig:FAP_all}.\\
\\

%132.970508 ($\pm$2.278348) days and 0.000732 ($\pm$0.000143) days
%%%%
\begin{deluxetable*}{lccccc}
\tablecaption{TOI-2046b Model Results\label{tab:TOI-2046-b_exopdot_result}}
\tablehead{
\colhead{Parameter} & \colhead{Symbol} & \colhead{Unit} & \colhead{Value} & \colhead{Lower Error} & \colhead{Upper Error}
}
\startdata
\multicolumn{6}{l}{\textbf{Constant period}} \\
Transit center & $t_{0}$ & $\mathrm{BJD}_{\mathrm{TDB}}$ & 2457792.2753 & 0.0002 & 0.0002 \\
Period & $P_{0}$ & days & 1.49718644 & $1.46 \times 10^{-7}$ & $1.51 \times 10^{-7}$ \\
Chi-square statistic & $\chi^{2}$ & & 446.9 & & \\
Bayesian information criterion & BIC ($\Delta$BIC)& & 456.3 (0.0) & \multicolumn{2}{l}{\checkmark\ Best Fit} \\
\\
\multicolumn{6}{l}{\textbf{Orbital decay}} \\
Transit center & $t_{0}$ & $\mathrm{BJD}_{\mathrm{TDB}}$ & 2457792.2775 & 0.0027 & 0.0033 \\
Period & $P_{0}$ & days & 1.4971836 & $4.16 \times 10^{-6}$ & $3.40 \times 10^{-6}$ \\
Decay rate & $dP/dE$ & days epoch$^{-1}$ & $1.71 \times 10^{-9}$ & $2.08 \times 10^{-9}$ & $2.56 \times 10^{-9}$ \\
Transit timing shift & PdT & sec & 36.08 & 43.8 & 54.04 \\
Chi-square statistic & $\chi^{2}$ & & 446.3 & & \\
Bayesian information criterion & BIC($\Delta$BIC) & & 460.4 (4.1)& \multicolumn{2}{l}{Positive Evidence} \\
\\
\multicolumn{6}{l}{\textbf{Apsidal precession}} \\
Transit center & $t_{0}$ & $\mathrm{BJD}_{\mathrm{TDB}}$ & 2457792.2753 & 0.0003 & 0.0003 \\
Sidereal period & $P_{s}$ & days & 1.49718645 & $1.10 \times 10^{-7}$ & $1.08 \times 10^{-7}$ \\
Eccentricity & $e$ & & 0.0002 & 0.0001 & 0.0019 \\
Argument of periastron & $\omega_{0}$ & rad & 2.94 & 1.89 & 2.29 \\
Precession rate & $\omega dE$ & rad epoch$^{-1}$ & $2.93 \times 10^{-5}$ & $2.58 \times 10^{-5}$ & $1.88 \times 10^{-4}$ \\
Chi-square statistic & $\chi^{2}$ & & 449.1 & & \\
Bayesian information criterion & BIC ($\Delta$BIC) & & 472.7 (16.4) & \multicolumn{2}{l}{Very Strongly Rejected} \\
\enddata
\tablecomments{According to the BIC metric, the Constant Period model is preferred, but the Orbital Decay model cannot be ruled out.}
\end{deluxetable*}

%%%%%%%%%%%%%
%%%%%%%%%%%%%
%%%%%%
%%%%%%
%%%%%%%
%
%%%%%

\section{Conclusions}
\noindent We conducted long-term observations of the exoplanets HAT-P-16b, TOI-1516b, and TOI-2046b using the 0.6 m telescope located at Adiyaman University, and the 1.0 m telescope at TÜBİTAK National Observatory. In total, we observed 11 new transit light curves for HAT-P-16b, 16 for TOI-1516b, and 13 for TOI-2046b. By combining these with TESS observations and previously published data, we refined the linear ephemeris for each system. The calculated planetary parameters including orbital periods and mid-transit times are consistent with previously published values, agreeing within 1–2$\sigma$ for all systems analyzed.\\
\\
Furthermore, we examined the potential presence of additional bodies by constructing O–C diagrams and analyzing TTVs for all three systems. The TTV's were modeled using the \texttt{ExoPdot} software with application of three distinct models: Constant Period, Orbital Decay, and Apsidal Precession. For each model, $\chi^2$, BIC, and $\Delta$BIC values were calculated to determine the best-fit scenario. The resulting BIC($\Delta$BIC) values indicate that the Orbital Decay model is statistically favored for HAT-P-16b and TOI-1516b, while the Constant Period model is preferred for TOI-2046b. Additionally, for systems where the orbital decay model yielded the best fit, the tidal quality factor ($Q$) of the host star was estimated. \\
\\
As part of the analysis, we applied the GLS periodogram to search for potential periodic signals in the transit timing residuals and assessed the statistical significance of the detected signals using the FAP metric. The FAP values obtained are as follows: 0.00014 (TOI-1516b), 0.00552 (HAT-P-16b) and 0.0196 (TOI-2046b) respectively. These results suggest a clear distinction in the significance of the signals among the systems. Specifically, the very low FAP value for TOI-1516b (FAP $<$ 0.001) indicates a highly significant periodic signal, suggesting the presence of potential dynamical interactions within the system, possibly due to an additional planetary or subplanetary companion. In contrast, the higher FAP value for HAT-P-16b (0.01 $>$ FAP $>$ 0.001) indicates that the extracted periodicity and the possible presence of an additional body are still relatively strong, but a result considerably less robust than for TOI-1516b. Further (continuous) photometric observations of both systems are recommended to achieve more robust results. Lastly, the much higher FAP value (0.0196) for TOI-2046b (0.1 $>$ FAP $>$ 0.01) implies that the observed signal in this system is very weak and most likely the result of noise. 
%% Please use the acknowledgment and contribution environments. This will 
%% be anonomyized when the "anonymous" style option is used. 
\begin{acknowledgments}
We thank the Adiyaman University Astrophysics Application and Research Center for their support in the acquisition of data with the ADYU60 telescope. We also thank the T{\"u}rkiye National Observatories for partial observational support using the T100 telescope with project number 18AT100-1300. Some of the data presented in this article were obtained from the Mikulski Archive for Space Telescopes (MAST) at the Space Telescope Science Institute. The specific observations analyzed can be accessed via \dataset[doi: 10.17909/txjq-nk46]{https://doi.org/10.17909/txjq-nk46}.
\end{acknowledgments}

\software{AstroImageJ (AIJ) \citep{collins2017astroimagej}, EXOFASTV2 \citep{eastman2013exofast,2017ascl.soft10003E,eastman2019exofastv2},
          EXOTIC \citep{2023ascl.soft02009Z,2020PASP..132e4401Z}, 
          ExoPdot \citep{Hagey..2022AJ....164..220H}
          }

\bibliography{sample7}{}
\bibliographystyle{aasjournalv7}

\appendix
\newpage
%\section{Appendix}

\begin{figure}
\centering
\includegraphics[width=0.48\textwidth]{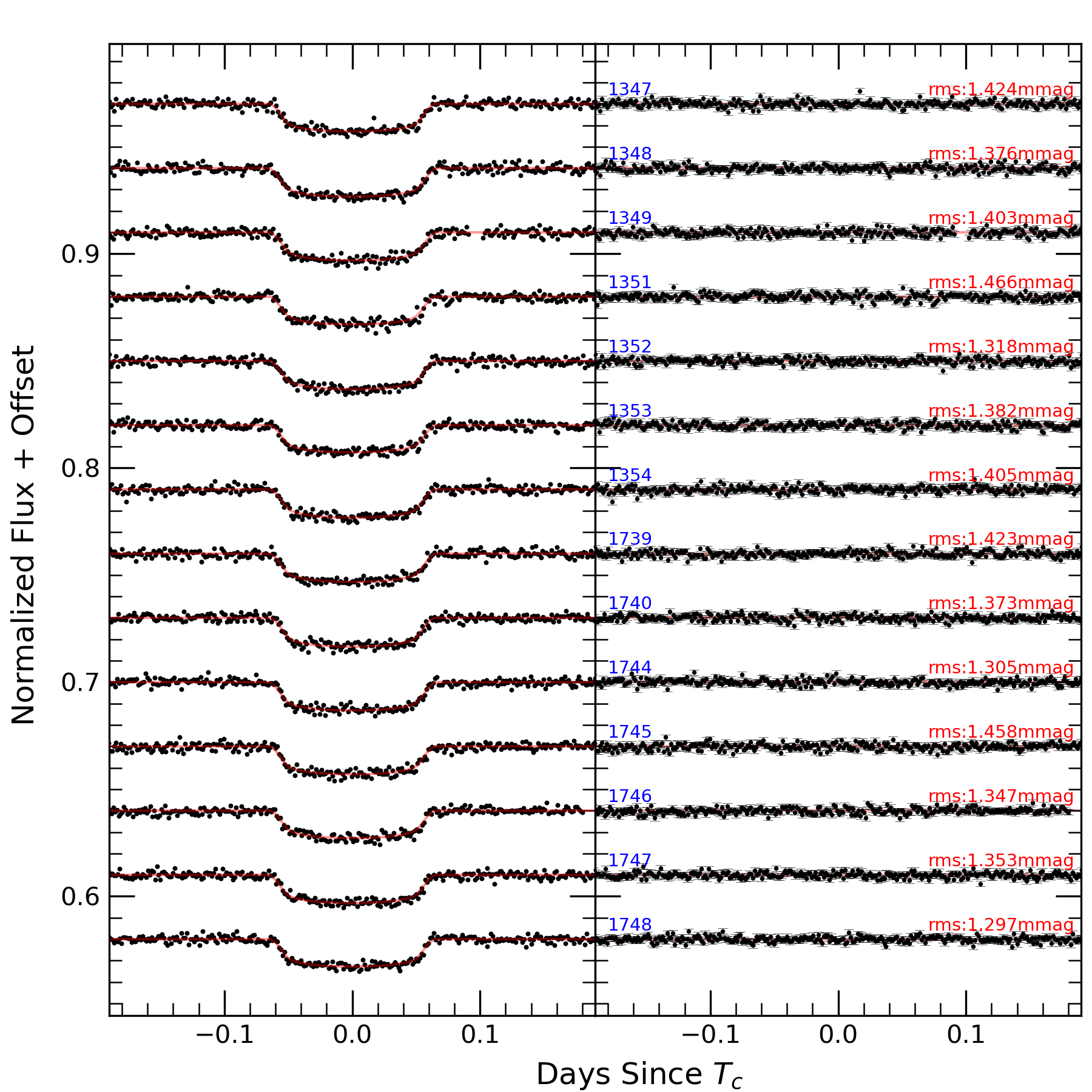}
\caption{Fourteen transit light curves of HAT-P-16b extracted from TESS observations along with the transit model fits (indicated as red lines) and the residuals. Other description is the same as in Fig. \ref{fig:hatp16adyu}}
\label{fig:hatp16_tess}
\end{figure}
%%%%
%%%%%%
\begin{figure}
\centering
\includegraphics[scale=0.65]{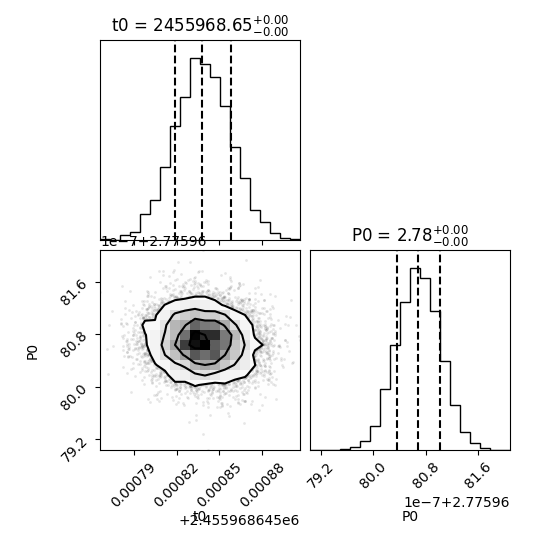}
\caption{The corner plot of the linear transit-timing model shown with 150 random draws from the MCMC posterior chains of HAT-P-16 b.}
\label{fig:hatp16b-linear-corner}
\end{figure}
%%%%%%%%%
\begin{figure}
\centering
\includegraphics[scale=0.50]{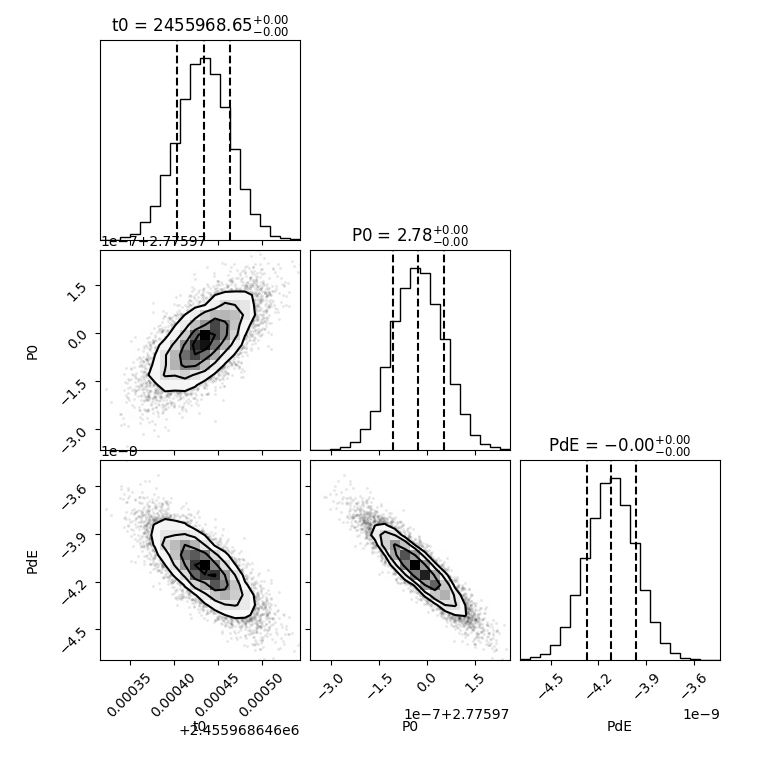}
\caption{The corner plot of the quadratic (orbital decay) model of HAT-P-16 b.}
\label{fig:hatp16b-decay-corner}
\end{figure}
%%%%%%
\begin{figure*}
\centering
\includegraphics[scale=0.6]{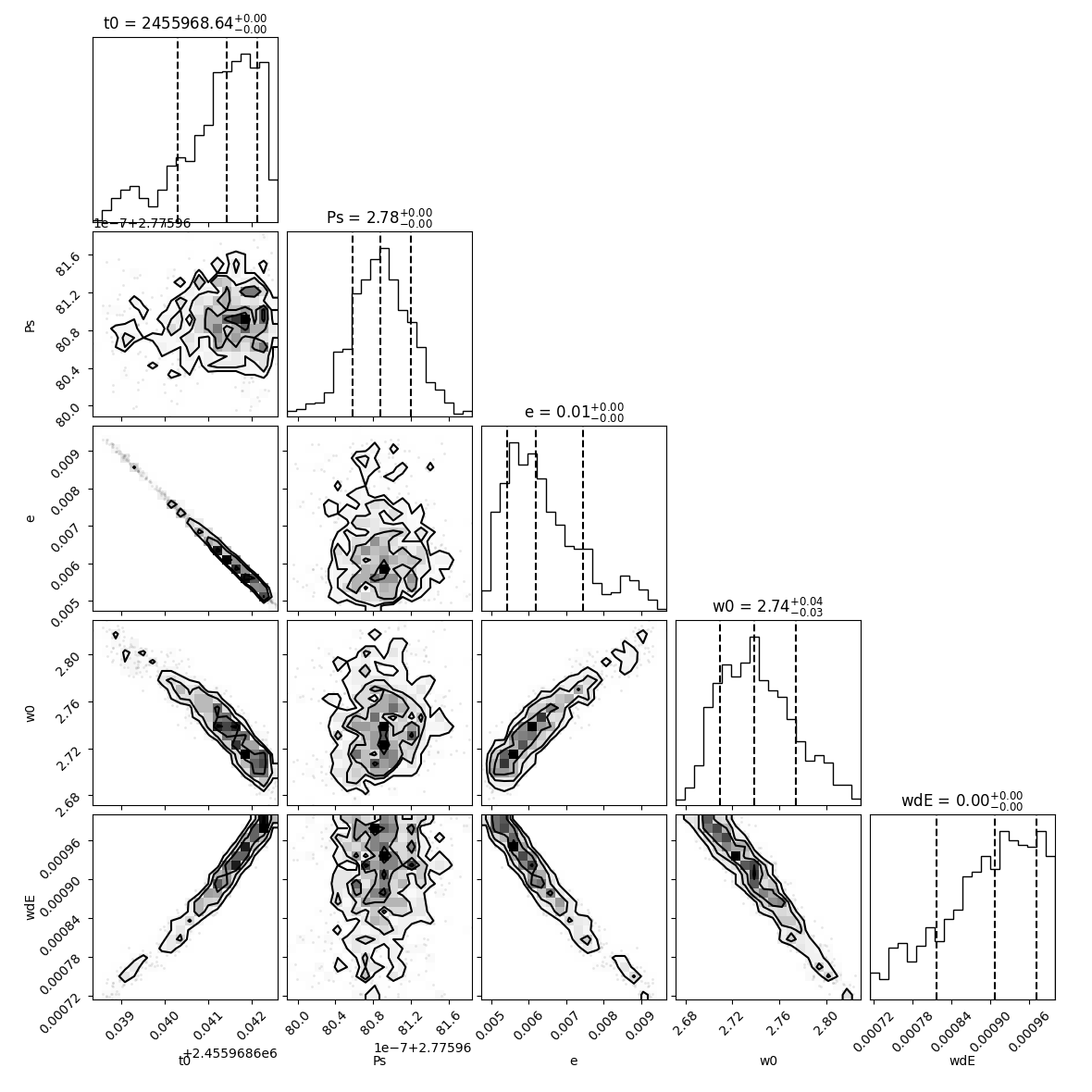}
\caption{The corner plot of the apsidal precession model of HAT-P-16b.}
\label{fig:hatp16b-precession_corner}
\end{figure*}
%%%%%%

%%%%%
\begin{figure*}
\centering
\includegraphics[scale=0.50]{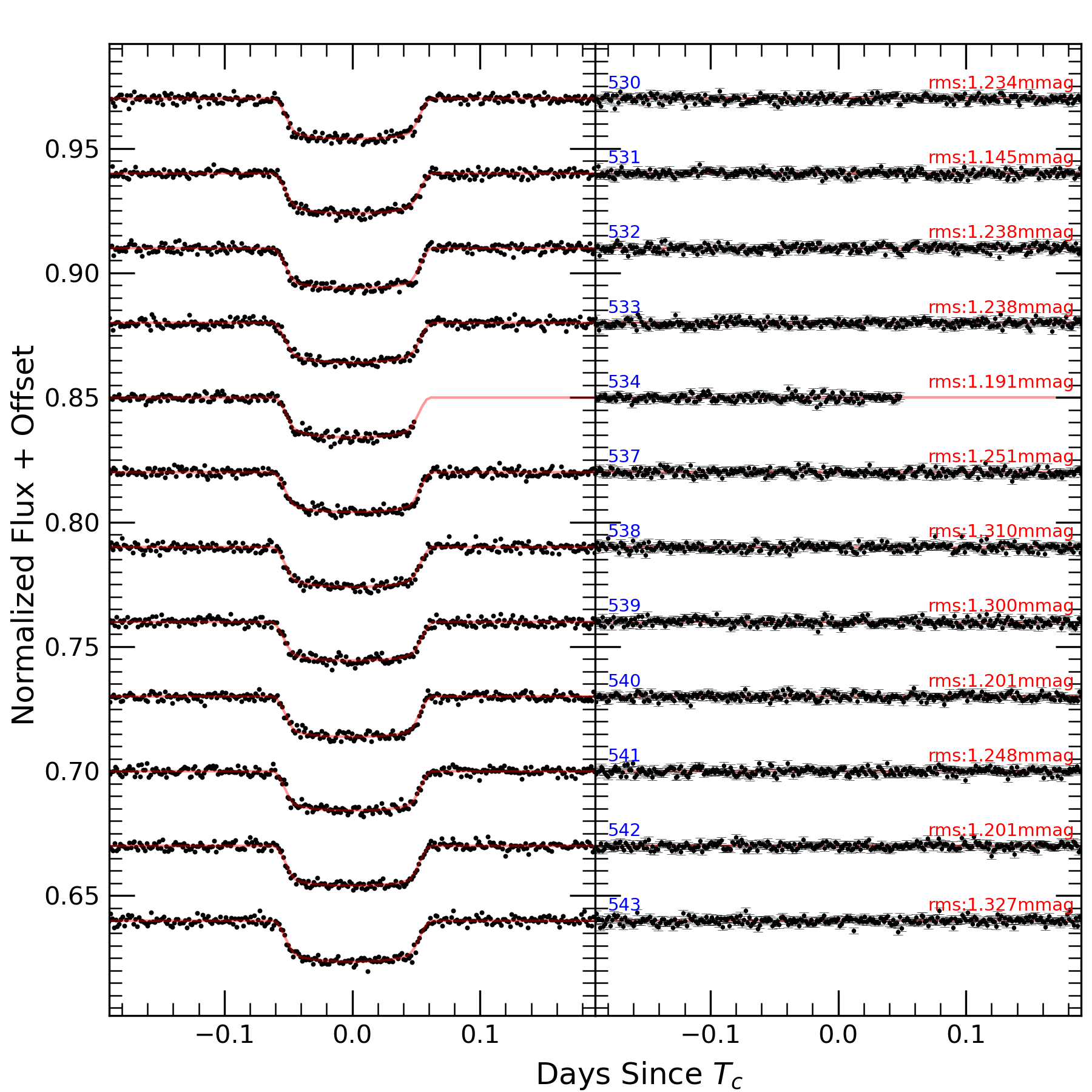}
\includegraphics[scale=0.50]{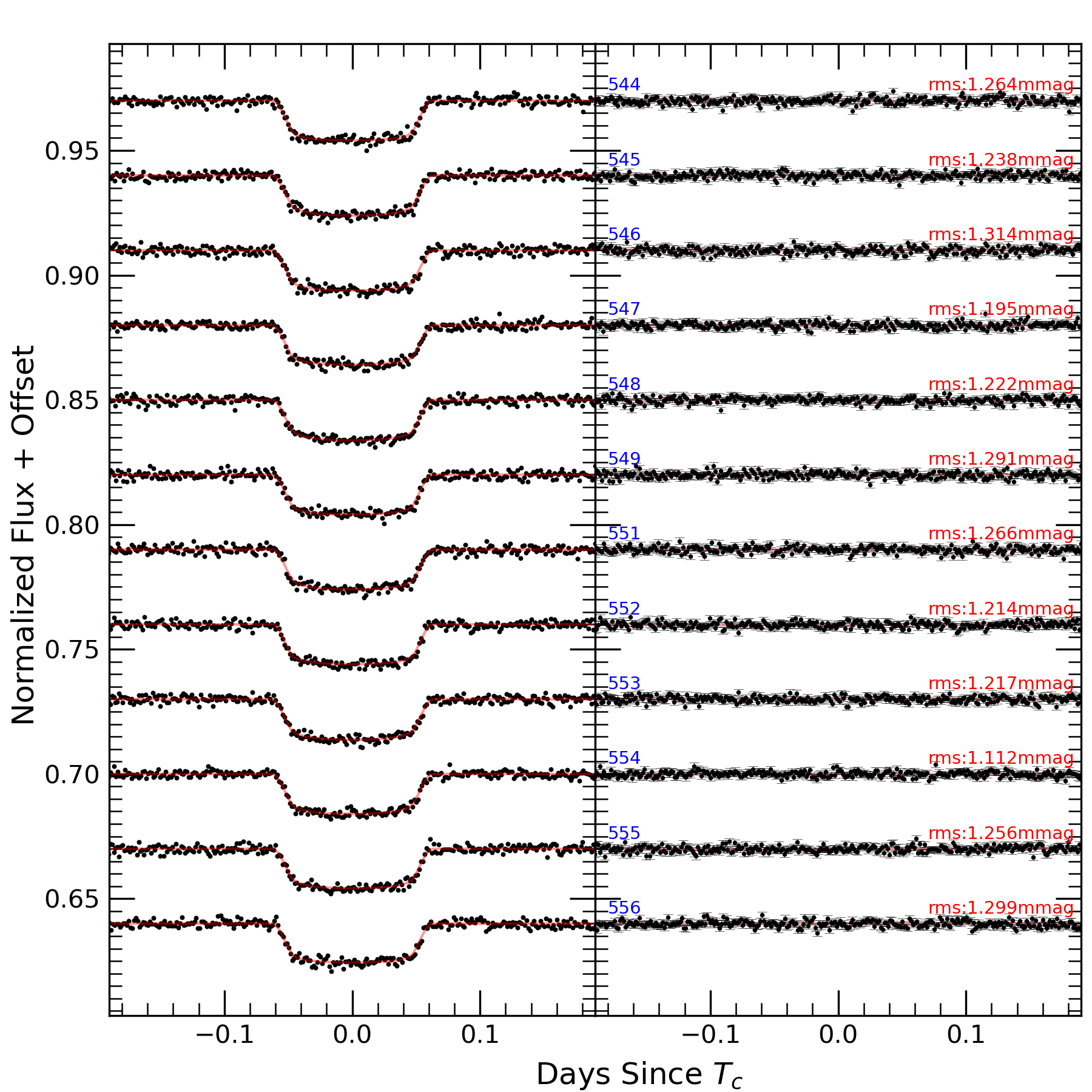}
\includegraphics[scale=0.50]{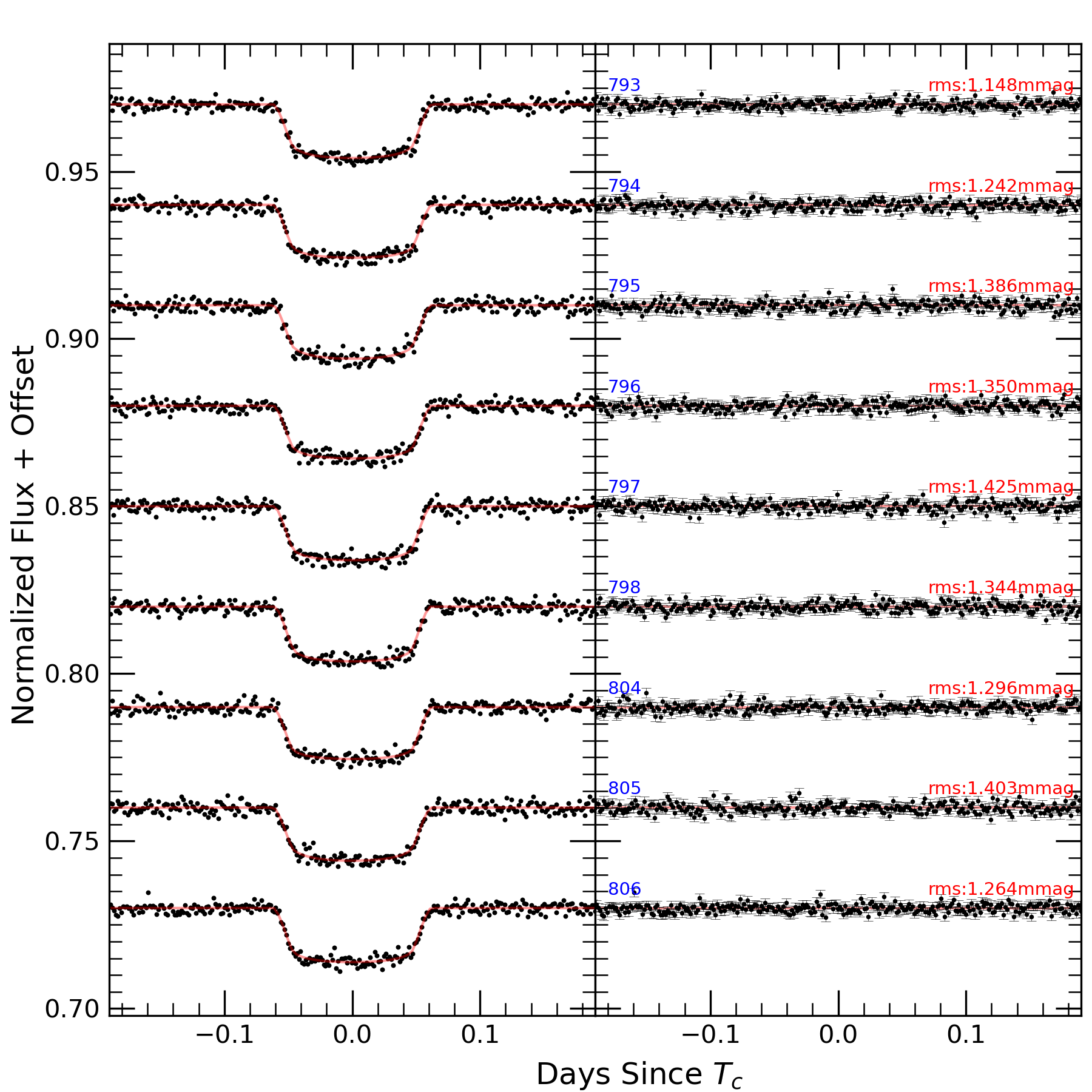}
\includegraphics[scale=0.50]{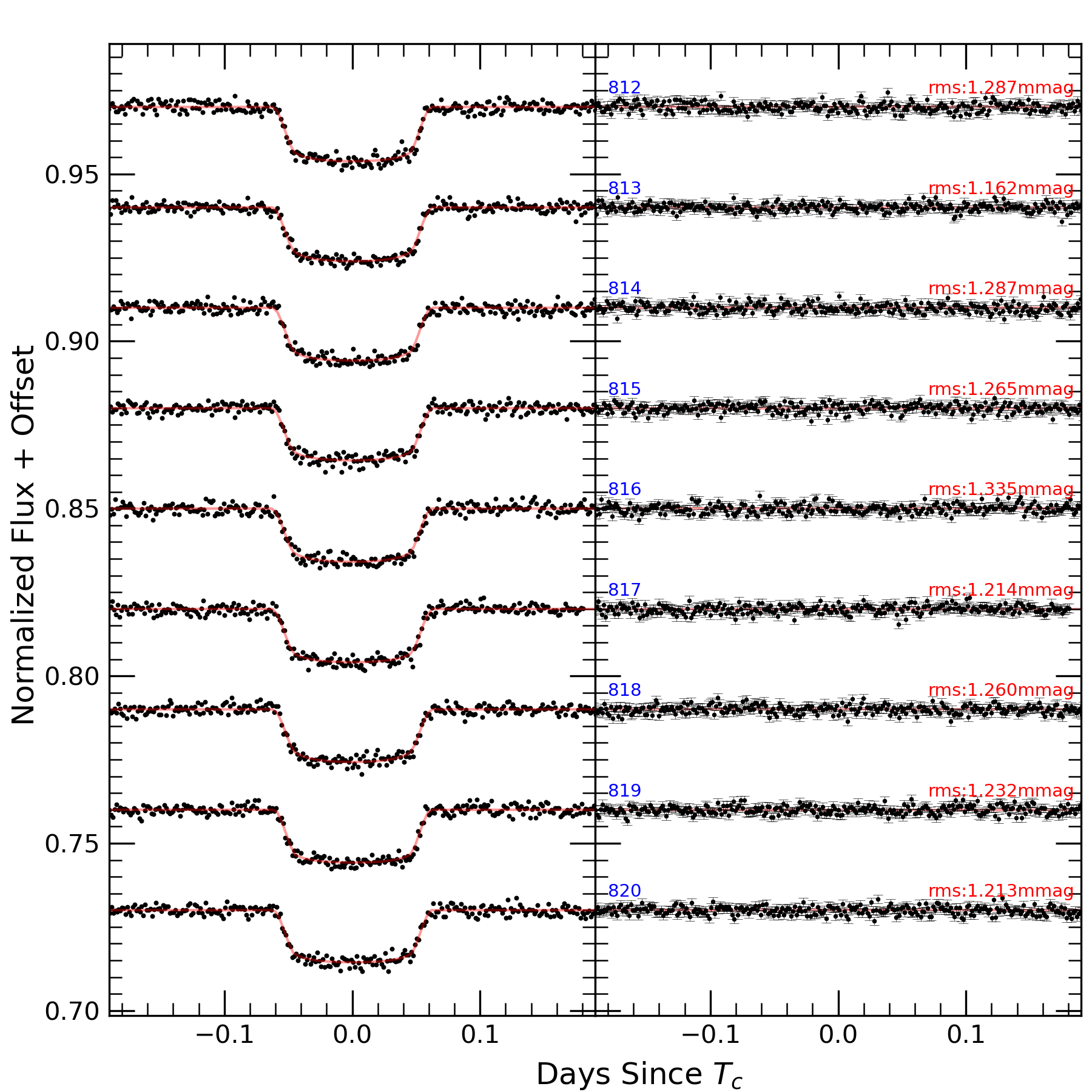}
\caption{Transit light curves of TOI-1516b produced from TESS {\bf Top Left:} sector 57, {\bf Top Right:} sector 58, {\bf Bottom Left:} sector 77, and {\bf Bottom Right:} sector 78} 
%along with the transit model fits (indicated as red lines) and the residuals.
\label{fig:toi1516_tess}
\end{figure*}
%%%%%%%
%%%%%
\begin{figure}
\centering
\includegraphics[scale=0.65]{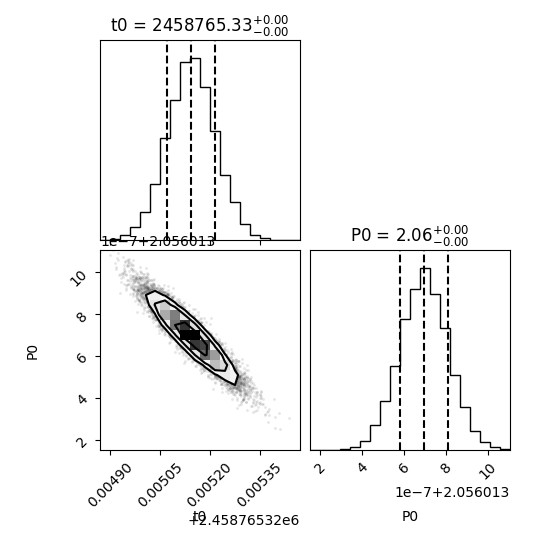}
\caption{The corner plot of the linear transit-timing model shown with 150 random draws from the MCMC posterior chains of TOI-1516b.}
\label{fig:toi1516b-linear-corner}
\end{figure}
%%%%%%
\begin{figure}
\centering
\includegraphics[scale=0.50]{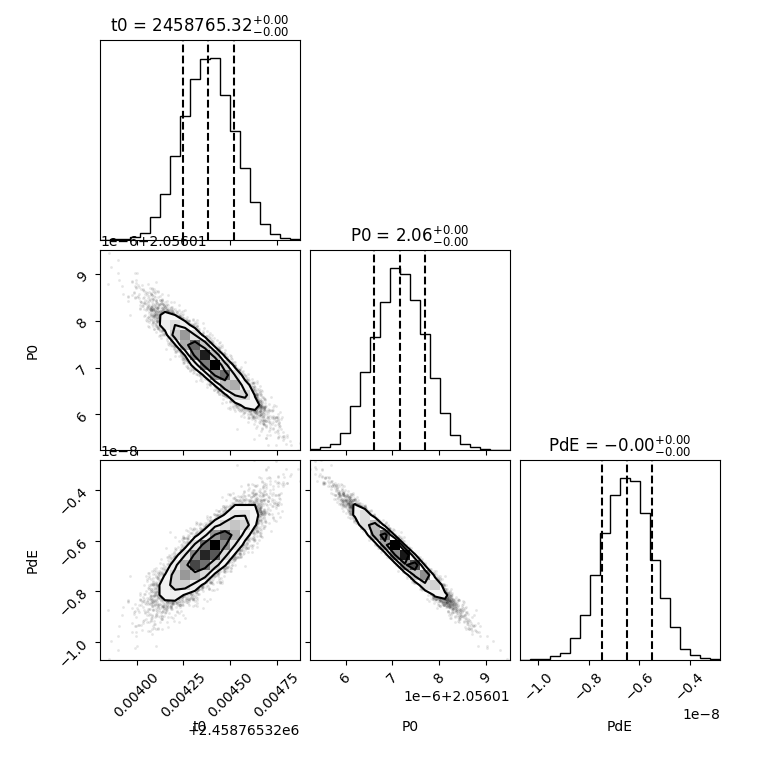}
\caption{The corner plot of the quadratic (orbital decay) model of TOI-1516b.}
\label{fig:toi1516b-decay-corner}
\end{figure}
%%%%%%%%
\begin{figure*}
\centering
\includegraphics[scale=0.6]{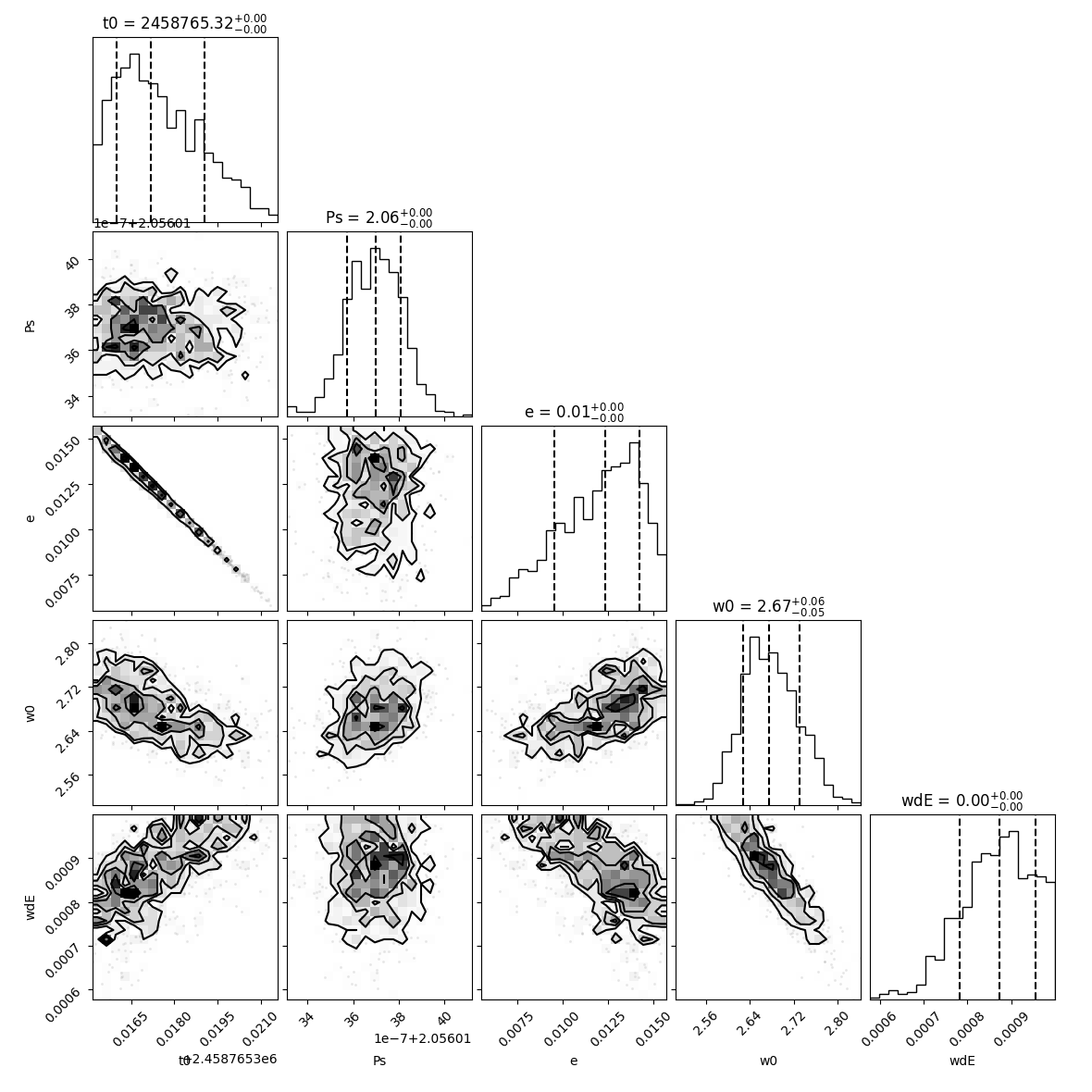}
\caption{The corner plot of the apsidal precession model of TOI-1516b.}
\label{fig:toi1516b-precession_corner}
\end{figure*}
%%%%%%%%
%%%%%
\begin{figure*}
\centering
\includegraphics[scale=0.55]{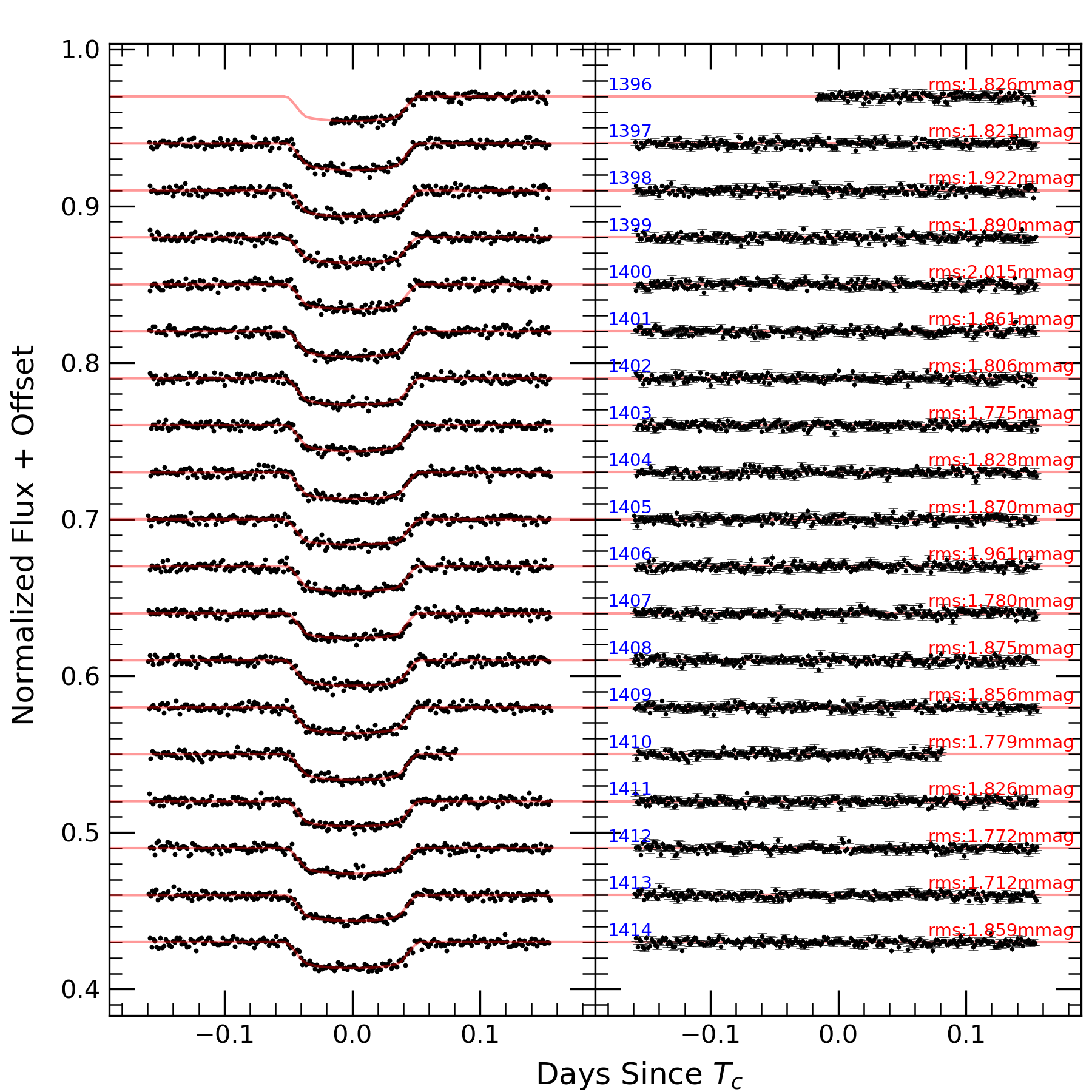}
\includegraphics[scale=0.55]{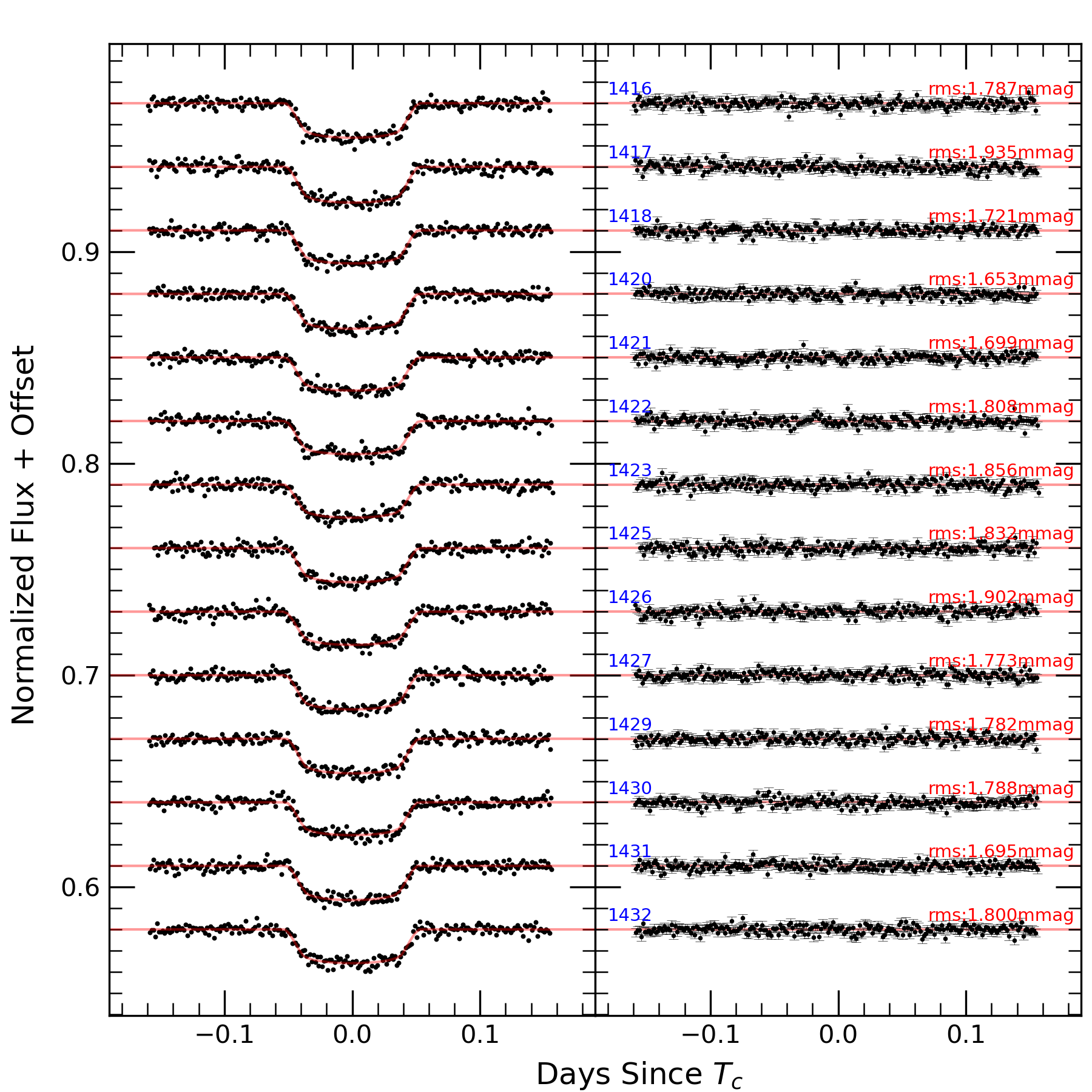}
\includegraphics[scale=0.55]{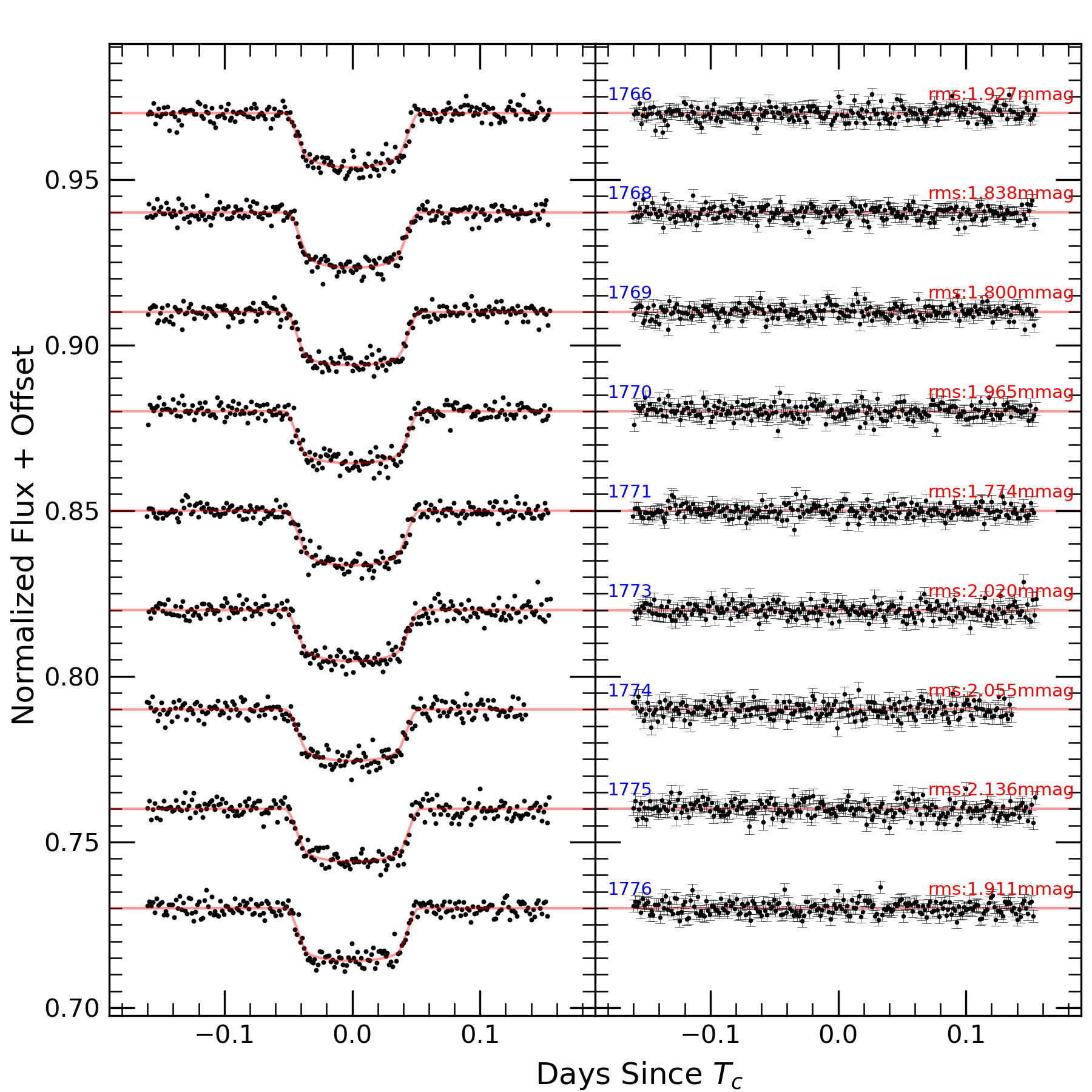}
\includegraphics[scale=0.55]{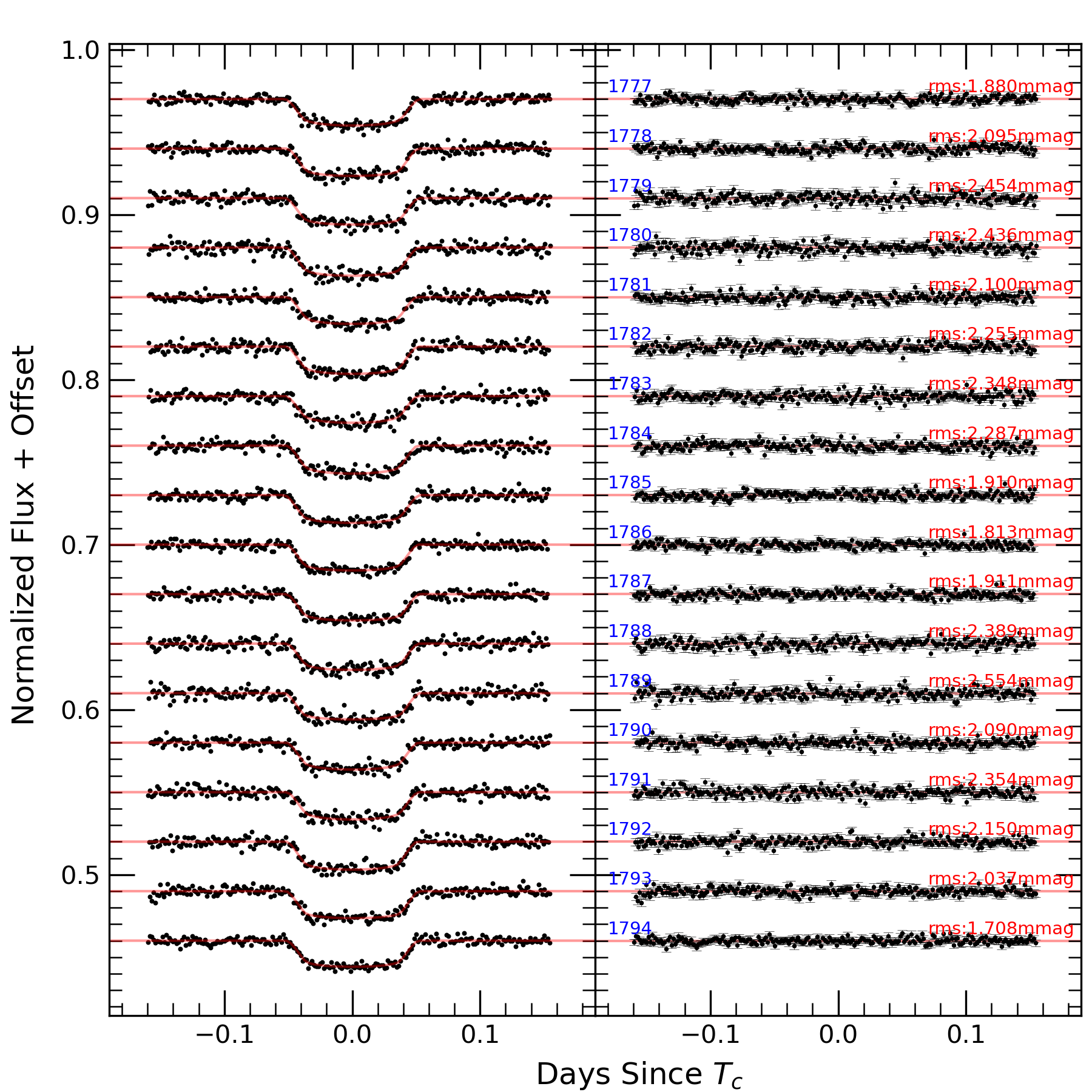}
\caption{Transit light curves of TOI-2046b produced from TESS {\bf Top Left:} sector 58, {\bf Top Right:} sector 59, {\bf Bottom Left:} sector 78, and {\bf Bottom Right:} sector 79} 
%along with the transit model fits (indicated as red lines) and the residuals.
\label{fig:toi2046_tess}
\end{figure*}
%%%%%%%
\begin{figure}
\centering
\includegraphics[scale=0.60]{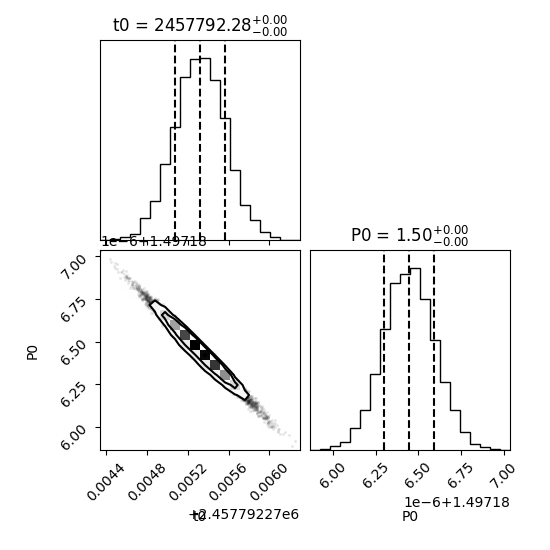}
\caption{The corner plot of the linear transit-timing model shown with 150 random draws from the MCMC posterior chains of TOI-2046b.}
\label{fig:toi2046b-linear-corner}
\end{figure}
%%%%%%%%%%
%%%%%%
%%%%%%
%%%%%%
\begin{figure}
\centering
\includegraphics[scale=0.50]{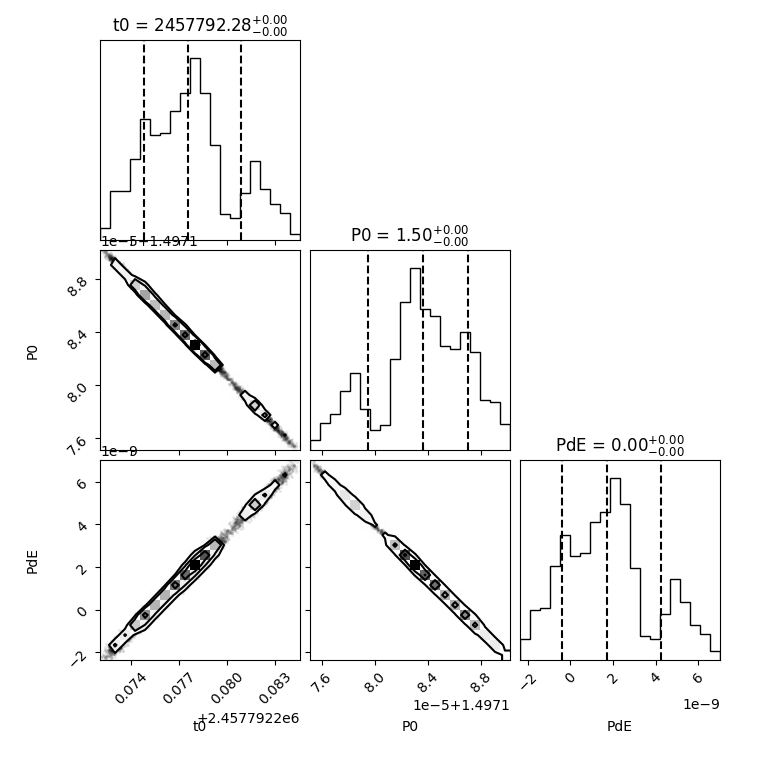}
\caption{The corner plot of the quadratic (orbital decay) model of TOI-2046b.}
\label{fig:toi2046b-decay-corner}
\end{figure}
%%%%%%%%%%
%%%%%%
%%%%%%
%%%%%%
\begin{figure*}
\centering
\includegraphics[scale=0.6]{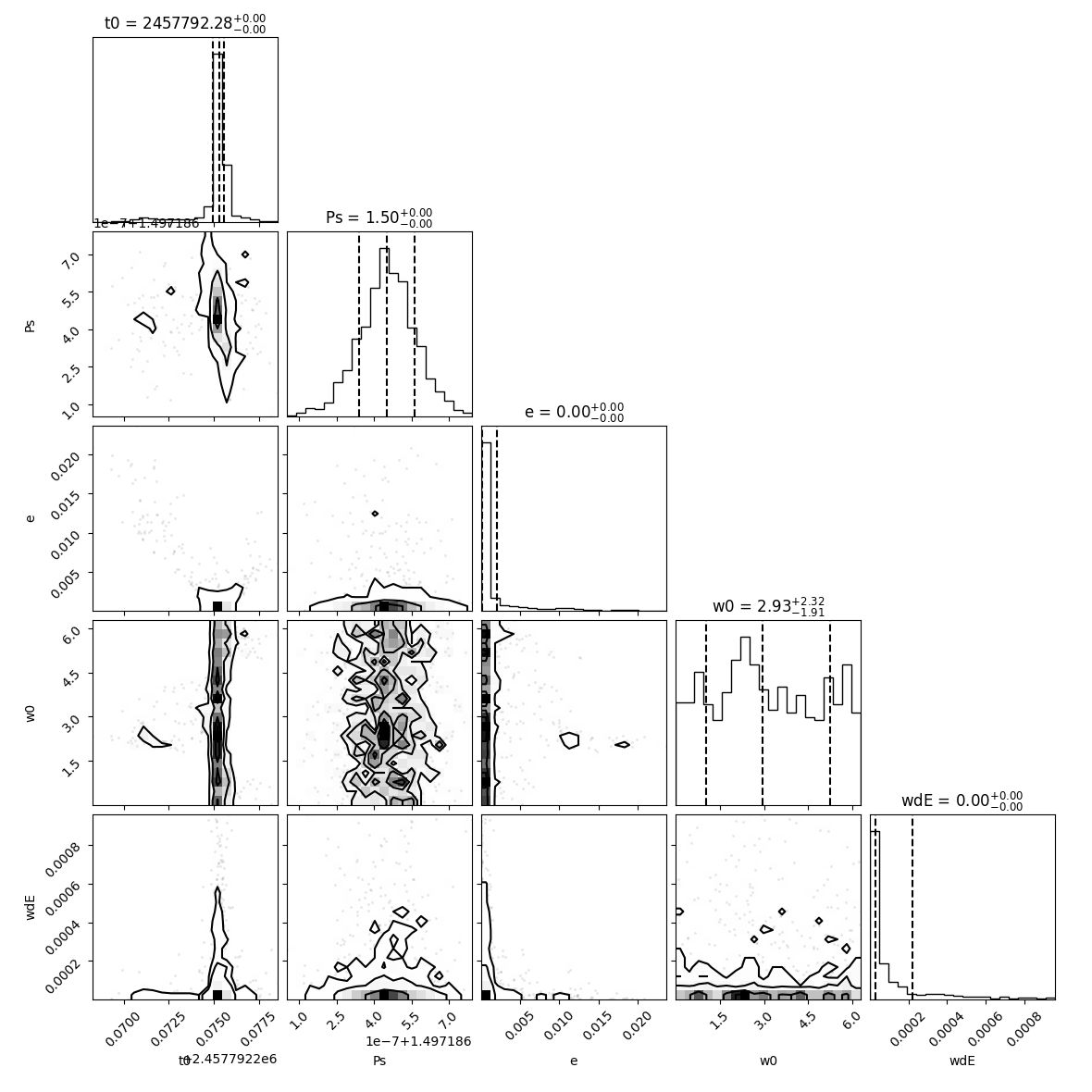}
\caption{The corner plot of the apsidal precession model of TOI-2046b.}
\label{fig:toi2046b-precession_corner}
\end{figure*}
%%%%%%%%%%%%%
\newpage
%%%%%%%
%Priors table

%%%%
\setlength{\tabcolsep}{17pt}
\captionsetup[longtable]{width=13.25cm, skip=3.5pt}
\renewcommand{\arraystretch}{1.08}
\begin{longtable}{lllll}
\caption{The Mid-Transit Times of HAT-P-16} \\
\midrule
\endfirsthead
\multicolumn{5}{@{}l}{\bfseries \tablename\ \thetable{} -- continued} \\
\toprule
Epoch & BJD & Error & $(O - C)$ (m) & Reference \\
\midrule
\endhead
\midrule
\endfoot
\endlastfoot
Epoch & BJD & Error & $(O - C)$ (m) & Reference \\[2pt]
\hline
-339 & 2455027.59293 & 0.00031 & 0.482496 & \cite{buchhave2010hat}$^{\textstyle a}$ \\
-318 & 2455085.88780 & 0.00049 & 0.008712 & \cite{ciceri2013simultaneous} \\
-318 & 2455085.88864 & 0.00006 & 1.448718 & \cite{Sun2023} \\
-314 & 2455096.99125 & 0.00009 & -1.247240 & \cite{Sun2023} \\
-304 & 2455124.75086 & 0.00009 & -0.787134 & \cite{Sun2023} \\
-300 & 2455135.85362 & 0.0005 & -2.043094 & \cite{ciceri2013simultaneous} \\
-300 & 2455135.85449 & 0.00006 & -2.043094 & \cite{Sun2023} \\
-209 & 2455388.46897 & 0.00123 & 0.703866 & ETD \\
-193 & 2455432.88669 & 0.00095 & 4.320036 & ETD \\
-191 & 2455438.43091 & 0.00095 & -7.107944 & ETD \\
-182 & 2455463.41931 & 0.0008 & -0.933849 & ETD \\
-182 & 2455463.42067 & 0.00049 & 1.946151 & ETD \\
-177 & 2455477.30172 & 0.00149 & 3.616204 & ETD \\
-175 & 2455482.85087 & 0.00066 & -0.611775 & ETD \\
-174 & 2455485.62913 & 0.0005 & 2.314236 & ETD \\
-169 & 2455499.50837 & 0.00019 & 1.104289 & \cite{ciceri2013simultaneous} \\
-165 & 2455510.61573 & 0.00095 & 7.048331 & ETD \\
-149 & 2455555.02710 & 0.00063 & 0.584499 & \cite{Sun2023} \\
-62 & 2455796.53707 & 0.00034 & 1.707416 & \cite{ciceri2013simultaneous} \\
-57 & 2455810.41478 & 0.00095 & -0.942530 & ETD \\
-50 & 2455829.84931 & 0.00059 & 2.259543 & ETD \\
-49 & 2455832.62549 & 0.00095 & 2.305554 & ETD \\
-48 & 2455835.40206 & 0.00091 & 3.791565 & ETD \\
-45 & 2455843.72852 & 0.00081 & 2.489596 & ETD \\
-27 & 2455893.69673 & 0.00065 & 3.317786 & \cite{sada2016exoplanet} \\
-23 & 2455904.79696 & 0.00065 & -2.258172 & ETD \\
-20 & 2455913.12765 & 0.00087 & 2.199859 & \cite{Sun2023} \\
-13 & 2455932.55600 & 0.00095 & -3.238068 & ETD \\
-12 & 2455935.32945 & 0.00115 & -7.512057 & ETD \\
0 & 2455968.64736 & 0.00106 & 1.680070 & ETD \\
80 & 2456190.72516 & 0.00059 & 2.480914 & \cite{sada2016exoplanet} \\
84 & 2456201.83039 & 0.000103 & 4.104956 & ETD \\
85 & 2456204.60421 & 0.00032 & 1.270966 & \cite{ciceri2013simultaneous} \\
85 & 2456204.60451 & 0.0003 & 2.710966 & \cite{ciceri2013simultaneous} \\
93 & 2456226.81096 & 0.00057 & 0.199051 & ETD \\
94 & 2456229.58988 & 0.00095 & 4.565061 & ETD \\
97 & 2456237.91484 & 0.00095 & 0.383093 & ETD \\
206 & 2456540.49484 & 0.00054 & -0.361758 & ETD \\
206 & 2456540.49863 & 0.00095 & 5.398241 & ETD \\
211 & 2456554.37172 & 0.00095 & -4.451705 & ETD \\
218 & 2456573.80636 & 0.00057 & -1.249632 & \cite{sada2016exoplanet} \\
224 & 2456590.46081 & 0.00095 & -2.413568 & ETD \\
227 & 2456598.79110 & 0.00107 & 0.604465 & \cite{turner2016ground} \\
227 & 2456598.79150 & 0.00079 & 2.044465 & ETD \\
228 & 2456601.56852 & 0.00066 & 3.530474 & ETD \\
229 & 2456604.34400 & 0.00084 & 2.136485 & ETD \\
238 & 2456629.32549 & 0.00095 & -1.769420 & ETD \\
331 & 2456887.49024 & 0.00095 & -1.810440 & ETD \\
349 & 2456937.45582 & 0.00095 & -3.862251 & ETD \\
358 & 2456962.44878 & 0.00095 & 9.511845 & ETD \\
390 & 2457051.27129 & 0.00095 & -3.415818 & ETD \\
465 & 2457259.47015 & 0.00099 & -1.405027 & ETD \\
492 & 2457334.42244 & 0.00078 & -0.162743 & ETD \\
496 & 2457345.52642 & 0.00095 & 0.021312 & ETD \\
510 & 2457384.39341 & 0.00095 & 4.985447 & ETD \\
519 & 2457409.37539 & 0.00095 & 3.959942 & ETD \\
519 & 2457409.37598 & 0.00095 & 2.519942 & ETD \\
599 & 2457631.44834 & 0.00095 & -3.879615 & ETD \\
599 & 2457631.45192 & 0.00095 & 1.880385 & ETD \\
599 & 2457631.45277 & 0.00095 & 3.320385 & ETD \\
599 & 2457631.45381 & 0.00095 & 4.760385 & ETD \\
603 & 2457642.55557 & 0.00062 & 2.064427 & ETD \\
608 & 2457656.43184 & 0.00095 & -3.465520 & ETD \\
621 & 2457692.52421 & 0.00092 & 2.892617 & ETD \\
626 & 2457706.40229 & 0.00032 & 0.242670 & \cite{Sun2023} \\
626 & 2457706.40419 & 0.00087 & 3.122670 & ETD \\
650 & 2457773.02393 & 0.00014 & -1.533076 & \cite{Sun2023} \\
745 & 2458036.74082 & 0.00068 & -1.482075 & ETD \\
756 & 2458067.27864 & 0.00095 & 1.904042 & ETD \\
858 & 2458350.42812 & 0.00062 & 2.277117 & ETD \\
867 & 2458375.41142 & 0.00063 & 1.251211 & ETD \\
903 & 2458475.34489 & 0.00082 & 0.027648 & ETD \\
907 & 2458486.45540 & 0.00095 & 8.851632 & ETD \\
907 & 2458486.45544 & 0.00095 & 8.851632 & ETD \\
978 & 2458683.54051 & 0.00095 & -2.281618 & ETD \\
996 & 2458733.51150 & 0.00073 & 2.866572 & ETD \\
1001 & 2458747.38887 & 0.00074 & -1.223377 & ETD \\
1001 & 2458747.39105 & 0.00095 & 1.656623 & ETD \\
1008 & 2458766.82093 & 0.00042 & -0.901302 & TESS \\
1008 & 2458766.82141 & 0.00041 & -0.901302 & \cite{Sun2023} \\
1009 & 2458769.59760 & 0.00042 & 0.584709 & TESS \\
1009 & 2458769.59792 & 0.00046 & 0.584709 & \cite{Sun2023} \\
1010 & 2458772.37316 & 0.00043 & -0.809281 & TESS \\
1010 & 2458772.37361 & 0.00043 & 0.630719 & \cite{Sun2023} \\
1012 & 2458777.92488 & 0.00046 & -0.717260 & TESS \\
1012 & 2458777.92492 & 0.00046 & -0.717260 & \cite{Sun2023} \\
1013 & 2458780.70159 & 0.00041 & 0.768751 & TESS \\
1013 & 2458780.70199 & 0.0004 & 0.768751 & \cite{Sun2023} \\
1013 & 2458780.70211 & 0.00068 & 0.768751 & ETD \\
1014 & 2458783.47726 & 0.00041 & -0.625239 & TESS \\
1014 & 2458783.47734 & 0.00041 & -0.625239 & \cite{Sun2023} \\
1015 & 2458786.25349 & 0.00043 & -0.579229 & TESS \\
1015 & 2458786.25400 & 0.00042 & 0.860771 & \cite{Sun2023} \\
1019 & 2458797.35862 & 0.00103 & 2.484814 & \cite{aladaug2021analysis} \\
1024 & 2458811.24399 & 0.00147 & 9.914867 & \cite{aladaug2021analysis} \\
1028 & 2458822.34006 & 0.00095 & -1.421091 & ETD \\
1028 & 2458822.34086 & 0.00055 & 0.018864 & ETD \\
1033 & 2458836.22609 & 0.00123 & 7.448962 & \cite{aladaug2021analysis} \\
1126 & 2459094.38499 & 0.00095 & -1.232058 & ETD \\
1130 & 2459105.49026 & 0.00049 & 0.391984 & ETD \\
1139 & 2459130.47387 & 0.00066 & 0.806079 & ETD \\
1139 & 2459130.47461 & 0.00052 & 2.246079 & \cite{aladaug2021analysis} \\
1144 & 2459144.34689 & 0.00076 & -9.043868 & \cite{aladaug2021analysis} \\
1148 & 2459155.45802 & 0.00062 & 1.220174 & ETD \\
1262 & 2459471.91470 & 0.00095 & -3.614625 & ETD \\
1268 & 2459488.56889 & 0.00095 & -6.218561 & ETD \\
1269 & 2459491.34883 & 0.00073 & -0.412550 & ETD \\
1282 & 2459527.43659 & 0.00046 & 0.185587 & ADYU60 \\
1283 & 2459530.21325 & 0.00095 & 0.231597 & ADYU60 \\
1287 & 2459541.31699 & 0.00053 & 0.415639 & ADYU60 \\
1287 & 2459541.31793 & 0.00044 & 1.855639 & ETD \\
1291 & 2459552.42074 & 0.0006 & 0.599682 & ETD \\
1292 & 2459555.19589 & 0.00052 & -0.794308 & ADYU60 \\
1296 & 2459566.29983 & 0.00072 & -0.610266 & ETD \\
1371 & 2459774.49881 & 0.00095 & 1.400525 & ETD \\
1400 & 2459855.00057 & 0.00041 & -0.145169 & TESS \\
1401 & 2459857.77686 & 0.00042 & -0.099216 & TESS \\
1405 & 2459868.88066 & 0.00038 & 0.084816 & TESS \\
1406 & 2459871.65696 & 0.00043 & 0.130896 & TESS \\
1407 & 2459874.43237 & 0.0004 & -1.263096 & TESS \\
1408 & 2459877.20710 & 0.0011 & -2.657085 & ADYU60 \\
1408 & 2459877.20803 & 0.00041 & -1.217085 & TESS \\
1409 & 2459879.98447 & 0.00038 & -1.171074 & TESS \\
1412 & 2459888.30917 & 0.00064 & -5.353043 & ADYU60 \\
1430 & 2459938.28067 & 0.00067 & 1.235147 & ADYU60 \\
1509 & 2460157.58007 & 0.0009 & -2.330021 & ETD \\
1523 & 2460196.44191 & 0.00095 & -4.565873 & ETD \\
1523 & 2460196.44406 & 0.00055 & -1.685873 & ETD \\
1546 & 2460260.28758 & 0.00062 & -6.387630 & ADYU60 \\
1559 & 2460296.38158 & 0.00095 & 2.850508 & ETD \\
1568 & 2460321.36196 & 0.00095 & -2.495398 & ETD \\
1639 & 2460518.45422 & 0.00095 & -4.988649 & ETD \\
1652 & 2460554.54710 & 0.0012 & 2.809488 & ADYU60 \\
1657 & 2460568.42272 & 0.00095 & 1.599540 & ADYU60 \\
1657 & 2460568.42287 & 0.00095 & -2.720460 & ETD \\
1657 & 2460568.42602 & 0.00071 & -2.720460 & ETD \\
1661 & 2460579.52851 & 0.00066 & 0.343583 & ADYU60 \\
1670 & 2460604.50915 & 0.00095 & -5.002322 & ETD \\
1670 & 2460604.51076 & 0.00045 & -2.122322 & ETD \\
1670 & 2460604.51076 & 0.00095 & -2.122322 & ETD \\
1675 & 2460618.38951 & 0.00057 & -3.332269 & ETD \\
1684 & 2460643.38005 & 0.00095 & 5.721825 & ETD \\
\bottomrule
\multicolumn{5}{@{}l}{\textit{$^{\textstyle a}$}: Mid-transit times calculated from observations of \cite{buchhave2010hat}}
\label{tab:oc_hatp16b}
\end{longtable}
%%%%%

%%%%%
\setlength{\tabcolsep}{16.8pt}
\captionsetup[longtable]{width=13.6cm, skip=3.5pt}
\renewcommand{\arraystretch}{1.08}
\begin{longtable}{lllll}
\caption{The Mid-Transit Times of TOI-1516: 1 point from \cite{fox2022neossat}} \\ %, 48 points from Exoplanet Transit Database (ETD), 42 points from TESS and 16 points from this work. Full version of the table is available online.} \\
\midrule
\endfirsthead
\multicolumn{5}{@{}l}{\bfseries \tablename\ \thetable{} -- continued} \\
\toprule
Epoch & BJD & Error & $(O - C)$ (m) & Reference \\
\midrule
\endhead
\midrule
\endfoot
\endlastfoot
 Epoch & BJD & Error & $(O - C)$ (m) & Reference \\[2pt]
\hline
0 & 2458765.32500 & 0.0001 & 0 & \cite{kabath2022toi}$^{\textstyle a}$ \\ 
147 & 2459067.56135 & 0.0005 & 2.647751 & ETD \\
161 & 2459096.34328 & 0.00032 & -0.507837 & ADYU60 \\
161 & 2459096.34647 & 0.0014 & 3.812162 & ETD \\
163 & 2459100.45952 & 0.0005 & 6.652793 & ETD \\
163 & 2459100.46052 & 0.00071 & 8.092793 & ETD \\
164 & 2459102.51079 & 0.0003 & -0.566893 & ADYU60 \\
171 & 2459116.90477 & 0.00085 & 2.175313 & ETD \\
177 & 2459129.23926 & 0.00034 & -0.822797 & ADYU60 \\
178 & 2459131.29623 & 0.00031 & 0.597518 & ADYU60 \\
179 & 2459133.35265 & 0.00038 & 2.017833 & ETD \\
180 & 2459135.40748 & 0.00032 & -0.881852 & ADYU60 \\
181 & 2459137.46344 & 0.00031 & -0.901536 & ADYU60 \\
195 & 2459166.24767 & 0.00028 & 0.262874 & ADYU60 \\
197 & 2459170.36012 & 0.00044 & 0.223504 & ADYU60 \\
250 & 2459279.32378 & 0.00041 & -6.579798 & ETD \\
250 & 2459279.32493 & 0.00085 & -5.139798 & ETD \\
251 & 2459281.38144 & 0.00123 & -5.159482 & ETD \\
252 & 2459283.43537 & 0.00036 & -8.059167 & ETD \\
270 & 2459320.44741 & 0.0005 & -2.654696 & ETD \\
289 & 2459359.51601 & 0.0008 & 4.172489 & ETD \\
305 & 2459392.41028 & 0.00042 & 1.4544 & \cite{fox2022neossat} \\
320 & 2459423.25986 & 0.00664 & 15.082256 & ETD \\
341 & 2459466.43082 & 0.00072 & 7.468873 & ETD \\
354 & 2459493.15221 & 0.0003 & -2.867032 & ETD \\
394 & 2459575.38995 & 0.0011 & -6.534429 & ETD \\
433 & 2459655.57499 & 0.00067 & -5.862143 & ETD \\
482 & 2459756.32211 & 0.00084 & -2.506704 & ADYU60 \\
484 & 2459760.43840 & 0.0006 & 4.653925 & ETD \\
484 & 2459760.43908 & 0.00039 & 3.213925 & ETD \\
487 & 2459766.60593 & 0.00055 & 3.154870 & ETD \\
503 & 2459799.50430 & 0.00054 & 5.719912 & ETD \\
504 & 2459801.55704 & 0.00073 & 5.700227 & ETD \\
504 & 2459801.55988 & 0.00055 & 1.380227 & ETD \\
504 & 2459801.56020 & 0.00041 & 5.700227 & ETD \\
519 & 2459832.39975 & 0.00046 & 5.404952 & ETD \\
530 & 2459855.01226 & 0.00028 & -0.571582 & TESS \\
531 & 2459857.06806 & 0.00027 & -0.591267 & TESS \\
532 & 2459859.12421 & 0.00027 & -0.610951 & TESS \\
533 & 2459861.18022 & 0.00029 & -0.630637 & TESS \\
534 & 2459863.23502 & 0.00049 & -2.090322 & TESS \\
536 & 2459867.34767 & 0.00099 & -0.689692 & ADYU60 \\
537 & 2459869.40430 & 0.001 & -0.709377 & ADYU60 \\
537 & 2459869.40473 & 0.00029 & 0.730623 & TESS \\
538 & 2459871.46074 & 0.0003 & 0.710938 & TESS \\
539 & 2459873.51627 & 0.0003 & -0.748747 & TESS \\
540 & 2459875.57285 & 0.00027 & 0.671568 & TESS \\
541 & 2459877.62873 & 0.00028 & 0.651884 & TESS \\
542 & 2459879.68442 & 0.00027 & -0.808041 & TESS \\
543 & 2459881.74076 & 0.00029 & 0.612513 & TESS \\
544 & 2459883.79604 & 0.00028 & -0.846772 & TESS \\
545 & 2459885.85266 & 0.00029 & 0.573144 & TESS \\
546 & 2459887.90862 & 0.0003 & 0.553458 & TESS \\
547 & 2459889.96452 & 0.00027 & 0.533974 & TESS \\
548 & 2459892.02139 & 0.00028 & 0.514089 & TESS \\
549 & 2459894.07661 & 0.00029 & 0.494404 & TESS \\
551 & 2459898.18719 & 0.00084 & -2.424966 & ADYU60 \\
551 & 2459898.18927 & 0.00029 & 0.455033 & TESS \\
552 & 2459900.24472 & 0.00027 & 0.435349 & TESS \\
553 & 2459902.30041 & 0.00029 & -1.024336 & TESS \\
554 & 2459904.35703 & 0.00026 & 0.395979 & TESS \\
555 & 2459906.41283 & 0.00029 & 0.376295 & TESS \\
556 & 2459908.46917 & 0.00028 & 0.356609 & TESS \\
590 & 2459978.37156 & 0.00089 & -1.752679 & ETD \\
627 & 2460054.44524 & 0.00057 & -1.041022 & ETD \\
681 & 2460165.46916 & 0.00064 & -2.104008 & ETD \\
682 & 2460167.52548 & 0.00069 & -2.123693 & ETD \\
682 & 2460167.52563 & 0.00074 & -0.683693 & ETD \\
700 & 2460204.53354 & 0.00089 & -1.038023 & ETD \\
733 & 2460272.37595 & 0.00113 & -10.327625 & ETD \\
734 & 2460274.43667 & 0.00089 & -3.147310 & ETD \\
752 & 2460311.44626 & 0.00086 & -2.061639 & ETD \\
793 & 2460395.74425 & 0.00027 & 0.011275 & TESS \\
794 & 2460397.79987 & 0.00029 & -0.008410 & TESS \\
795 & 2460399.85574 & 0.00033 & -0.028080 & TESS \\
796 & 2460401.91167 & 0.00032 & -0.047808 & TESS \\
797 & 2460403.96822 & 0.00031 & -0.067392 & TESS \\
798 & 2460406.02452 & 0.0003 & 1.352853 & TESS \\
804 & 2460418.35995 & 0.0003 & -0.205258 & TESS \\
805 & 2460420.41576 & 0.00036 & -0.224941 & TESS \\
806 & 2460422.47251 & 0.00029 & 1.195373 & TESS \\
812 & 2460434.80866 & 0.00029 & 1.077263 & TESS \\
813 & 2460436.86423 & 0.00027 & -0.382421 & TESS \\
814 & 2460438.92025 & 0.0003 & -0.402107 & TESS \\
815 & 2460440.97678 & 0.00028 & 1.018208 & TESS \\
816 & 2460443.03268 & 0.00032 & 0.998524 & TESS \\
817 & 2460445.08767 & 0.00027 & -0.461162 & TESS \\
818 & 2460447.14447 & 0.00031 & -0.480846 & TESS \\
819 & 2460449.20001 & 0.00029 & -0.500530 & TESS \\
820 & 2460451.25664 & 0.00029 & 0.919784 & TESS \\
823 & 2460457.42144 & 0.00172 & -4.899270 & ETD \\
841 & 2460494.43097 & 0.00064 & -2.373600 & ADYU60 \\
842 & 2460496.48688 & 0.0009 & -2.393284 & ETD \\
842 & 2460496.48793 & 0.0007 & -0.953284 & ETD \\
858 & 2460529.38505 & 0.0015 & 0.171756 & ETD \\
859 & 2460531.43939 & 0.00055 & -2.727929 & ADYU60 \\
860 & 2460533.49840 & 0.0011 & 1.572387 & ADYU60 \\
860 & 2460533.50165 & 0.0015 & 7.332387 & ETD \\
861 & 2460535.55309 & 0.00055 & 0.112752 & ETD \\
877 & 2460568.44738 & 0.00103 & -3.082258 & ETD \\
878 & 2460570.50226 & 0.00035 & -4.541943 & ADYU60 \\
894 & 2460603.40296 & 0.00059 & 2.343098 & ETD \\
895 & 2460605.45946 & 0.00054 & 2.323413 & ETD \\
896 & 2460607.51041 & 0.00055 & -4.896272 & ETD \\
911 & 2460638.33794 & 0.006151 & 2.008454 & ETD \\
911 & 2460638.35494 & 0.00046 & -22.471546 & ETD \\
912 & 2460640.40885 & 0.00027 & -0.891230 & ETD \\
947 & 2460712.36931 & 0.000558 & -1.580204 & ETD \\
\bottomrule
\multicolumn{5}{l}{\textit{$^{\textstyle a}$}: Mid-transit times calculated from observations of \cite{kabath2022toi}}
\label{tab:oc_toi1516b}
\end{longtable}
%\twocolumn
%%%%%%

%%%%%%
\setlength{\tabcolsep}{21.3pt}
\captionsetup[longtable]{width=14.4cm, skip=3.5pt}
\renewcommand{\arraystretch}{1.08}
\begin{longtable}{lllll}
\caption{The Mid-Transit Times of TOI-2046: 1 point from \cite{fox2022neossat}}\\
%48 points from Exoplanet Transit Database (ETD), 42 points from TESS and 11 points from this work. Full version of the table is available online.} \\
\midrule
\endfirsthead
\multicolumn{5}{@{}l}{\bfseries \tablename\ \thetable{} -- continued} \\
\toprule
Epoch & BJD & Error & $(O - C)$ (m) & Reference \\
\midrule
\endhead
\midrule
\endfoot
\endlastfoot
 Epoch & BJD & Error & $(O - C)$ (m) & Reference \\[2pt]
\hline
0 & 2457792.27670 & 0.0023 & 0 & \cite{kabath2022toi}$^{\textstyle a}$ \\
901 & 2459141.24135 & 0.00077 & 1.005679 & ADYU60 \\
1185 & 2459566.446300 & 0.0016 & 9.10512 & TUG-T100 \\
1185 & 2459566.447080 & 0.001 & 10.22832 & TUG-T100 \\
1090 & 2459424.21418 & 0.000854 & 9.648 & \cite{fox2022neossat} \\
1396 & 2459882.34880 & 0.0026 & 2.025705 & TESS \\
1397 & 2459883.84525 & 0.00043 & 0.317222 & TESS \\
1398 & 2459885.34088 & 0.00121 & -1.391263 & ETD \\
1398 & 2459885.34237 & 0.00045 & 0.048672 & TESS \\
1399 & 2459886.83923 & 0.00043 & -0.219749 & TESS \\
1400 & 2459888.33641 & 0.00044 & -0.488233 & TESS \\
1401 & 2459889.83418 & 0.00041 & 0.683281 & TESS \\
1402 & 2459891.33106 & 0.00039 & 0.414796 & TESS \\
1403 & 2459892.82733 & 0.00038 & -1.293687 & TESS \\
1404 & 2459894.32521 & 0.0004 & -0.122112 & TESS \\
1405 & 2459895.82310 & 0.00045 & 1.049342 & TESS \\
1406 & 2459897.31867 & 0.00044 & -0.659142 & TESS \\
1407 & 2459898.81693 & 0.00041 & 0.512374 & TESS \\
1408 & 2459900.31471 & 0.00042 & 1.683888 & TESS \\
1409 & 2459901.81065 & 0.00043 & -0.024624 & TESS \\
1410 & 2459903.30717 & 0.0004 & -1.733082 & TESS \\
1411 & 2459904.80540 & 0.0004 & -0.561566 & TESS \\
1412 & 2459906.30220 & 0.00041 & -0.830050 & TESS \\
1413 & 2459907.80009 & 0.0004 & 0.341464 & TESS \\
1414 & 2459909.29663 & 0.00044 & 0.073008 & TESS \\
1416 & 2459912.29218 & 0.0004 & 0.976009 & TESS \\
1417 & 2459913.78851 & 0.00045 & 0.707525 & TESS \\
1418 & 2459915.28560 & 0.0004 & 0.439040 & TESS \\
1420 & 2459918.28034 & 0.00037 & -0.097920 & TESS \\
1421 & 2459919.77778 & 0.00039 & 1.073586 & TESS \\
1422 & 2459921.27400 & 0.0004 & -0.634899 & TESS \\
1423 & 2459922.77034 & 0.00048 & -2.343384 & TESS \\
1425 & 2459925.76636 & 0.00042 & -0.000352 & TESS \\
1426 & 2459927.26341 & 0.00048 & -0.268838 & TESS \\
1427 & 2459928.76035 & 0.00043 & -0.537323 & TESS \\
1429 & 2459931.75508 & 0.00041 & 0.365707 & TESS \\
1430 & 2459933.25201 & 0.00044 & 0.097200 & TESS \\
1430 & 2459933.25379 & 0.00052 & 2.977224 & ADYU60 \\
1431 & 2459934.74878 & 0.00039 & -0.171262 & TESS \\
1432 & 2459936.24588 & 0.00045 & -0.439747 & TESS \\
1432 & 2459936.24618 & 0.00066 & -0.439747 & ADYU60 \\
1436 & 2459942.23363 & 0.00091 & -1.513686 & ADYU60 \\
1440 & 2459948.21985 & 0.00062 & -0.781564 & ADYU60 \\
1444 & 2459954.21202 & 0.00068 & -2.929442 & ADYU60 \\
1452 & 2459966.18800 & 0.0018 & -0.586413 & ADYU60 \\
1454 & 2459969.18410 & 0.0027 & 0.414194 & ETD \\
1461 & 2459979.66476 & 0.00085 & 0.243285 & ETD \\
1467 & 2459988.64767 & 0.0008 & -4.418532 & ETD \\
1479 & 2460006.61146 & 0.00086 & -0.245197 & ETD \\
1501 & 2460039.55227 & 0.001 & 4.465106 & ETD \\
1521 & 2460069.49862 & 0.00077 & -4.150650 & ETD \\
1537 & 2460093.44803 & 0.00154 & -0.051840 & ETD \\
1539 & 2460096.44071 & 0.00048 & 0.704510 & ETD \\
1561 & 2460129.386800 & 0.001 & 8.01187 & TUG-T100 \\
1561 & 2460129.386500 & 0.001 & 7.57987 & TUG-T100 \\
1583 & 2460162.321980 & 0.00071 & 3.87562 & TUG-T100 \\
1583 & 2460162.321000 & 0.00085 & 2.46442 & TUG-T100 \\
1688 & 2460319.52553 & 0.00051 & 2.876632 & ETD \\
1712 & 2460355.45894 & 0.00037 & -3.420337 & ETD \\
1720 & 2460367.43804 & 0.00124 & -0.882154 & ETD \\
1722 & 2460370.42848 & 0.00041 & -4.836095 & ETD \\
1734 & 2460388.39557 & 0.00045 & 0.606332 & TESS \\
1738 & 2460394.38157 & 0.00053 & 1.509362 & TESS \\
1766 & 2460436.30666 & 0.00042 & -0.199123 & TESS \\
1768 & 2460439.30204 & 0.00042 & -0.467608 & TESS \\
1769 & 2460440.79824 & 0.00041 & 2.143908 & TESS \\
1770 & 2460442.29541 & 0.00044 & 0.166938 & TESS \\
1771 & 2460443.79367 & 0.00041 & 1.338454 & TESS \\
1773 & 2460446.78660 & 0.00047 & 1.069969 & TESS \\
1774 & 2460448.28504 & 0.00051 & 0.801484 & TESS \\
1775 & 2460449.78183 & 0.00051 & 0.532999 & TESS \\
1776 & 2460451.27904 & 0.00044 & 0.264514 & TESS \\
1777 & 2460452.77550 & 0.00043 & -0.003974 & TESS \\
1778 & 2460454.27326 & 0.00046 & -0.272455 & TESS \\
1779 & 2460455.77038 & 0.00057 & 0.899060 & TESS \\
1780 & 2460457.26694 & 0.00052 & 0.630575 & TESS \\
1781 & 2460458.76453 & 0.00049 & 0.362091 & TESS \\
1782 & 2460460.26162 & 0.00052 & 0.093600 & TESS \\
1783 & 2460461.75896 & 0.00058 & -0.174879 & TESS \\
1784 & 2460463.25577 & 0.00053 & 0.996637 & TESS \\
1785 & 2460464.75294 & 0.00044 & 0.728152 & TESS \\
1786 & 2460466.25133 & 0.00041 & 0.459667 & TESS \\
1787 & 2460467.74834 & 0.00043 & 0.191182 & TESS \\
1788 & 2460469.24546 & 0.00057 & -1.517303 & TESS \\
1789 & 2460470.74239 & 0.00056 & 1.094208 & TESS \\
1790 & 2460472.23825 & 0.00046 & 0.825728 & TESS \\
1791 & 2460473.73709 & 0.00053 & -0.882758 & TESS \\
1792 & 2460475.23368 & 0.00049 & -1.151242 & TESS \\
1793 & 2460476.73034 & 0.00049 & 4.461489 & ADYU60 \\
1794 & 2460478.22745 & 0.00041 & -0.029376 & ETD \\
1875 & 2460599.50260 & 0.00067 & 0.873611 & ADYU60 \\
1881 & 2460608.48317 & 0.00055 & 1.776642 & ADYU60 \\
1883 & 2460611.47786 & 0.00048 & -3.080328 & ETD \\
1885 & 2460614.47297 & 0.00089 & -3.080328 & ETD \\
1887 & 2460617.46405 & 0.000491 & -1.640328 & ETD \\
1887 & 2460617.46406 & 0.00049 & -0.200328 & ETD \\
1887 & 2460617.46515 & 0.000809 & -0.200328 & ETD \\
1887 & 2460617.46551 & 0.000463 & -0.200328 & ETD \\
1887 & 2460617.46578 & 0.000683 & -0.200328 & ETD \\
1887 & 2460617.46585 & 0.000383 & -0.200328 & ETD \\
1887 & 2460617.46585 & 0.000625 & 1.239672 & ETD \\
1887 & 2460617.46630 & 0.001405 & -0.371238 & ADYU60 \\
1887 & 2460617.46702 & 0.000665 & -2.348206 & ADYU60 \\
1893 & 2460626.44918 & 0.00086 & 0.726945 & ETD \\
1895 & 2460629.44232 & 0.0006 & -3.934872 & ETD \\
1905 & 2460644.41610 & 0.000537 & 0.214219 & ETD \\
1917 & 2460662.37942 & 0.000438 & 1.654219 & ETD \\
1923 & 2460671.36471 & 0.000462 & -2.836690 & ETD \\
1923 & 2460671.36567 & 0.000597 & -1.396690 & ETD \\
1925 & 2460674.36335 & 0.000989 & -3.373660 & ETD \\
1929 & 2460680.34641 & 0.000497 & 0.946341 & ETD \\
1929 & 2460680.34728 & 0.000602 & 0.946341 & ETD \\
1931 & 2460683.34048 & 0.001096 & -0.127584 & ETD \\
1931 & 2460683.34265 & 0.000347 & -0.664568 & ETD \\
1931 & 2460683.34327 & 0.000293 & 3.118463 & ETD \\
1935 & 2460689.33057 & 0.000523 & -2.446385 & ETD \\
1937 & 2460692.32463 & 0.000553 & -1.006385 & ETD \\
1939 & 2460695.32178 & 0.000455 & 1.873614 & ETD \\
1949 & 2460710.28680 & 0.001847 & 3.313614 & ETD \\
1949 & 2460710.29012 & 0.000583 & 3.313614 & ETD \\
\bottomrule
\multicolumn{5}{@{}l}{\textit{$^{\textstyle a}$}: Mid-transit times calculated from observations of \cite{kabath2022toi}}
\label{tab:oc_toi2046b}
\end{longtable}



\end{document}